\documentclass{article}

\usepackage{PRIMEarxiv}

\usepackage[utf8]{inputenc} 
\usepackage[T1]{fontenc}    
\usepackage{hyperref}       
\usepackage{url}            
\usepackage{booktabs}       
\usepackage{amsfonts}       
\usepackage{nicefrac}       
\usepackage{microtype}      
\usepackage{lipsum}
\usepackage{amsmath}
\usepackage{fancyhdr}       
\usepackage{graphicx}       
\usepackage{amssymb}

\usepackage{algorithmic}
\usepackage{algorithm} 
\usepackage{float}
\graphicspath{{media/}}     

\title{CICDWOA: A Collective Cognitive Sharing Whale Optimization Algorithm with Cauchy Inverse Cumulative Distribution for 2D/3D Path Planning and Engineering Design Problems
\thanks{\textit{\underline{Citation}}: 
\textbf{Authors. Title. Pages.... DOI:000000/11111.}} 
}

\author{
  Junhao Wei, Yanxiao Li \\
  Faculty of Applied Sciences \\
  Macao Polytechnic University \\
  Macao, China\\
  \texttt{\{p2312195,p2525981\}@mpu.edu.mo}  
  \And
   Seyedali Mirjalili   \\
   Torrens University\\
   Brisbane, Australia\\
   \texttt{ali.mirjalili@gmail.com} 
   \And
   Dexing Yao, Yifu Zhao, Haochen Li, Xudong Ye  \\
   Faculty of Applied Sciences \\
   Macao Polytechnic University \\
   Macao, China\\
   \texttt{\{p2522978,p2523269,p2523372,p2525593\}@mpu.edu.mo}  
   \And
   Zikun Li  \\
   School of Economics and Management \\
   South China Normal University \\
   Guangzhou, China\\
   \texttt{18520610821@163.com} 
   \And
   Qingyang Xu \\
   Faculty of Applied Sciences \\
   Macao Polytechnic University \\
   Macao, China\\
   \texttt{p2311371@mpu.edu.mo} 
   \And
   Baili Lu \\
   College of Animal Science and Technology \\
   Zhongkai University of Agriculture and Engineering \\
   Guangzhou, China\\
   \texttt{18023304003@163.com} 
   \And
   Ngai Cheong \\
   Faculty of Applied Sciences \\
   Macao Polytechnic University \\
   Macao, China\\
   \texttt{ncheong@mpu.edu.mo} 
   \And
   Dengcheng Yang\\
   Information Engineering School \\
   Nanchang University \\
   Nanchang, China\\
   \texttt{yangdingcheng@ncu.edu.cn} 
   \And
   Sio-Kei Im\\
   Macao Polytechnic University \\
   Macao, China\\
   \texttt{marcusim@mpu.edu.mo} \\
   \And
   Yapeng Wang, Xu Yang*\\
   Faculty of Applied Sciences \\
   Macao Polytechnic University \\
   Macao, China\\
   \texttt{\{yapengwang,xuyang\}@mpu.edu.mo} \\
}

\begin{document}
\maketitle

\begin{abstract}
The Whale Optimization Algorithm (WOA) has shown strong optimization ability but still suffers from premature convergence and weak search diversity. To address these issues, this paper proposes an enhanced WOA variant called CICDWOA. The proposed algorithm introduces a Good Nodes Set (GNS) method for uniform population initialization, a Collective Cognitive Sharing (CCS) mechanism to enhance group collaboration, and an Enhanced Spiral Updating strategy based on the Cauchy Inverse Cumulative Distribution (CICD) to strengthen global exploration and local exploitation balance. In addition, a nonlinear convergence factor and a Hybrid Gaussian-Cauchy mutation based on Differential Evolution (DE) further improve convergence efficiency and population diversity. CICDWOA was evaluated on 23 benchmark functions, 2D robot path planning problems, 3D UAV path planning tasks and 10 engineering design problems. Statistical experiment results show that CICDWOA achieves faster convergence, higher accuracy, and better robustness than classical WOA and other advanced metaheuristic algorithms. CICDWOA gained average Friedman value of 1.6790, ranking first among the SOTA algorithms. And the results of engineering simulations confirm that CICDWOA provides an effective and general framework for solving complex optimization and engineering problems. The code of CICDWOA are available on \href{URL}{https://github.com/JunhaoWei-mpu/ROBIS-Lab/tree/CICDWOA}.
\end{abstract}


\section{Introduction}
In modern science and engineering, optimization is a fundamental and central research task. Whether minimizing loss functions and tuning hyperparameters in machine learning, optimizing strength, weight, and energy consumption in mechanical and structural design, or managing resource scheduling and cost control in industrial production, optimization problems are embedded throughout the analysis and design of complex systems. The goal of optimization is to seek the optimal or near-optimal solution of an objective function under given constraints. However, most real-world optimization problems exhibit characteristics such as high dimensionality, nonlinearity, multi-modality, and uncertainty, making them computationally expensive and difficult to solve. Traditional analytical methods often fail to provide satisfactory solutions under such conditions.\par
In early practices, optimization problems were often addressed through empirical rules, trial-and-error procedures, or exhaustive searches. Although intuitive, these methods are extremely inefficient and become infeasible as problem dimensionality increases. With the advancement of computer science and numerical analysis, researchers developed various deterministic algorithms based on mathematical models, such as the Steepest Descent (SD), Conjugate Gradient (CG), and Quasi-Newton (QN) methods. These algorithms rely on gradient or Hessian information and iteratively approach the optimal solution, exhibiting high convergence efficiency for differentiable and convex objective functions. However, many real-world optimization problems do not satisfy these ideal assumptions. When faced with non-convex, discontinuous, noisy, or black-box objective functions, the performance of traditional methods deteriorates sharply or even fails completely, often being trapped in local optima or stagnating.\par
To overcome these limitations, researchers began to draw inspiration from natural and social phenomena, leading to the development of a new class of intelligent optimization techniques known as Metaheuristic Algorithms (MAs). These algorithms eliminate the dependence on gradient information and analytical formulations, emphasizing heuristic search and population-based cooperation to approximate the global optimum. Metaheuristic algorithms typically start with a randomly generated set of initial solutions and iteratively update the population through information sharing, adaptive adjustment, and strategy balancing, progressively approaching the global optimum. Due to their independence from explicit objective formulations, strong global search ability, and ease of implementation, metaheuristic algorithms have become powerful tools for solving complex nonlinear optimization problems. They have been widely applied in path planning, structural optimization, energy scheduling, hyperparameter tuning, and complex system modeling.\par
As research deepened, metaheuristic algorithms evolved into diverse theoretical and application branches. Based on their underlying inspiration mechanisms, they can generally be categorized into four groups: Evolutionary Algorithms (EAs), Swarm Intelligence Algorithms (SIAs), Physics-Based Algorithms (PBAs), and Human-Based Algorithms (HBAs). Figure ~\ref{standardMAs} gives the details of the representative algorithms of each group.\par
Evolutionary Algorithms are inspired by Darwin's theory of natural selection. Representative examples include the Genetic Algorithm (GA) \cite{GA}, Differential Evolution (DE) \cite{DE}, and Evolution Strategy (ES), which evolve populations through selection, crossover, and mutation operators, demonstrating strong global exploration capability.\par
Swarm Intelligence Algorithms emulate the social cooperation behaviors of animal groups. For instance, Particle Swarm Optimization (PSO) mimics the collaborative foraging behavior of bird flocks \cite{PSO}; the Ant Colony Optimization (ACO) algorithm simulates pheromone-based communication among ants \cite{ACO}; the Grey Wolf Optimizer (GWO) models the hierarchical hunting strategy of wolf packs \cite{GWO}; and the Whale Optimization Algorithm (WOA) is inspired by the bubble-net hunting behavior of humpback whales, showing remarkable convergence speed and robustness \cite{WOA}. Recently proposed algorithms further expand this category. The Philoponella Prominens Optimizer (PPO) is inspired by the unique mating and predatory behaviors of the spider Philoponella prominens, simulating post-mating escape strategies, sexual cannibalism, and dynamic game mechanisms during hunting. The Kangaroo Escape Optimizer (KEO) draws on the survival and escape strategies of kangaroos in complex environments, modeling their jumping, avoidance, and energy-adjustment behaviors to balance global exploration and local exploitation \cite{KEO}. The Tetragonula Carbonaria Optimization Algorithm (TGCOA) is inspired by the spiral nest-building and collective thermoregulation behaviors of the stingless bee Tetragonula carbonaria, simulating the colony's adaptive coordination and structural expansion under environmental changes to achieve dynamic optimization stability \cite{TGCOA}.\par
Physics-Based Algorithms are rooted in natural laws of energy transfer and physical dynamics. For example, Simulated Annealing (SA) controls the trade-off between global and local search through a temperature-cooling process \cite{SA}; the Black Hole (BH) algorithm mimics gravitational attraction mechanisms \cite{BH}; the Attraction-Repulsion Optimization Algorithm (AROA) models gravitational and repulsive forces guiding solutions toward optima; and the Cloud Drift Optimization (CDO) algorithm draws inspiration from the dynamic drifting process of clouds in atmospheric flow fields, simulating the collective movement of cloud particles under airflow, gravity, and environmental perturbations to achieve adaptive balance between exploration and exploitation \cite{CDO}.\par
On the other hand, Human-Based Algorithms are inspired by social learning and cooperative behaviors. The Harmony Search (HS) algorithm imitates the improvisational process of musicians \cite{HS}; the Teaching-Learning-Based Optimization (TLBO) algorithm models classroom teaching and self-learning processes \cite{TLBO}; and the Soccer League Competition Algorithm (SLCA) is based on competitive team dynamics \cite{SLCA}. The Hiking Optimization Algorithm (HOA) takes inspiration from hikers navigating complex terrains to reach mountain peaks, simulating individuals dynamically adjusting their steps and directions based on terrain steepness and path feasibility to achieve efficient local and global search \cite{HOA}. The Hannibal Barca Optimizer (HBO), inspired by the strategic thinking of the ancient Carthaginian general Hannibal Barca, incorporates the concept of pincer tactics and dynamic learning mechanisms to simulate tactical coordination and information balance in multidimensional decision-making, effectively addressing complex optimization problems \cite{HBO}. Through diverse inspiration sources and adaptive search mechanisms, these algorithms have achieved remarkable results across a wide range of scientific and engineering fields.\par
\begin{table}[htbp]
	\centering
	\caption{Current research on standard metaheuristic algorithms.}
	\resizebox{\textwidth}{!}{ 
	\begin{tabular}{lllll}
	\toprule
	Algorithm & Type & Year & Author(s) & Source of Inspiration \\
	\midrule
Genetic Algorithm (GA) \cite{GA} & EAs & 1992 & Holland et al. & Theory of evolution and Mendelian genetics. \\
Differential Evolution (DE) \cite{DE} & EAs & 1995 & Storn et al. & Mutation, crossover, and selection mechanisms. \\ \midrule
Ant Colony Optimization (ACO) \cite{ACO} & SIAs & 1992 & Dorigo et al. & Foraging behavior of ants. \\
Particle Swarm Optimization (PSO) \cite{PSO} & SIAs & 1995 & Kennedy et al. & Social foraging behavior of birds. \\
Grey Wolf Optimizer (GWO) \cite{GWO} & SIAs & 2014 & Mirjalili & Social hierarchy and hunting behavior of \\
 &  &  &  &  grey wolves. \\
Whale Optimization Algorithm (WOA) \cite{WOA} & SIAs & 2016  & Mirjalili et al. & Hunting behavior of humpback whales.\\
Philoponella Prominens Optimizer (PPO) \cite{PPO} & SIAs & 2025 & Y Gao et al. & Mating behavior of the P. prominens. \\
Kangaroo Escape Optimizer (KEO) \cite{KEO} & SIAs & 2025 & SZ Almutairi et al. & Survival-driven escape behavior of Kangaroos. \\
Tetragonula Carbonaria Optimization  & SIAs & 2025 & MGM Gámez et al. & Stingless bee Tetragonula carbonaria builds  \\
Algorithm (TGCOA) \cite{TGCOA} & & & & and regulates temperature in the hive.\\\midrule
Simulated Annealing (SA) \cite{SA} & PBAs & 1987 & Van Laarhoven et al. & Physical annealing process. \\
Black Hole (BH) \cite{BH} & PBAs & 2013 & S Kumar et al. & Black hole phenomenon. \\
Attraction-Repulsion Optimization  & PBAs & 2024 & Cymerys et al. & Gravitational attraction and repulsive forces. \\
Algorithm (AROA) \cite{AROA} &  &  &  &  \\
Cloud Drift Optimization (CDO)  & PBAs & 2025 & M Alibabaei Shahraki et al. & Dynamic behavior of cloud particles influenced \\ 
algorithm \cite{CDO} &  &  &  &  by atmospheric forces. \\\midrule
Harmony Search (HS) \cite{HS}  & HBAs & 2001 & Zong Woo Geem et al. & Process of musicians improvising to find  \\
  &  &  &  &  a pleasing harmony. \\
Teaching-Learning-Based Optimization (TLBO) \cite{TLBO} & HBAs & 2011 & Rao et al. & Teaching and self-learning processes. \\
Soccer League Competition Algorithm (SLCA) \cite{SLCA} & HBAs & 2014 & Moosavian et al. & Competitive dynamics of soccer leagues. \\
Hiking Optimization Algorithm (HOA) \cite{HOA} & HBAs & 2024 & SO Oladejo et al. & Hiking. \\
Hannibal Barca Optimizer (HBO) \cite{HBO} & HBAs & 2025 & MW Ouertani et al. & Ancient Carthaginian general's strategic ingenuity. \\ 
\bottomrule
	\end{tabular}
	}
	\label{standardMAs}
\end{table}

However, according to the No Free Lunch (NFL) Theorem, no optimization algorithm can achieve superior performance across all problem domains \cite{NFL}. The performance of each metaheuristic algorithm depends heavily on the characteristics of the problem and its ability to maintain an appropriate balance between global exploration and local exploitation. Excessive exploration enables wide coverage of the solution space but may result in low search efficiency and slow convergence. Conversely, excessive exploitation can accelerate local convergence but often leads to premature stagnation and loss of population diversity. Thus, achieving a dynamic and adaptive balance between exploration and exploitation has become a central issue in the design and enhancement of metaheuristic algorithms.\par
In recent years, researchers have proposed numerous improvement strategies from various perspectives. In 2020, Shengliang Wang et al. proposed the Heterogeneous Comprehensive Learning and Dynamic Multi-Swarm Particle Swarm Optimization (HCLDMS-PSO), which integrates comprehensive learning and nonlinear dynamic multi-swarm mechanisms \cite{HCLDMS}. By adopting adaptive inertia weights and mutation operators, the algorithm significantly enhanced global exploration and the ability to escape local optima. In 2021, Amin Abdollahi Dehkordi et al. introduced the Nonlinear Chaotic Harris Hawks Optimization (NCHHO) algorithm, which incorporates chaotic maps and nonlinear control parameters to strengthen global search ability and effectively balance exploration and exploitation \cite{NCHHO}. In 2022, Zhang et al. developed the Locally Mutated and Randomly Accelerated Arithmetic Optimization Algorithm (LMRAOA), combining multi-leader wandering search, random fast jumping, and adaptive lens opposition-based learning strategies, thereby improving global optimization accuracy and convergence efficiency \cite{LMRAOA}. In 2023, J. Geng et al. proposed the Modified Adaptive Sparrow Search Algorithm (MASSA), which introduces chaotic reverse learning, dynamic adaptive weights, and spiral search strategies to maintain population diversity while enhancing convergence speed and global performance \cite{MASSA}. In 2024, Wei et al. presented the Improved Particle Swarm Optimization (IPSO), integrating Tent chaotic mapping, Lévy flight, and an adaptive t-distribution mechanism, achieving faster convergence and higher optimization precision \cite{IPSO}. In 2025, Lu et al. proposed the Multi-Rule-Based Monarch Optimization (MRBMO), which strengthens the search-for-food strategy and incorporates a Lens-Imaging Opposition-Based Learning (LIOBL) mechanism, substantially enhancing global exploration capabilities and avoiding premature convergence \cite{MRBMO}. These studies demonstrate that introducing mutation mechanisms, opposition-based learning, dynamic weighting, and multi-strategy hybridization can significantly improve the global exploration efficiency and adaptability of metaheuristic algorithms—representing one of the major research directions in intelligent optimization over recent years.\par
Among various metaheuristic algorithms, the Whale Optimization Algorithm (WOA) has attracted extensive attention due to its simple structure, few control parameters, and fast convergence speed. WOA simulates the bubble-net hunting behavior of humpback whales, utilizing three main mechanisms—encircling prey, spiral updating, and random search—to perform global and local exploration dynamically within the solution space. Nevertheless, the standard WOA still exhibits several shortcomings: it tends to become trapped in local optima when handling complex multi-modal problems, suffers from insufficient late-stage search accuracy, and lacks flexibility in balancing exploration and exploitation. These limitations are particularly pronounced in high-dimensional, multi-constrained, and dynamic environments, thereby restricting its further practical applications.\par
To overcome these limitations, numerous researchers have proposed various WOA variants and extensions in recent years, leading to continuous improvements in algorithmic performance. In 2020, Ruiye Jiang et al. proposed the Whale Optimization Algorithm with Reinforced Armed Forces (WAROA), which introduces a strategic 'armed forces scheme' to classify and coordinate search agents, thereby improving efficiency in solving complex, large-scale, and constrained problems \cite{WAROA}. In 2021, Vamsi Krishna Reddy Aala Kalananda et al. developed the Hybrid Social Whale Optimization Algorithms (HS-WOA and HS-WOA+), integrating the global exploration capability of WOA with the fast convergence of Social Group Optimization (SGO), achieving a dynamic balance between exploration and exploitation \cite{HSWOA}. In 2022, W. Yang et al. proposed the Multi-Strategy Whale Optimization Algorithm (MSWOA), incorporating chaotic Logistic mapping, Gaussian weight regulation, Levy flight, and Evolutionary Population Dynamics (EPD) \cite{MSWOA}. This hybrid framework improved population diversity and search efficiency, demonstrating outstanding performance in engineering design and classification problems. In 2023, J. Anitha et al. introduced the Modified Whale Optimization Algorithm (MWOA), which employs cosine regulation functions and position correction factors to enhance the equilibrium between exploration and exploitation, resulting in improved accuracy and stability for image multilevel threshold segmentation tasks \cite{MWOA}. In 2024, L. Wu developed the Nonlinear Random Reuse Mutation Whale Optimization Algorithm (NRRMWOA), integrating nonlinear adaptive parameter control, random reuse, and late-stage perturbation mechanisms to strengthen local refinement and maintain population diversity, thereby avoiding premature convergence \cite{NRRMWOA}. In 2025, Wei et al. proposed LSEWOA, which combines a Leader-Followers Search-for-Prey Strategy with an Enhanced Spiral Updating Strategy, supported by a nonlinear adaptive parameter mechanism, leading to simultaneous improvement in convergence rate and global exploration capability \cite{LSEWOA}. In summary, existing WOA improvements have primarily focused on chaotic mapping, opposition-based learning, dynamic parameter control, multi-strategy collaboration, and population interaction mechanisms. These developments have significantly enhanced global exploration and convergence stability, providing valuable theoretical and practical foundations for further research. Table \ref{improved} provides details of the above metaheuristic algorithm variants.\par
\begin{table}[htbp]
	\centering
	\caption{Current research on improved metaheuristic algorithms.}
	\resizebox{\textwidth}{!}{ 
	\begin{tabular}{llll}
	\toprule
	Algorithm & Year & Authors & Source of Inspiration \\
	\midrule
        HCLDMS-PSO  & 2020 & Wang et al. & Modified dynamic multi-swarm (DMS) strategy, \\ 
        \cite{HCLDMS} &  &  & Gaussian mutation operator and inertia weight. \\ 
        \midrule
        WAROA \cite{WAROA} & 2020 & R Jiang et al. & Armed force program and sub-population. \\ 
        \midrule
        NCHHO \cite{NCHHO} & 2021 & AA Dehkordi et al. & Chaotic and nonlinear control parameters. \\ 
        \midrule
        HS-WOA \cite{HSWOA} & 2021 & VKRA Kalananda et al. & Social group optimization (SGO). \\ 
        \midrule
        LMRAOA \cite{LMRAOA} & 2022 & YJ Zhang et al. & Multi-Leader Wandering Around Search Strategy,   \\ 
         &  &  & Random High-Speed Jumping Strategy (RHSJ)  \\ 
          &  &  &  and adaptive LIOBL. \\ 
        \midrule
        MSWOA \cite{MSWOA} & 2022 & W Yang et al. & Logistic mapping, adaptive weights, dynamic  \\ 
         &  &  & convergence factors and Evolutionary Population  \\ 
          &  &  & Dynamics mechanism. \\ 
        \midrule
        MASSA \cite{MASSA} & 2023 & J Geng et al. & Chaotic reverse learning, dynamic adaptive weight   \\ 
         &  &  & and adaptive spiral search technique. \\ 
        \midrule
        MWOA \cite{MWOA} & 2023 & J Anitha et al. & Cosine function and correction factors. \\ 
        \midrule
        IPSO \cite{IPSO} & 2024 & J Wei et al. & Tent mapping and Levy flight. \\ 
        \midrule
        NRRMWOA  & 2024 & L Wu et al. & Nonlinear adaptive parameter, random reuse   \\ 
        \cite{NRRMWOA} &  &  & strategy and late disturbance mutation strategy. \\ 
        \midrule
        MRBMO \cite{MRBMO} & 2025 & B Lu et al. & Good Nodes Set initialization, Levy flight and  \\ 
         &  &  & LIOBL. \\ 
        \midrule
        LSEWOA \cite{LSEWOA} & 2025 & Wei et al. & Good Nodes Set initialization and Tangent flight. \\ 
	\bottomrule
	\end{tabular}
	}
	\label{improved}
\end{table}

WOA struggles to balance exploration and exploitation, and the population quality tends to deteriorate significantly over iterations, leading to insufficient global exploration and premature convergence to local optima. To address these limitations, this paper proposes a Collective Cognitive Sharing Whale Optimization Algorithm with Cauchy Inverse Cumulative Distribution, termed CICDWOA. The algorithm reconstructs multiple layers of mechanisms in the standard WOA to simultaneously improve global exploration, population diversity, and convergence stability. First, during population initialization, CICDWOA adopts a Good Nodes Set (GNS) initialization strategy to ensure a more uniform distribution of initial individuals in high-dimensional search spaces. This reduces clustering effects and blind search regions, thereby providing a high-quality starting point for subsequent global exploration. Second, in the search behavior modeling stage, a Collective Cognitive Sharing (CCS) mechanism is introduced. This mechanism allows individuals to be guided by both the global best solution and the population's collective cognition, transforming the traditional 'random-dependent' update into a collective-cooperative learning process. As a result, information utilization and cooperative search efficiency are greatly improved. Third, CICDWOA introduces an Adaptive Exponential Spiral (AES) nonlinear model during the encircling phase. By combining an exponential modulation term with a spiral trajectory, AES breaks the linear limitation of the original WOA's encircling mechanism, allowing individuals to approach optimal solutions with higher motion flexibility and adaptive search capacity. Furthermore, a Cauchy Inverse Cumulative Distribution (CICD) mechanism is applied to perturb the leader position. Leveraging the heavy-tailed property of the Cauchy distribution, the algorithm incorporates long-distance jumps, which substantially enhance the probability of escaping local optima while maintaining fine local exploitation. To strengthen mid- and late-stage search elasticity, CICDWOA employs a Hybrid Gaussian-Cauchy Mutation strategy inspired by Differential Evolution (DE). This hybrid perturbation model enables cross-scale exploration and maintains dynamic balance between local refinement and global exploration. Finally, to overcome the rapid linear decay of the convergence factor $a$ in the original WOA, a nonlinear adaptive convergence control mechanism is introduced. By adjusting the shape parameter, this mechanism adaptively regulates the rate of the convergence factor $a$'s variation, allowing the algorithm to automatically tune exploration and exploitation intensities across different search stages, thereby enhancing global optimization capability and convergence precision.\par
The proposed CICDWOA is comprehensively evaluated using 23 benchmark functions, 2D robot path planning, 3D UAV path planning and 10 engineering design optimization problems. Experimental results demonstrate that CICDWOA achieves superior optimization performance, convergence stability, and robustness, confirming its generality and efficiency in solving complex real-world optimization problems.\par

\section{Whale Optimization Algorithm}
The Whale Optimization Algorithm (WOA), proposed by Mirjalili et al. in 2016, is a metaheuristic optimization algorithm inspired by the hunting behavior of humpback whales \cite{WOA}. In the WOA, the behaviors of 'Encircling Prey' and 'Spiral Updating' exhibited by humpback whales during hunting are simulated.

\subsection{Encircling Prey}
Humpback whales are capable of identifying the position of their prey and encircling it. Since the optimal position in the search space is unknown, the WOA assumes that the current best candidate solution is either the target prey or close to the optimal solution. Once the best search agent is defined, other agents will attempt to update their positions towards the best search agent. This behavior is represented by the following equations:
\begin{equation}
    {D} = \left| {C} \cdot {X}^*(t) - {X}(t) \right|
    \label{eq1}
\end{equation}
\begin{equation}
    {X}(t+1)={X}^*(t)-{A}\cdot {D}
    \label{eq2}
\end{equation}
Where $t$ is the current iteration number, $A$ and $C$ are coefficient vectors, $X^*$ is the current optimal solution, and $X$ represents the whale's position. If a better solution is found during each iteration, i.e., if the fitness value at the current position is less than the fitness value of $X^*$, the current position vector is set as the new $X^*$.\par
The calculations for $A$ and $C$ are as follows:
\begin{equation}
    {A}=2{a}\cdot {r}-{a}
    \label{eq3}
\end{equation}
\begin{equation}
    C=2\cdot {r}
    \label{eq4}
\end{equation}
Where $r$ is a random number vector between 0 and 1, and convergence factor $a$ decreases linearly from 2 to 0 during the iterations. \par
\begin{figure}[H]
    \centering
    \includegraphics[width=0.6\textwidth]{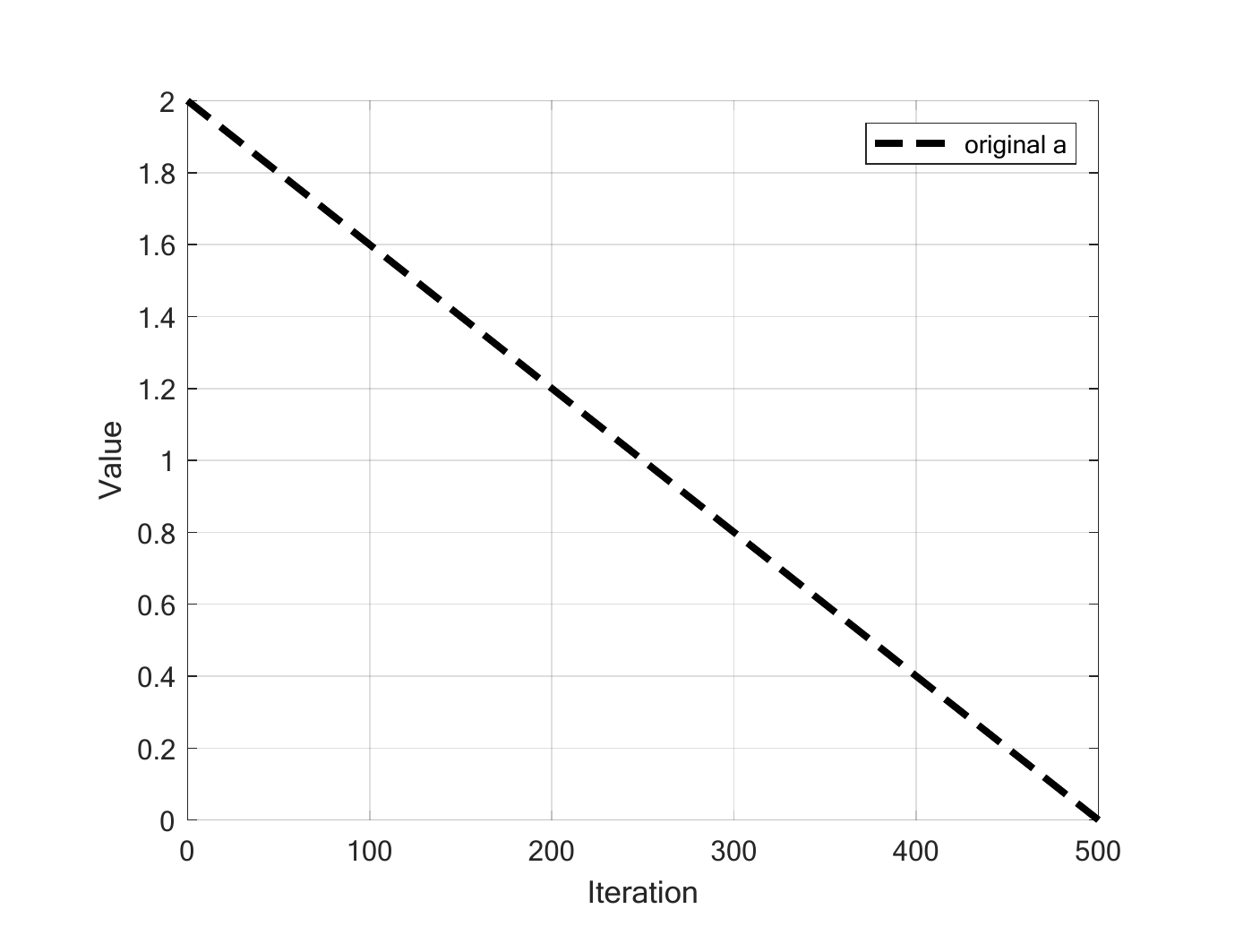}
    \caption{The variation of the convergence factor $a$ with iterations.}
    \label{original_a}
\end{figure}

The convergence factor $a$ is calculated using the formula:
\begin{equation}
    {a}=2-2\cdot\frac tT
    \label{eq5}
\end{equation}
Where $t$ is the current iteration number, and $T$ is the maximum number of iterations. Figure \ref{original_a} illustrates the variation of the convergence factor $a$.

\subsection{Attacking Prey}
In addition to encircling prey, whales use bubble-net feeding to attack their prey. They expel circular bubbles underwater to create a net that traps their prey. At around 15 meters underwater, whales spiral upward while releasing bubbles of varying sizes, surrounding the prey and drawing them closer to the center. At this point, the whale will consume the prey trapped within the bubble-net in an almost vertical posture. The bubble-net attack method consists of two strategies: contraction encircling prey and spiral ascending.
\subsubsection{Encircling Prey Mechanism}
This behavior is achieved by decreasing the value of A in \eqref{eq3}, with the position update as shown in \eqref{eq2}.\par
\subsubsection{Spiral Position Updating}
This method first calculates the straight-line distance $D'$ between the whale's position $X$ and the prey's position $X^*$, as shown in \eqref{eq7}, and then creates a spiral equation to simulate the whale's spiral ascent while surrounding the prey, as given by \eqref{eq6}:
\begin{equation}
    {X}(t+1)={X}^*(t)+{D'}\cdot e^{bl}\cdot\cos(2\pi l)
    \label{eq6}
\end{equation}
\begin{equation}
    {D'}=|{X}^*(t)-{X}(t)|
    \label{eq7}
\end{equation}
\begin{equation}
    l=(a_1-1)\cdot Rand+1
    \label{eq8}
\end{equation}
\begin{equation}
    a_1=-1-\frac tT
    \label{eq9}
\end{equation}
where $b$ is a constant that defines the logarithmic spiral shape, typically set to 1; $a_1$ is a parameter that varies linearly between [-2, -1]; $Rand$ is a random number between 0 and 1, and the spiral coefficient $l$ takes values between [-2, 1].

\subsection{Searching for Prey}
When a whale moves beyond the position of its prey, it will abandon its previous direction and randomly search in other directions for more prey to avoid getting trapped in a local optimum. The modeling of this search behavior is as follows:
\begin{equation}
    {D}^{\prime\prime}=|{C}\cdot{X}_{rand}-{X}(t)|
    \label{eq10}
\end{equation}
\begin{equation}
    {X}(t+1)={X}_{rand}-{A}\cdot {D}''
    \label{eq11}
\end{equation}
where ${X}_{rand}$ is a random whale selected from the current whale population.

\subsection{Population Initialization}
The original WOA uses a Pseudo-Random Number method for population initialization:
\begin{equation}
    {X}_{i,j}=(ub-lb)\cdot Rand+lb
    \label{eq12}
\end{equation}
Where ${X}_{i,j}$ represents the position of the whale individual, $ub$ and $lb$ represent the upper and lower bounds of the problem, and $Rand$ is a random number between 0 and 1.

\subsection{Pseudo-code and Analysis of WOA}
As shown in Algorithm \ref{alg1}, the WOA has a relatively simple structure with few parameters, making it easy to understand and implement. By dynamically adjusting the convergence factor a, WOA emphasizes global exploration during the early stages of search and focuses on local exploitation during the later stages, preventing premature convergence to local optima. However, despite WOA's ability to balance exploration and exploitation, it may not fully explore the search space in multi-modal optimization problems, especially in complex search spaces, leading to early convergence at local optima. In some specific problems, WOA exhibits slower convergence, particularly in the later stages of the algorithm when the population diversity significantly decreases, leading to stagnation. Moreover, WOA faces difficulties in balancing exploration and exploitation. Therefore, this paper introduces the CICDWOA to adjust the intensities of exploration and exploitation, further enhancing the algorithm's global search performance and convergence stability.

\begin{algorithm}[htbp]
\caption{Whale Optimization Algorithm (WOA)}
\label{alg1}
\begin{algorithmic}
\STATE 
\STATE \textbf{Begin}
\STATE \hspace{0.5cm} (\textit{Pseudo-random number initialization})
\STATE \hspace{0.5cm} Initialize the population using pseudo-random number method;
\STATE \hspace{0.5cm} Initialize parameters $(T, N, p, \text{etc.})$;
\STATE \hspace{0.5cm} Calculate the fitness of each search agent;
\STATE \hspace{0.5cm} The best search agent is $X^*$;
\STATE 
\STATE \textbf{while} $t < T$
\STATE \hspace{0.5cm}\textbf{for each} search agent
\STATE \hspace{1cm} Update $a$, $A$, $C$, $l$, and $p$;
\STATE \hspace{1cm}\textbf{if} $p < 0.5$
\STATE \hspace{1.5cm}\textbf{if} $|A| < 1$ \hspace{0.2cm}(\textit{Search for prey})
\STATE \hspace{2cm} Update the position by \eqref{eq11};
\STATE \hspace{1.5cm}\textbf{else} \hspace{0.2cm}(\textit{Encircling prey})
\STATE \hspace{2cm} Update the position by \eqref{eq2};
\STATE \hspace{1.5cm}\textbf{end if}
\STATE \hspace{1cm}\textbf{else} \hspace{0.2cm}(\textit{Spiral updating})
\STATE \hspace{1.5cm} Update the position by \eqref{eq6};
\STATE \hspace{1cm}\textbf{end if}
\STATE \hspace{0.5cm}\textbf{end for}
\STATE \hspace{0.5cm} Check if any agent exceeds the boundary and amend it;
\STATE \hspace{0.5cm} Calculate fitness of all agents;
\STATE \hspace{0.5cm} Update $X^*$ if a better solution is found;
\STATE \hspace{0.5cm} $t = t + 1$;
\STATE \textbf{end while}
\STATE 
\STATE \textbf{return} $X^*$
\STATE \textbf{End}
\end{algorithmic}
\end{algorithm}

\section{The Proposed CICDWOA}
\subsection{Good Nodes Set}
\begin{figure}[H]
    \centering
    \includegraphics[width=\textwidth]{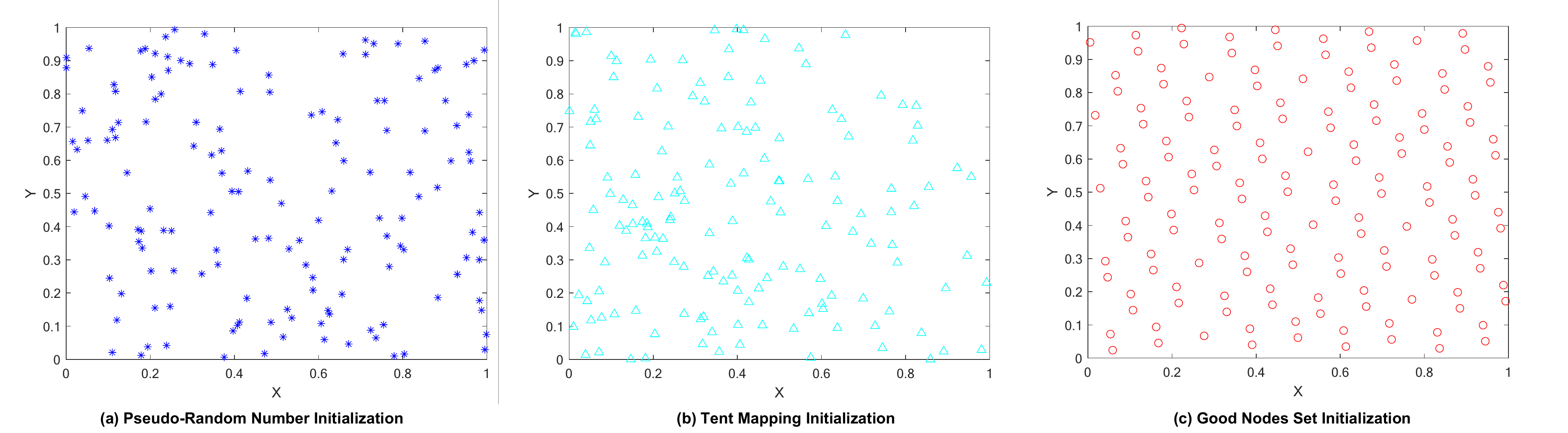}
    \caption{Comparison of different initialization methods (2D perspective). (a) Population initialized using Pseudo-Random Numbers when $N$=150; (b) Population initialized using Tent Mapping when $N$=150; (c) Population generated using Good Nodes Set initialization when $N$=150.} 
    \label{fig3}
\end{figure}

\begin{figure}[H]
    \centering
    \includegraphics[width=\textwidth]{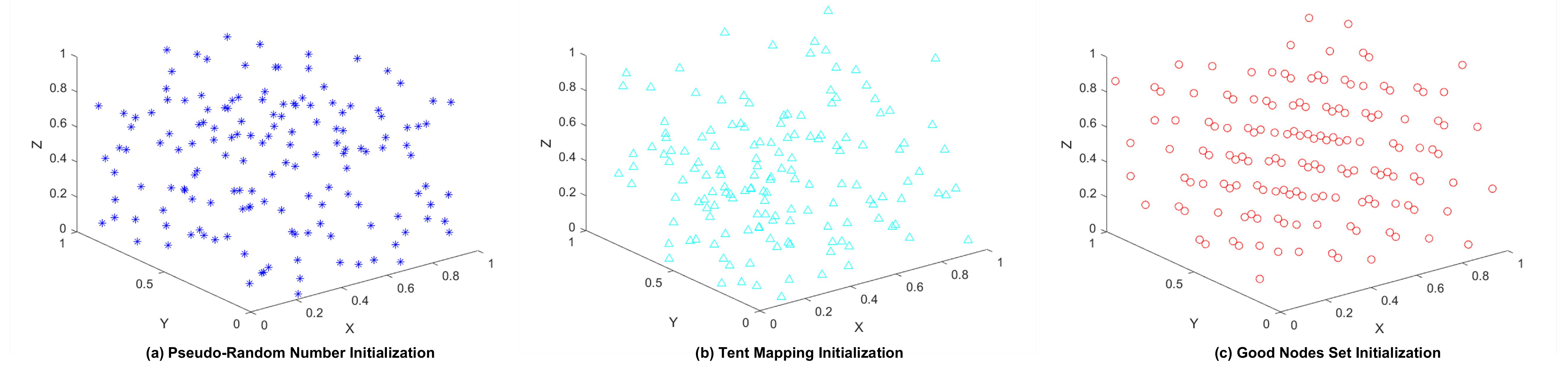}
    \caption{Comparison of different initialization methods (3D perspective). (a) Population initialized using Pseudo-Random Numbers when $N$=150; (b) Population initialized using Tent Mapping when $N$=150; (c) Population generated using Good Nodes Set initialization when $N$=150.} 
    \label{fig4}
\end{figure}

The traditional Whale Optimization Algorithm (WOA) initializes its population using Pseudo-Random Number method. This method is straightforward and possesses strong randomness; however, the distribution of individuals generated in the search space is often uneven. Such non-uniformity can lead to the aggregation of individuals, resulting in poor population diversity. Consequently, the WOA may suffer from premature convergence and reduced exploration efficiency during the optimization process. As illustrated in Figure~\ref{fig3}(a) and Figure~\ref{fig4}(a), when the population size is $N$=150, population initialized via Pseudo-Random Number method exhibit evident clustering and sparse regions in the search space.\par
To address the limitations of Pseudo-Random Number Initialization, various chaotic mapping strategies have been proposed. Among them, the Tent Mapping method is a widely adopted and simple chaotic mapping technique, characterized by strong ergodicity, sensitivity to initial conditions, and a high degree of chaos. The Tent Mapping function is defined as shown in \eqref{eq13}.
\begin{equation}
    x_{n+1} = \begin{cases} 
    2x_n, & \text{if } x_n < 0.5 \\
    2(1 - x_n), & \text{if } x_n \geq 0.5
    \end{cases}
    \label{eq13}
\end{equation}\par
The sequence generated by the Tent Mapping can theoretically traverse the entire interval [0,1], thereby enhancing the diversity of the initial population to a certain extent. Compared with Pseudo-Random Number initialization, Tent Mapping can generate more uniformly distributed individuals, effectively mitigating over-concentration and facilitating a more efficient global search during the early optimization stage. Figure~\ref{fig3}(b) and Figure~\ref{fig4}(b) depicts the population distribution obtained with Tent Mapping when $N$=150, where a more uniform spatial distribution can be clearly observed. Nevertheless, despite its improvement in randomness and diversity, Tent Mapping still suffers from local non-uniformity—that is, the uneven distribution density in certain subregions of the search space. This issue becomes more pronounced as the dimensionality increases. Furthermore, as a chaotic-sequence-based Pseudo-Random Number approach, Tent Mapping cannot guarantee strict uniformity in high-dimensional spaces.\par
To overcome these drawbacks, this study employs the Good Nodes Set (GNS) initialization method to further enhance population uniformity in the search space \cite{RWOA}. The concept of the Good Nodes Set was originally proposed by the Chinese mathematician Luogeng Hua, who introduced it as a mathematical technique for generating uniformly distributed point sets in high-dimensional Euclidean spaces. The core idea of GNS is to construct a dimension-independent sequence such that its projection on any subspace exhibits uniformity, thereby significantly improving the population's initial coverage and quality.\par
Let $U^D$ denote a D-dimensional unit hypercube, and $r$$\in$$U^D$. The Good Nodes Set $P_r^M$ is defined as follows:
\begin{equation}
    P_r^M=\{p(k)=(\{kr\},\{kr^2\},...,\{kr^D\})|k=1,2,...,M\}
    \label{eq14}
\end{equation}\par
where ${x}$ denotes the fractional part of $x$, $M$ is the total number of points, and $r$ is a positive deviation parameter. The constant $C(r,\varepsilon)$, which depends only on $r$ and $\varepsilon$, is a positive real number.\par
The set $P_r^M$ is termed the Good Nodes Set, and each point $p(k)$ is referred to as a good node. Assuming the upper and lower bounds of the $i^{th}$ dimension of the search space are  $x_{max}^i$ and  $x_{min}^i$, respectively, the mapping from the GNS to the actual search space can be expressed as:
\begin{equation}
  x_k^i=x_{min}^i+p_i(k)\cdot (x_{max}^i-x_{min}^i)    
  \label{eq15}
\end{equation}


\subsection{Collective Cognitive Sharing (CCS)}
In the standard WOA, the Search-for-Prey phase is a key component for global exploration. During this stage, each whale randomly selects another individual as a reference, updates its position based on their distance, and thereby achieves population diffusion and exploration. Although such random interactions increase population diversity and mitigate premature convergence to local optima, they also introduce several implicit issues. Firstly, because the reference individual is chosen randomly, interactions between whales lack stability, causing frequent oscillations in their movement trajectories. Especially in later iterations, this randomness weakens the population's coordination, slows convergence, and increases the likelihood of stagnation or premature convergence. Moreover, the prey-searching mechanism relies solely on information from a single random individual, ignoring the potential for collective information sharing within the population. This decentralized update strategy underutilizes global knowledge, leading to reduced adaptability and robustness in complex high-dimensional search spaces.\par
\begin{figure}[htbp]
    \centering
    \includegraphics[width=0.6\textwidth]{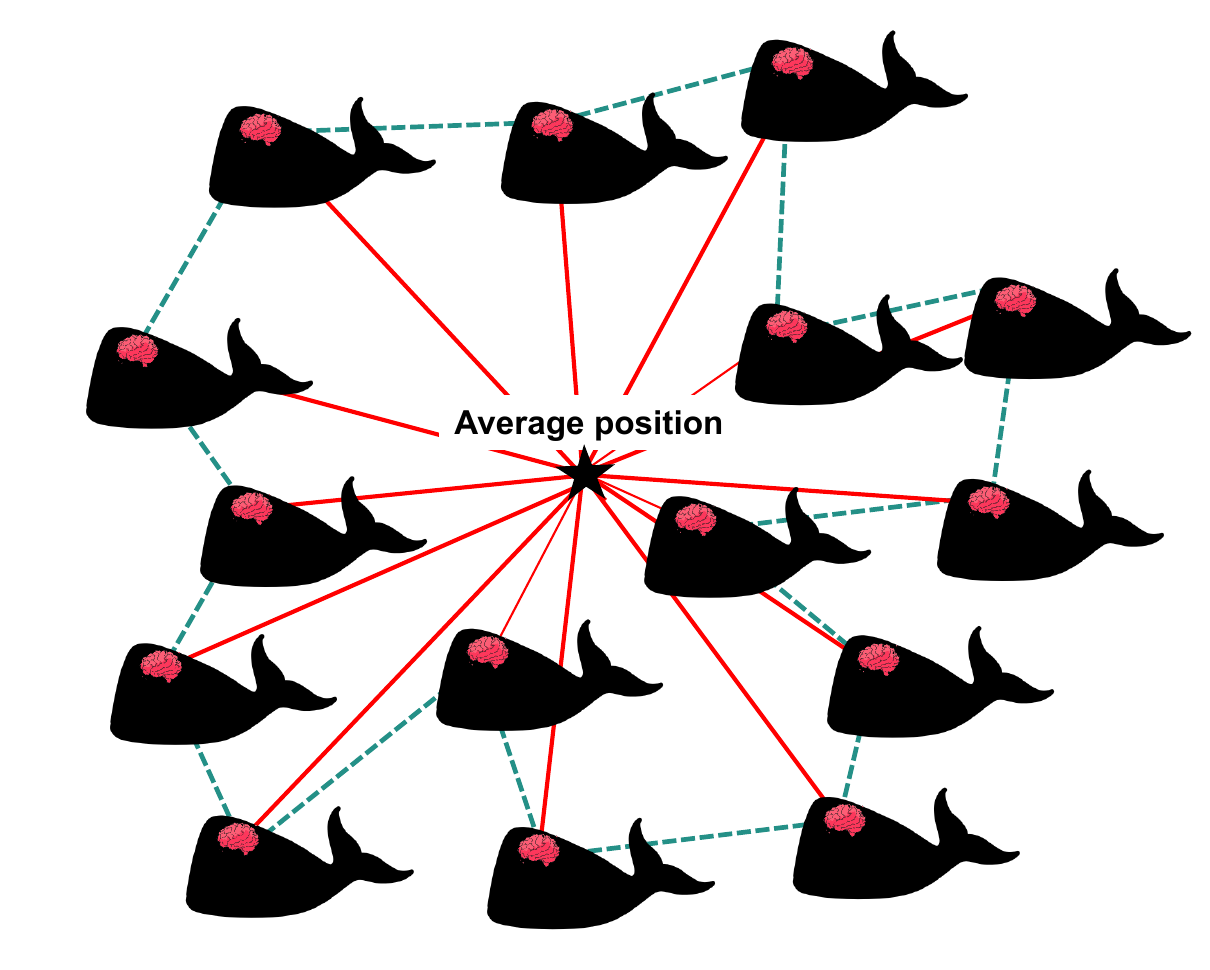}
    \caption{Collective Cognitive Sharing (CCS) mechanism} 
    \label{fig5}
\end{figure}

To overcome these issues, we propose a novel Collective Cognitive Sharing (CCS) mechanism for the search-for-prey phase. As shown in Figure \ref{fig5}, CCS introduces the concept of collective cognition, enabling each individual to update its position by considering both the global best (Leader) and the whale population's mean position. This transforms the algorithm's behavior from random dependence to collective coordination. The core principle of CCS is to construct a shared cognitive structure where each whale's update is guided by a weighted combination of the leader's position and the population's mean position. An adaptive regulation factor $G$ and a random coefficient $\alpha$ are introduced to dynamically balance exploration intensity and convergence rate. The Collective Cognitive Sharing mechanism is modeled below. 
\begin{equation}
  X(t+1)=\frac{1}{2}\left(X(t)+{X}_{mean}(t)\right)+G(t)\cdot|\alpha\cdot X(t)-X^*(t)|  
  \label{eq16}
\end{equation}
\begin{equation}
  g_1=2\cdot rand-1
  \label{eq17}
\end{equation}
\begin{equation}
  G(t)=2\left(1-\frac{t}{T}\right)\cdot g_1    
  \label{eq18}
\end{equation}
\begin{equation}
  \alpha=2\left(1-rand\right)  
  \label{eq19}
\end{equation}
\begin{equation}
    {X}_{mean}(t)=\frac1N\sum_{i=1}^N{X}_i(t)
     \label{eq20}
\end{equation}
where ${X}_{mean}(t)$ is the average position of whole whale population; $rand$ is a uniformly distributed random number in the range [0,1]; $g_1$ is used to introduce random directionality; $G(t)$ is the population sharing coefficient, which linearly decays with iterations and includes a sign, combining both a deterministic decay factor and a random sign to ensure strong perturbation in the early stages of exploration and gradual stability in the later stages; $\alpha$ is a random factor used to adjust the relative deviation of the individual's current position from the Leader, either amplifying or reducing it.

Fundamentally, CCS introduces a population mean term in the individual update process, enabling whale individuals to not only learn from the leader's behavior but also perceive and absorb the average movement trend of the entire population. This facilitates a more directional and collective search process. In the early stages of the algorithm, due to the large value of $G$, individuals are strongly driven by shared collective cognition, allowing for extensive exploration and expanding the search range, which helps to escape local optima traps. In later iterations, as $G$ gradually decreases, the search behavior stabilizes, with smaller individual update steps, thereby enhancing local exploitation capabilities and improving solution precision and convergence speed. This gradual convergence characteristic naturally creates a dynamic balance between global exploration and local exploitation, maintaining diversity in the early stages while ensuring high-precision convergence in the later stages.\par
Moreover, the CCS mechanism significantly improves upon the limitations of the original WOA in terms of information utilization. In traditional WOA, each individual updates its position based solely on the information from a randomly selected neighbor, resulting in low information exchange efficiency. CCS addresses this by introducing a process of sharing the collective average information, allowing each individual to indirectly acquire the global statistical characteristics of the entire population during updates. This significantly enhances the coherence and information flow within the population. As a result, the movement direction of each whale individual is not only attracted to the leader but is also constrained by the overall distribution structure of the population, thereby reducing the phenomena of excessive dispersion or concentration among individuals and ensuring the smoothness and stability of the search process. Through collective cognitive sharing, the population forms a 'collaborative consensus-driven' behavior pattern in the search space, characterized by mutual cooperation, information complementarity, and dynamic convergence toward the global objective.\par
The Collective Cognitive Sharing (CCS) mechanism represents a fundamental transition from random individual-driven behavior to collective cognitive collaboration in the original WOA. By incorporating global shared information and adaptive adjustment factors, the algorithm enhances convergence directionality while maintaining diversity, significantly improving search stability and global optimization capability. The introduction of this mechanism effectively overcomes the deficiencies in the original WOA, such as excessive reliance on randomness during prey searching, insufficient population coherence, and unstable convergence. It provides a new theoretical foundation and practical basis for improving the robustness and performance of the algorithm in complex optimization problems.

\subsection{Adaptive Exponential Spiral (AES) Mechanism }
In the standard Whale Optimization Algorithm (WOA), the core idea of the prey encirclement phase is to simulate the behavior of humpback whales narrowing the encirclement around their prey during hunting. The position update mechanism mainly relies on the Euclidean distance between the current individual and the global best solution. This process is controlled by \eqref{eq3} and \eqref{eq4}, where the position change is determined by vectors $A$ and $C$. While this linear contraction model effectively guides the population to quickly converge toward the optimal solution in the early stages, its linear nature also leads to overly simplistic search behavior in the later stages of convergence. When $|A|<1$, whale individuals tend to linearly approach only near the optimal solution, lacking sufficient randomness and directional diversity, which increases the likelihood of getting trapped in local optima, especially in high-dimensional, multi-modal optimization problems. Moreover, the original WOA models the 'shrinking encirclement' process at a fixed rate, lacking an adaptive control mechanism, making it difficult for the algorithm to balance global exploration and local exploitation at different search stages.\par
\begin{figure}[htbp]
    \centering
    \includegraphics[width=0.5\textwidth]{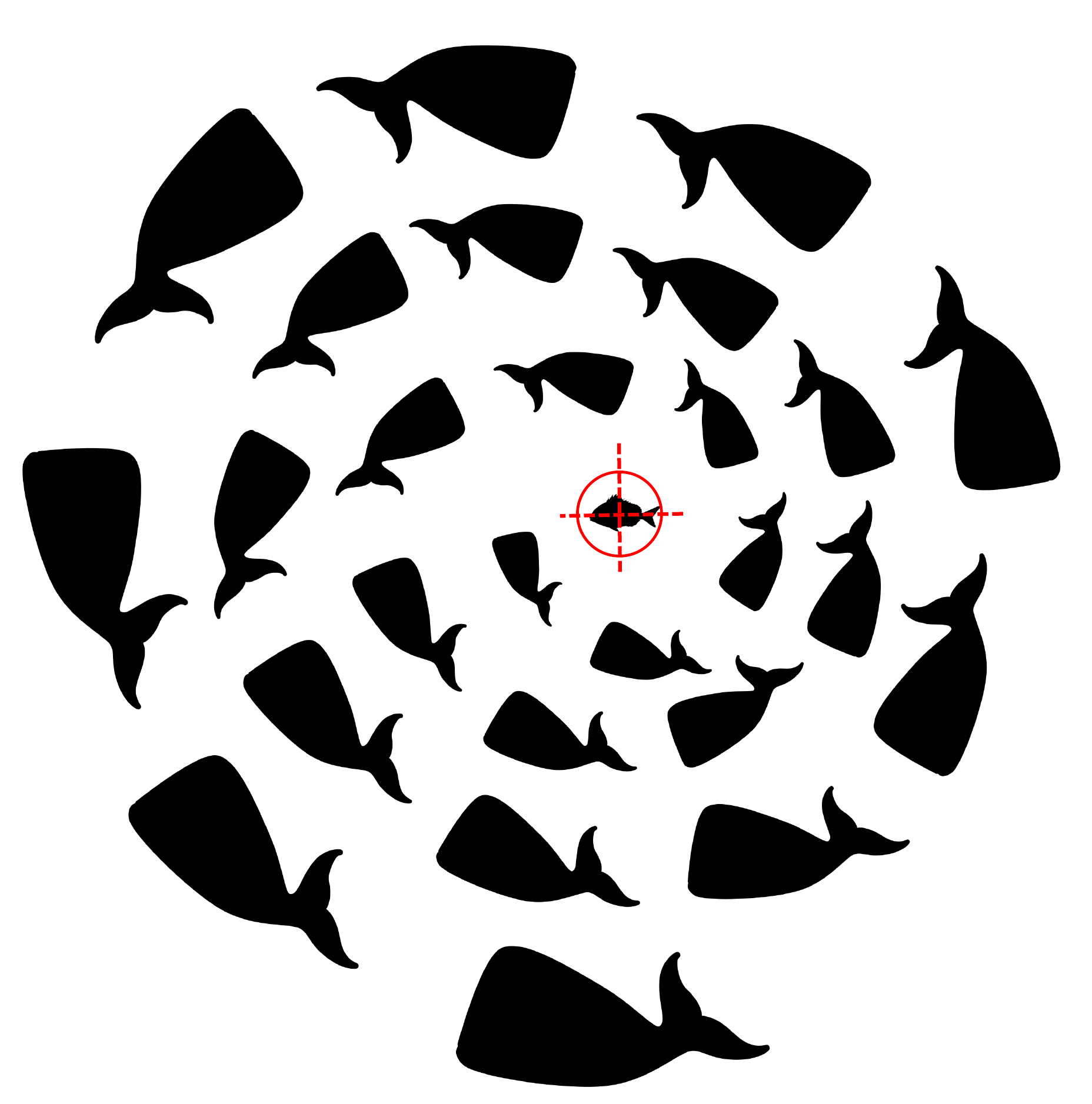}
    \caption{Adaptive Exponential Spiral mechanism} 
    \label{fig6}
\end{figure}

To address the aforementioned issues, this paper proposes an improved Adaptive Exponential Spiral mechanism, as shown in Figure~\ref{fig6}, to replace the traditional linear encirclement model in WOA. The design inspiration for this mechanism comes from the three-dimensional spiral predation behavior of whales in nature \cite{LSWOA}. In actual hunting, humpback whales encircle their prey along a spiral trajectory, adjusting the spiral radius and angle to modify their approach path. This nonlinear encirclement method allows them to control the hunting process more flexibly in complex marine environments. Inspired by this, we introduce a mathematical model of the spiral flight trajectory in the algorithm, allowing whale individuals to update their positions along a nonlinear path as they approach the optimal solution, thereby enhancing the algorithm's global search capability and its ability to escape local optima. In the Adaptive Exponential Spiral model, we introduce parameters $j$ and a dynamic exponential term $Z$ to control the nonlinear amplitude of the spiral encirclement and the search radius. The core position update formula for the new mechanism is given by:
\begin{equation}
	{X}(t+1)={X}^*(t)+h(s)\cdot|{A}\cdot {D}|
	 \label{eq21}
 \end{equation}
\begin{equation}
	 h(s)=e^{Z\cdot j}\cdot cos(2\pi j)
	 \label{eq22}
\end{equation}
\begin{equation}
	{D}=|{C}\cdot {X}^*(t)-{X}(t)|
	 \label{eq23}
\end{equation}
\begin{equation}
	Z=e^{cos(\pi\cdot(1-\frac tT))}
	 \label{eq24}
\end{equation}
\begin{equation}
	j=2\cdot rand-1
	 \label{eq25}
\end{equation}
where $A$ and $C$ are coefficient vectors, calculated by \eqref{eq3} and \eqref{eq4}; $h(s)$ is the spiral flight step size; $j$ and $Z$ are spiral coefficients; $rand$ is a random number between 0 and 1; and $D$ represents the distance between the individual and the current optimal solution, calculated by \eqref{eq1}.\par
Through this modeling, whale individuals no longer simply shrink in linear distance while encircling the prey, but instead approach it along a spiral trajectory determined by both the exponential and cosine terms. This nonlinear spiral update mechanism provides individuals with higher degrees of freedom in their movement, enabling them to not only maintain convergence toward the global optimal solution but also introduce periodic deviations in their path, thereby expanding the local search range. Compared to the original WOA, the Adaptive Exponential Spiral mechanism significantly enhances the diversity of the population, effectively preventing premature convergence caused by individuals clustering in narrow neighborhoods during the later search stages. Through the exponential modulation term $e^{Z\cdot j}$, the movement amplitude of individuals is dynamically adjusted with the number of iterations, achieving an adaptive transition from global exploration to local exploitation. Additionally, the introduction of the spiral factor enables individuals to approach the target region from multiple angles and levels as they close in on the prey, making it easier to escape local optimal traps.\par
The Adaptive Exponential Spiral mechanism, by incorporating the spatial dynamic properties of spiral flight and exponential nonlinear control, breaks through the linear convergence limitations of the original WOA during the prey encirclement phase. It significantly improves the algorithm's global optimization capability and solution diversity in complex optimization problems. Whale individuals no longer move along the shortest path when approaching the prey but adaptively adjust their trajectory in a spiral manner near the target, enabling the algorithm to maintain a fast convergence speed while strengthening its ability to escape local optima. This provides a more stable and efficient search mechanism for the application of WOA in multi-modal, high-dimensional, and dynamic optimization problems.

\subsection{Enhanced Spiral Updating Mechanism}
In WOA, the design inspiration for the spiral ascent phase comes from the bubble-net feeding behavior of humpback whales. The whales move in a spiral path around their prey, gradually approaching the target position. The standard WOA is based on this behavior and models the position update of the individual by \eqref{eq6}. This process simulates the individual's spiral approach around the leader, thus enhancing the algorithm's local exploitation capability. However, this mechanism has notable limitations in complex multi-modal environments.\par
First, the search step size and rotation amplitude in the standard spiral model are fixed, lacking adaptive adjustment capabilities. This results in insufficient global exploration in the early stages of the algorithm and excessive oscillation or unstable convergence in the later stages. Second, the spiral ascent process relies entirely on the current position of the leader, with individuals only moving locally around it, lacking the ability for random long jumps, which makes them prone to getting trapped in local optima. Third, since all individuals in the population are attracted to the same leader, the algorithm may quickly lose diversity in the early iterations, leading to premature convergence. To address these issues, this paper proposes an Enhanced Spiral Updating mechanism. By introducing the Cauchy Inverse Cumulative Distribution and inertia weight w, the algorithm achieves adaptive adjustment of the search range and enhanced random disturbance, forming a dynamic balance between global exploration and local exploitation.\par
\subsubsection{Cauchy Inverse Cumulative Distribution (CICD)}
\begin{figure}[htbp]
    \centering
    \includegraphics[width=1\textwidth]{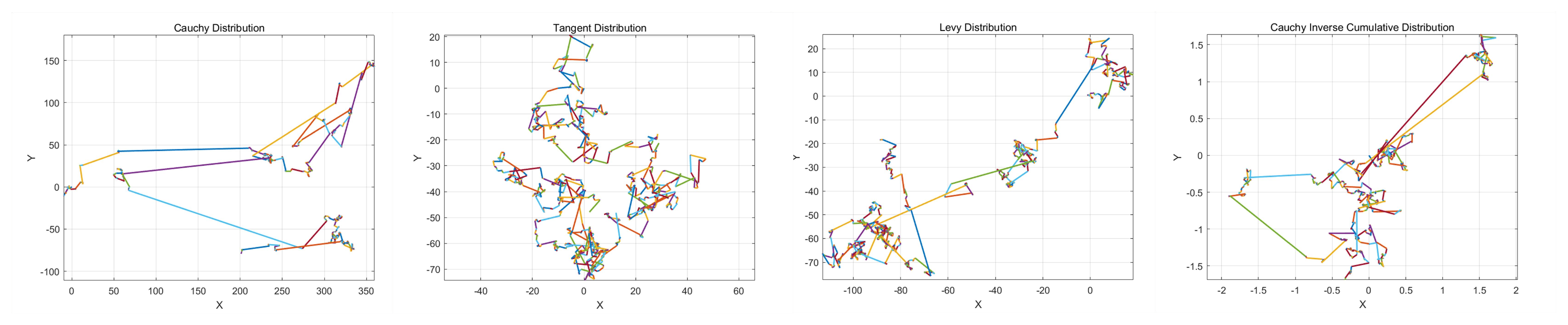}
    \caption{Comparison between regular distributions (Cauchy Distribution, Tangent Distribution and Levy Distribution) and Cauchy Inverse Cumulative Distribution.} 
    \label{fig7}
\end{figure}

The Cauchy Inverse Cumulative Distribution (CICD) is a distribution with heavy-tail characteristics, widely used in metaheuristic optimization algorithms to improve search diversity and the ability to escape local optima. Figure \ref{fig7} shows a simulation of tangent flight. Its mathematical expression is as follows:
\begin{equation}
	CICD=a+b\cdot tan(\pi \cdot (p-\frac{1}{2}))
	 \label{eq26}
\end{equation}
where $a$ is the location parameter, $b$ is the scale parameter, and $p\in [0,1]$ is a random number uniformly distributed.\par
In this study, the parameters are set as $a$=0 and $b$=0.01 to ensure that the generated samples primarily exhibit small perturbations while maintaining a small proportion of large random fluctuations, thereby forming a 'majority of minor perturbations, minority of long jumps' search characteristic.\par
Unlike the Gaussian distribution, the heavy-tail property of the Cauchy distribution means it occasionally generates extreme values, allowing the algorithm to make abrupt jumps during local convergence, thus escaping local optima traps. The random factor generated by the CICD is denoted as s and is applied to the scaling term of the leader's position:
\begin{equation}
	X^*(t)'=X^*(t) \cdot CICD
	 \label{eq27}
\end{equation}
where $X^*(t)'$ is the leader's position after the CICD perturbation.\par
This operation not only introduces slight nonlinear shifts during individual updates but also enables the algorithm to escape the current search region at critical moments, facilitating global-level random jumps. Compared to traditional linear perturbations, CICD enhances the population diversity and global search capability through its heavy-tail characteristics, thereby improving the robustness and escape performance of the algorithm. Thus, the introduction of CICD serves as a 'global disturbance source' in enhancing the spiral ascent mechanism. Through nonlinear scaling of the leader, it expands the reachable space of the spiral search, maintaining a degree of randomness and diversified search ability in the later stages, providing stable support for global convergence.\par

\subsubsection{Inertia Weight}
The inertia weight $\omega$ is one of the key parameters that regulate the search intensity and convergence speed of the algorithm. To enable the spiral search process to possess stage-aware capabilities, this study introduces an inertia weight function that adaptively changes with the number of iterations:
\begin{equation}
    \omega=\frac{1}{1+e^{-{s_1}(\frac tT-0.5)}}
    \label{eq28}
\end{equation}
where $t$ is the current iteration number, $T$ is the maximum number of iterations and $s_1$ is the scaling parameter of the Sigmoid function.\par
\begin{figure}[htbp]
    \centering
    \includegraphics[width=0.6\textwidth]{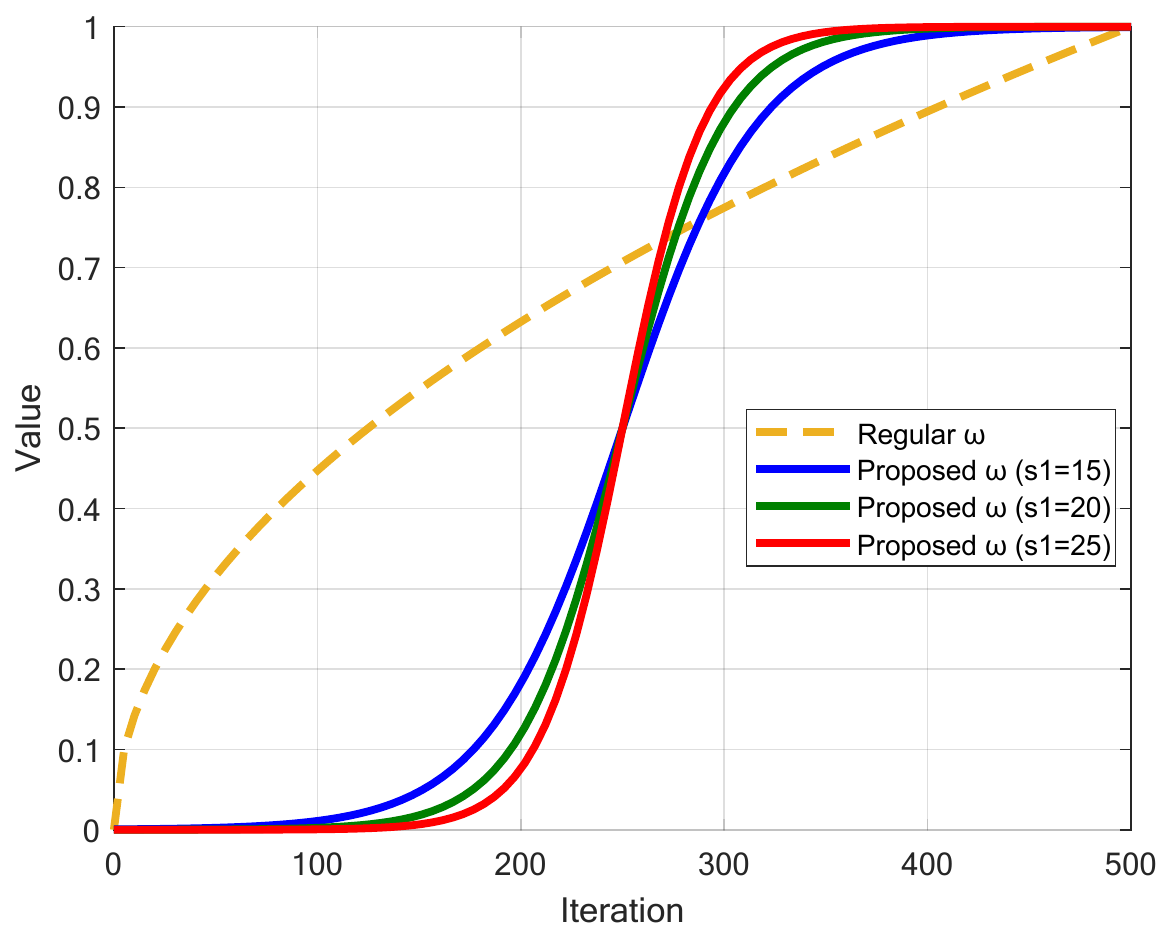}
    \caption{Comparison of different inertia weights.} 
    \label{fig8}
\end{figure}
The above function adopts an 'S'-shaped Sigmoid form, ensuring that $\omega$ remains small in the early stages of the algorithm, rapidly increases in the middle and later stages, and asymptotically approaches 1. Figure~\ref{fig8} is a comparison of different inertia weights. The core idea behind this design is as follows: In the early stages of the search, a smaller $\omega$ suppresses the convergence strength of the spiral term ${D'}\cdot e^{bl}\cdot\cos(2\pi l)$, reducing the individual's reliance on the leader and maintaining the randomness and diversity of global exploration. In contrast, during the middle and later stages, as $\omega$ increases rapidly, the attraction of the spiral term gradually intensifies, encouraging individuals to converge faster towards the optimal region, thus improving convergence speed and solution accuracy. Compared to the fixed-step spiral update in traditional WOA, the introduction of the Sigmoid-based inertia weight allows the algorithm's search behavior to adaptively adjust based on the iteration stage: global exploration is favored in the early stages, while local refinement is emphasized in the later stages. This dynamic adjustment mechanism effectively addresses the issue in standard WOA, where the balance between exploration and exploitation remains constant and the transition to the convergence phase is not well-defined. As a result, the search process becomes smoother, more intelligent, and more efficient.\par
The entire Enhanced Spiral Updating Mechanism is modeled below:
\begin{equation}
        {X}(t+1)={X}^*(t)'+\omega \cdot {D'}\cdot e^{bl}\cdot\cos(2\pi l)
    \label{eq29}
\end{equation}

\subsection{Hybrid Gaussian-Cauchy Mutation based on DE}
In the standard Whale Optimization Algorithm (WOA), the position update process primarily relies on bio-inspired mechanisms, such as prey encirclement and spiral ascent updates. While these mechanisms effectively guide individuals toward the optimal region in the early iterations, as the algorithm progresses into the middle and later stages, the population's positions tend to converge, and the differences between individuals decrease significantly. This leads to a lack of search diversity and a gradual decline in exploration capability. Due to the algorithm's heavy reliance on the guidance of the current best individual, when the best individual becomes trapped in a local optima, the entire population is likely to stagnate, making it difficult to further improve fitness values. Therefore, to enhance the global escape ability of WOA in the later stages, this study introduces a Hybrid Gaussian-Cauchy Mutation Strategy based on Differential Evolution (DE) in CICDWOA, which strengthens the population's random disturbance and directional diversity, thereby achieving adaptive corrections in the search process.\par
In this strategy, after performing the conventional position update, each individual will further generate an intermediate candidate solution through the differential evolution mechanism. Let the current individual be $X_i$, and four different individuals $X_D$, $X_E$, $X_F$ and $X_G$ be randomly selected from the population. The intermediate solution can be represented as:
\begin{equation}
    X_{new1}(t)=X_i(t)+F\cdot((X_E(t)-X_D(t))+(X_G(t)-X_F(t)))
    \label{eq30}
\end{equation}
where $X_{new1}(t)$ is an intermediate solution; $F$ is the differential scaling factor, used to control the amplification ratio of the differential vector. 
\par
Unlike traditional DE, this study uses the Cauchy distribution to generate the value of $F$, enhancing the dynamic adaptability of the step size:
\begin{equation}
    F=1+tan(\pi\cdot(Rand-0.5)) 
    \label{eq31}
\end{equation}
The Cauchy distribution has typical heavy-tail characteristics, which allow F to generate small scaling factors in most cases, ensuring a fine search within local regions, while occasionally generating larger disturbances, thereby forming irregular long jumps. This characteristic significantly improves the exploration depth of the algorithm, enabling individuals to escape local optima even in the later stages. To further enhance the search flexibility and directional diversity of the intermediate solution, a Hybrid Gaussian-Cauchy Mutation Operator is introduced to $X_{new2}$ after the differential operation. This operator combines the Gaussian distribution and the Cauchy distribution in a weighted manner, forming a multi-scale random disturbance model. Its mathematical expression is:
\begin{equation}
    X_{new2}(t)=X_{new1}(t)\cdot(1+\alpha\cdot Gaussian+(1-\alpha)\cdot Cauchy)
    \label{eq32}
\end{equation}
where $Gaussian\sim N(0,\sigma)$, $Cauchy\sim Cauchy(0,1)$, and $\alpha$ is the mixing weight factor. In this study, we set $\sigma$=0.1, and $\alpha$=0.5.\par
In this model, the Gaussian term $Gaussian$ with small standard deviation is used to guide individuals to make subtle adjustments within their current neighborhood, ensuring the stability of local exploitation. Meanwhile, the Cauchy term $Cauchy$ introduces heavy-tail disturbances, enabling individuals to make larger jumps when necessary, thereby expanding the search range. The linear combination of the two distributions gives the mutation process both 'fine-grained search' and 'coarse-grained jump' characteristics, significantly improving search flexibility and directional diversity. Additionally, to ensure that the generated new solution $X_{new2}$ remains within the feasible search space, a boundary correction mechanism is employed. When an individual exceeds the boundary, its position is adjusted to the corresponding boundary value:\par
\begin{equation}
    X_{new}(t) = \begin{cases} 
    ub, & \text{if}  X_{new2}(t) > ub\\
    lb, & \text{if}  X_{new2}(t) < lb \\
    X_{new2}(t), & \text{otherwise}
    \end{cases}
    \label{eq33}
\end{equation}\par
This operation effectively prevents search space overflow and solution degeneration caused by excessive mutation. Finally, the fitness function is used to compare the original individual and the mutated individual. If the new individual has a better fitness value, it replaces the original individual; otherwise, the original solution is retained. Through this design, the differential evolution structure and the Hybrid Gaussian-Cauchy mutation in CICDWOA work in synergy: the differential operator provides search directions based on population information, and the mixed distribution disturbance endows the algorithm with cross-scale jump capabilities, achieving an adaptive balance between local convergence and global exploration. This mechanism not only maintains the algorithm's convergence accuracy in the middle and later stages but also effectively overcomes the issue of diminished exploration ability and premature convergence in standard WOA, providing a stronger global search capability and stability foundation for the algorithm in complex optimization problems.

\subsection{Convergence Factor $a$}
This paper improves the core convergence factor $a$ in the standard WOA, proposing an adaptive convergence factor update mechanism based on a nonlinear control function to dynamically adjust the algorithm's search behavior at different stages. The mechanism aims to make the evolution of the convergence factor $a$ more aligned with the natural progression of the optimization search, thereby achieving a balance between global exploration and local exploitation, and addressing the issue in the standard WOA where the transition between search stages is too rigid due to linear decay. In the standard WOA, the convergence factor $a$ follows a linear decay, gradually decreasing from an initial value of 2 to 0, with the specific calculation given by \eqref{eq5}.\par
\begin{figure}[htbp]
    \centering
    \includegraphics[width=0.6\textwidth]{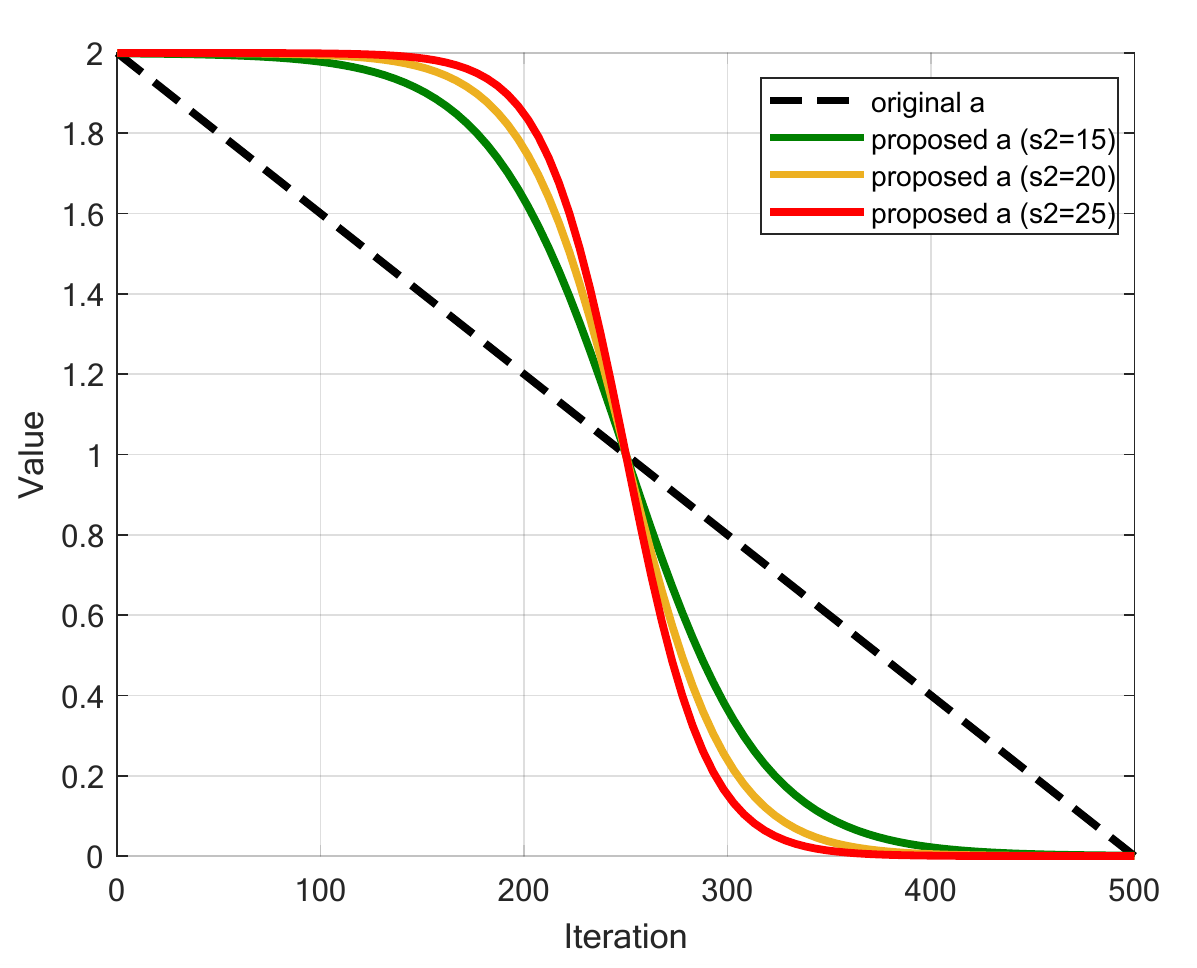}
    \caption{The comparison of the curves between the proposed convergence factors $a$($s_2$=15,$s_2$=20,$s_2$=25) and the original linearly decreasing convergence factor $a$ clearly demonstrates that as the parameter $s_2$ increases, the convergence rate in the middle stage is significantly accelerated, thereby providing a more balanced search dynamic across different optimization phases.} 
    \label{compare_a}
\end{figure}

While simple, this linear form results in a rapid decrease of convergence factor $a$ in the early stages, causing the algorithm to prematurely converge to local optima before fully exploring the global search space. In later stages, the slow decay rate of convergence factor $a$ makes it difficult for the algorithm to perform fine-tuned searches. This linear decay model fails to capture the nonlinear nature of search intensity over time in complex optimization problems, often leading to premature convergence or insufficient convergence speed. To address this limitation, this paper proposes an adaptive coefficient control method based on a nonlinear convergence curve in the CICDWOA. This method introduces a scaling parameter $s_2$ as a nonlinear function to create a differential rate of change for convergence factor $a$ at different stages, as shown by the following equation:
\begin{equation}
    a=2-\frac2{1+e^{-{s_2}(\frac tT-0.5)}}
    \label{eq34}
\end{equation}
where $s_2$ is the nonlinear adjustment factor, controlling the rate of change in the middle section of the curve. Figure~\ref{compare_a} shows a comparison of the convergence factor $a$ ($s_2$=15, $s_2$=20, $s_2$=25) and the standard linear decay form.

\subsection{Pseodo-code of CICDWOA}
The pseudo-code of the CICDWOA is provided in Algorithm~\ref{alg2} and Algorithm~\ref{alg3}.

\begin{algorithm}[H]
\caption{CICDWOA}
\label{alg2}
\begin{algorithmic}
\STATE 
\STATE \textbf{Begin}
\STATE \hspace{0.5cm} Initialize parameters $(T, N, p, \text{etc.})$;
\STATE \hspace{0.5cm} Calculate the fitness of each search agent;
\STATE \hspace{0.5cm} The best search agent is $X^*$;
\STATE 
\STATE \textbf{while} $t < T$
\STATE \hspace{0.5cm}\textbf{for each} search agent
\STATE \hspace{1cm} Update $a$, $A$, $C$, $l$, and $p$;
\STATE \hspace{1cm}\textbf{if} $p < 0.5$
\STATE \hspace{1.5cm}\textbf{if} $|A| < 1$ \hspace{0.2cm}(\textit{Adaptive Exponential Spiral Mechanism})
\STATE \hspace{2cm} Update position by \eqref{eq21};
\STATE \hspace{1.5cm}\textbf{else} \hspace{0.2cm}(\textit{Collective Cognitive Sharing mechanism})
\STATE \hspace{2cm} Update position by \eqref{eq16};
\STATE \hspace{1.5cm}\textbf{end if}
\STATE \hspace{1cm}\textbf{else} \hspace{0.2cm}(\textit{Enhanced Spiral Updating Mechanism})
\STATE \hspace{1.5cm} Update position by \eqref{eq29};
\STATE \hspace{1cm}\textbf{end if}
\STATE \hspace{0.5cm}\textbf{end for}
\STATE 
\STATE \hspace{1cm} (\textit{Hybrid Gaussian-Cauchy Mutation based on DE})
\STATE 
\STATE \hspace{0.5cm} Calculate the fitness of each search agent;
\STATE \hspace{0.5cm} Update $X^*$ if a better solution is found;
\STATE \hspace{0.5cm} $t = t + 1$;
\STATE \textbf{end while}
\STATE 
\STATE \textbf{return} $X^*$
\STATE \textbf{End}
\end{algorithmic}
\end{algorithm}

\begin{algorithm}[htbp]
\caption{Hybrid Gaussian-Cauchy Mutation based on DE}
\label{alg3}
\begin{algorithmic}
\STATE 
\STATE \textbf{for each} search agent
\STATE \hspace{0.5cm} Generate intermediate solution $X_{{new1}}$ by Differential Evolution via \eqref{eq30};
\STATE \hspace{0.5cm} Generate another intermediate solution $X_{{new2}}$ by \eqref{eq32};
\STATE \hspace{0.5cm}\textbf{if} $X_{{new2}}$ within the boundary
\STATE \hspace{1cm} $X_{{new}} = X_{{new2}}$;
\STATE \hspace{0.5cm}\textbf{else if} $X_{{new2}}$ exceeds upper boundary
\STATE \hspace{1cm} Set to upper boundary $ub$;
\STATE \hspace{0.5cm}\textbf{else if} $X_{{new2}}$ exceeds lower boundary
\STATE \hspace{1cm} Set to lower boundary $lb$;
\STATE \hspace{0.5cm}\textbf{end if}
\STATE \textbf{end for}
\end{algorithmic}
\end{algorithm}

\section{Experiments}
The experimental environment for this study is as follows: Windows 11 (64-bit), Intel(R) Core(TM) i5-8300H CPU @ 2.30GHz processor, 8GB RAM, and Matlab R2023a as the simulation platform. To verify the performance and effectiveness of the CICDWOA algorithm, the following experiments were designed:\par
\begin{itemize}
  \item Experiment 1 (Parameter Sensitivity Analysis Experiment):\\ Test nine combinations of scaling factors $s_1$($s_1$=15,$s_1$=20 and $s_1$=25) and scaling factors $s_2$( $s_2$=15,  $s_2$=20,  $s_2$=25) on the selected 23 benchmark functions to select the values of $s_1$ and  $s_2$ that best balance the exploration and exploitation of CICDWOA;
  \item Experiment 2 (Ablation Study):\\ Remove six improvement strategies respectively from CICDWOA, and perform ablation study on the benchmark functions to explore the effect of each improvement strategy;
  \item Experiment 3 (Qualitative Analysis Experiment):\\ Conduct qualitative analysis experiments by applying CICDWOA to benchmark functions, comprehensively evaluating the performance, robustness, search behaviors and exploration-exploitation balance of CICDWOA on different types of problems. The evaluation includes convergence behavior, trajectory of a search agent, population diversity, and exploration-exploitation capabilities;
  \item Experiment 4 (Comparative Experiment):\\ Compare the CICDWOA with standard metaheuristic algorithms, excellent WOA variants, and state-of-the-art (SOTA) improved metaheuristic algorithms on benchmark functions to verify the superiority of CICDWOA.
\end{itemize}

\subsection{Parameter Sensitivity Analysis Experiment}
As shown in \eqref{eq28} and \eqref{eq34}, the evolution curves of the inertia weight w and convergence factor a are controlled by the scaling parameters $s_1$ and $s_2$, respectively. Here, $s_1$ represents the scaling factor for the inertia weight $\omega$, primarily influencing the extent to which the whale individuals retain historical position information during the iteration process. On the other hand, $s_2$ serves as the scaling factor for the convergence factor $a$, determining the balance between the exploration and exploitation phases. Therefore, the proper selection of $s_1$ and $s_2$ values has a critical impact on the overall search performance of the CICDWOA.\par
In the proposed CICDWOA, the inertia weight $\omega$ adopts a nonlinear decreasing form based on the Sigmoid function:\par
\begin{equation}
    \omega=\frac{1}{1+e^{-{s_1}(\frac tT-0.5)}}
    \label{eq35}
\end{equation}
The variation trend of the inertia weight $\omega$ determines the extent to which whale individuals are guided by the leader during the position update phase. Larger $s_1$ values result in a steeper descent of the Sigmoid function, which leads to stronger inertia in the early iterations and causes individuals to retain more historical information, thus enhancing population diversity. Conversely, smaller $s_1$ values lead to a smoother change in the inertia weight, facilitating a more stable convergence in the later stages of the algorithm. At the same time, the variation of the convergence factor a is given by the following equation:\par
\begin{equation}
    a=2-\frac2{1+e^{-{s_2}(\frac tT-0.5)}}
    \label{eq36}
\end{equation}
The parameter $s_2$ controls the rate at which the search space contracts, thus affecting the balance between exploration and exploitation. When $s_2$ is smaller, the Sigmoid curve is relatively flat, meaning that the convergence factor $a$ decreases slowly in the early iterations, allowing the algorithm to conduct a more extensive global search. Conversely, larger $s_2$ values cause the convergence factor $a$ to decrease rapidly, prompting the algorithm to focus on local regions during the middle and late stages to accelerate convergence.\par
To systematically analyze the influence of $s_1$ and $s_2$ on the performance of CICDWOA, this study conducted parameter sensitivity experiments on 23 standard benchmark test functions shown in Table~\ref{funcdetails}. Nine different parameter combinations were tested, as shown in Table~\ref{tab:s1s2_combinations}. In these experiments, the population size was set to $N$=30, and the maximum number of iterations was $T$=500. Each parameter combination was independently run 30 times on each test function, and the results were statistically compared using the Friedman test.\par

\begin{table}[htbp]
\centering
\caption{Parameter combinations of $s_1$ and $s_2$ used in sensitivity analysis.}
\begin{tabular}{ccc}
\toprule
Combination Index & $s_1$ & $s_2$   \\
\midrule
Comb 1 & 15 & 15  \\
Comb 2 & 15 & 20  \\
Comb 3 & 15 & 25  \\
Comb 4 & 20 & 15  \\
Comb 5 & 20 & 20  \\
Comb 6 & 20 & 25  \\
Comb 7 & 25 & 15  \\
Comb 8 & 25 & 20  \\
Comb 9 & 25 & 25  \\
\bottomrule
\end{tabular}
\label{tab:s1s2_combinations}
\end{table}

\begin{figure*}[htbp]
    \centering
    \includegraphics[width=1\textwidth]{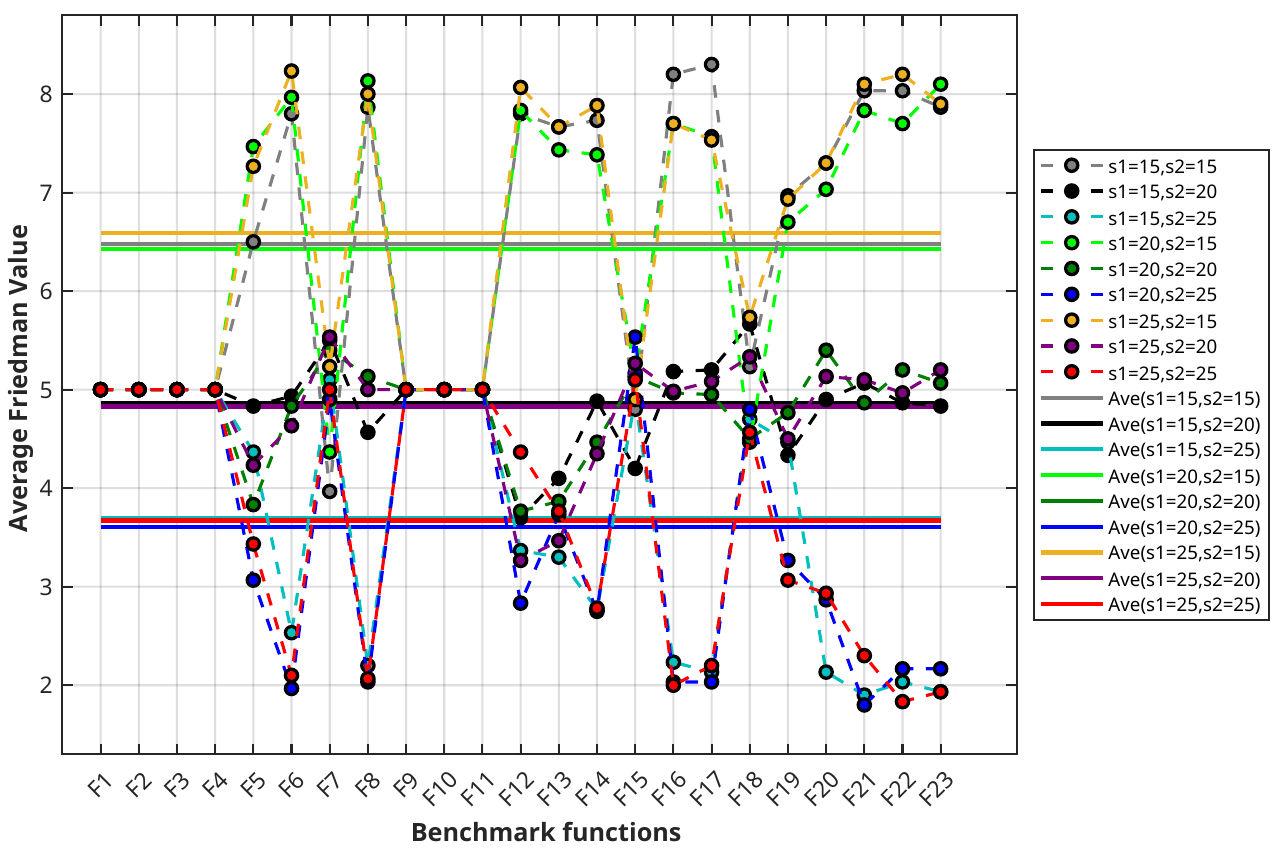}
    \caption{The curves of the $Ave$ and $Std$ performance of CICDWOAs under different parameter combinations on the 23 benchmark functions are presented. The solid lines represent the Average Friedman Value of each algorithm, while the dashed lines indicate the individual rankings of each algorithm across the 23 benchmark functions.} 
    \label{zhexiantu}
\end{figure*}

Table~\ref{tab:sensitivity_cicdwoa} and Figure~\ref{zhexiantu} present the average performance metrics and overall rankings of CICDWOA on the 23 benchmark functions under different parameter combinations. The results indicate that the parameter combinations significantly influence the algorithm's convergence behavior and global optimization ability. Based on the overall average values and ranking results, when $s_1$=20 and $s_2$=25, CICDWOA achieved the best performance, with an average Friedman value of 3.6065, ranking first overall. The second-best performance was observed for $s_1$=25 and $s_2$=25 (average value 3.6717, ranked second), followed by $s_1$=15 and $s_2$=25 (average value 3.6985, ranked third).\par
As $s_2$ increases (i.e., the convergence coefficient Sigmoid function becomes steeper), the algorithm demonstrates faster convergence and stronger local search capabilities on most complex multi-modal functions (such as F5, F6, F8, and F16-F23). This suggests that larger $s_2$ values help CICDWOA focus on high-quality solutions in the later stages. On the other hand, when $s_1$ is smaller, the inertia weight decreases more slowly, leading to higher population diversity during the early exploration phase, which helps avoid the risk of getting stuck in local optima. Therefore, the combination of $s_1$ and $s_2$ has a synergistic effect on the algorithm's performance: $s_2$ adjusts the search inertia, while $s_2$ balances the exploration intensity. Together, they shape the dynamic search behavior of CICDWOA.

\begin{table*}[htbp]
\centering
\caption{Sensitivity analysis results of different $(s_1,s_2)$ parameter combinations on 23 benchmark functions.}
\resizebox{\textwidth}{!}{
\begin{tabular}{cccccccccc}
\toprule
Function & Comb 1 & Comb 2 & Comb 3 & Comb 4 & Comb 5 & Comb 6 & Comb 7 & Comb 8 & Comb 9 \\ 
\midrule
F1  & 5.0000 & 5.0000 & 5.0000 & 5.0000 & 5.0000 & 5.0000 & 5.0000 & 5.0000 & 5.0000 \\ 
F2  & 5.0000 & 5.0000 & 5.0000 & 5.0000 & 5.0000 & 5.0000 & 5.0000 & 5.0000 & 5.0000 \\ 
F3  & 5.0000 & 5.0000 & 5.0000 & 5.0000 & 5.0000 & 5.0000 & 5.0000 & 5.0000 & 5.0000 \\ 
F4  & 5.0000 & 5.0000 & 5.0000 & 5.0000 & 5.0000 & 5.0000 & 5.0000 & 5.0000 & 5.0000 \\ 
F5  & 6.5000 & 4.8333 & 4.3667 & 7.4667 & 3.8333 & 3.0667 & 7.2667 & 4.2333 & 3.4333 \\ 
F6  & 7.8000 & 4.9333 & 2.5333 & 7.9667 & 4.8333 & 1.9667 & 8.2333 & 4.6333 & 2.1000 \\ 
F7  & 3.9667 & 5.5000 & 5.1000 & 4.3667 & 5.4000 & 4.9000 & 5.2333 & 5.5333 & 5.0000 \\ 
F8  & 7.8667 & 4.5667 & 2.2000 & 8.1333 & 5.1333 & 2.0333 & 8.0000 & 5.0000 & 2.0667 \\ 
F9  & 5.0000 & 5.0000 & 5.0000 & 5.0000 & 5.0000 & 5.0000 & 5.0000 & 5.0000 & 5.0000 \\ 
F10 & 5.0000 & 5.0000 & 5.0000 & 5.0000 & 5.0000 & 5.0000 & 5.0000 & 5.0000 & 5.0000 \\ 
F11 & 5.0000 & 5.0000 & 5.0000 & 5.0000 & 5.0000 & 5.0000 & 5.0000 & 5.0000 & 5.0000 \\ 
F12 & 7.8000 & 3.7000 & 3.3667 & 7.8333 & 3.7667 & 2.8333 & 8.0667 & 3.2667 & 4.3667 \\ 
F13 & 7.6667 & 4.1000 & 3.3000 & 7.4333 & 3.8667 & 3.7333 & 7.6667 & 3.4667 & 3.7667 \\ 
F14 & 7.7333 & 4.8833 & 2.7667 & 7.3833 & 4.4667 & 2.7500 & 7.8833 & 4.3500 & 2.7833 \\ 
F15 & 4.8000 & 4.2000 & 4.9000 & 5.1667 & 5.1333 & 5.5333 & 4.9000 & 5.2667 & 5.1000 \\ 
F16 & 8.2000 & 5.1833 & 2.2333 & 7.7000 & 4.9667 & 2.0333 & 7.7000 & 4.9833 & 2.0000 \\ 
F17 & 8.3000 & 5.2000 & 2.1333 & 7.5667 & 4.9500 & 2.0333 & 7.5333 & 5.0833 & 2.2000 \\ 
F18 & 5.2333 & 5.6667 & 4.7000 & 4.4667 & 4.5000 & 4.8000 & 5.7333 & 5.3333 & 4.5667 \\ 
F19 & 6.9667 & 4.3333 & 4.4667 & 6.7000 & 4.7667 & 3.2667 & 6.9333 & 4.5000 & 3.0667 \\ 
F20 & 7.3000 & 4.9000 & 2.1333 & 7.0333 & 5.4000 & 2.8667 & 7.3000 & 5.1333 & 2.9333 \\ 
F21 & 8.0333 & 5.0667 & 1.9000 & 7.8333 & 4.8667 & 1.8000 & 8.1000 & 5.1000 & 2.3000 \\ 
F22 & 8.0333 & 4.8667 & 2.0333 & 7.7000 & 5.2000 & 2.1667 & 8.2000 & 4.9667 & 1.8333 \\ 
F23 & 7.8667 & 4.8333 & 1.9333 & 8.1000 & 5.0667 & 2.1667 & 7.9000 & 5.2000 & 1.9333 \\ \midrule
Average & 6.4812 & 4.8594 & 3.6985 & 6.4283 & 4.8326 & \textbf{3.6065} & 6.5935 & 4.8283 & 3.6717 \\ \midrule
Rank & 7 & 6 & 3 & 8 & 5 & \textbf{1} & 9 & 4 & 2 \\ 
\bottomrule
\end{tabular}
}
\label{tab:sensitivity_cicdwoa}
\end{table*}

\subsection{Ablation Study}

To clearly evaluate the contribution of each proposed strategy, a component-removal-based ablation methodology was adopted in this study.
Specifically, instead of incrementally adding individual components to the original WOA, we start from the complete CICDWOA framework and remove one improvement strategy at a time while keeping all other components unchanged.
The removed strategy is replaced by its corresponding original mechanism in WOA, allowing the performance degradation caused by the absence of each component to be directly observed.

This leave-one-out ablation design offers two key advantages.
First, it isolates the individual contribution of each strategy under the same optimization environment, avoiding the confounding effects introduced by strong interdependencies among components.
Second, since the proposed strategies are jointly designed to cooperate within a unified framework, incremental addition may underestimate the importance of certain components that only become effective when combined with others.
Therefore, the adopted ablation scheme provides a more faithful assessment of how each improvement contributes to the overall performance of CICDWOA.

Based on this methodology, six ablated variants were constructed, each removing one specific strategy from the complete algorithm, as described below:\par

\begin{figure*}[htbp]
    \centering
    \includegraphics[width=\textwidth]{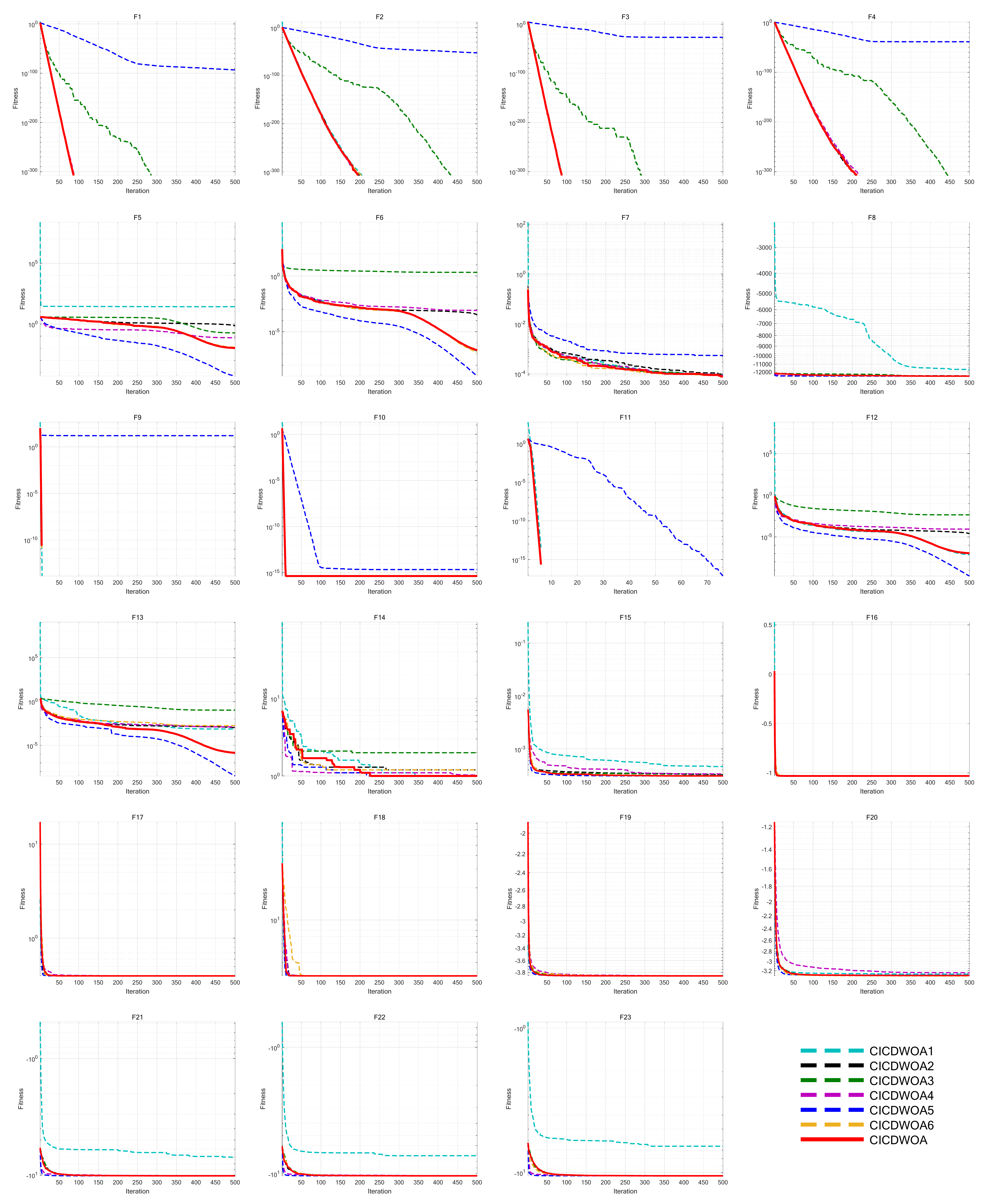}
    \caption{Iteration Curves of different CICDWOAs in Ablation Study.} 
    \label{abla_fig}
\end{figure*}

\begin{itemize}
  \item We replaced the Good Nodes Set Initialization with Pseudo-Random Number initialization, resulting in a modified version of CICDWOA, referred to as CICDWOA1;
  \item We replaced the newly proposed update method for convergence factor $a$ with the original linear update method in WOA, leading to a modified version known as CICDWOA2;
  \item We replaced the Collective Cognitive Sharing mechanism with the original WOA's prey search strategy, resulting in CICDWOA3;
  \item We replaced the Adaptive Exponential Spiral Strategy with the original WOA's prey encircling strategy, resulting in CICDWOA4;
  \item We replaced the Enhanced Spiral Updating Strategy with the original WOA's spiral ascent mechanism, leading to CICDWOA5;
  \item We removed the Hybrid Gaussian-Cauchy Mutation based on DE, resulting in CICDWOA6.
\end{itemize}
Additionally, the number of iterations was set to $T$=500 and the population size was set to $N$=30. Each algorithm was run 30 times on 23 benchmark functions to validate the effectiveness of each improvement strategy. The iteration curves are shown in Figure ~\ref{abla_fig}.\par
\begin{figure*}[htbp]
    \centering
    \includegraphics[width=\textwidth]{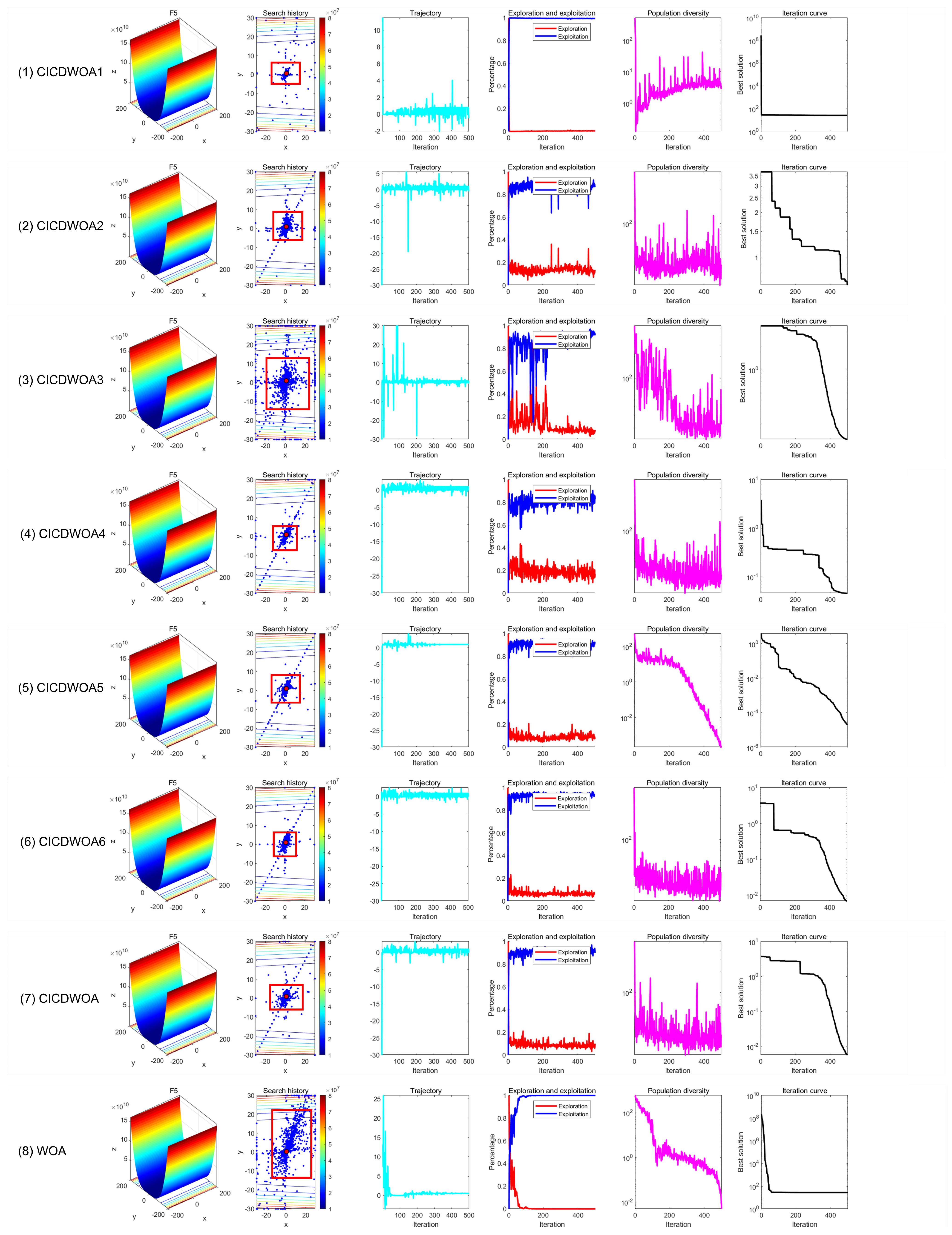}
    \caption{Results of qualitative analysis experiment (F5).} 
    \label{quaF5}
\end{figure*}

\begin{figure*}[htbp]
    \centering
    \includegraphics[width=\textwidth]{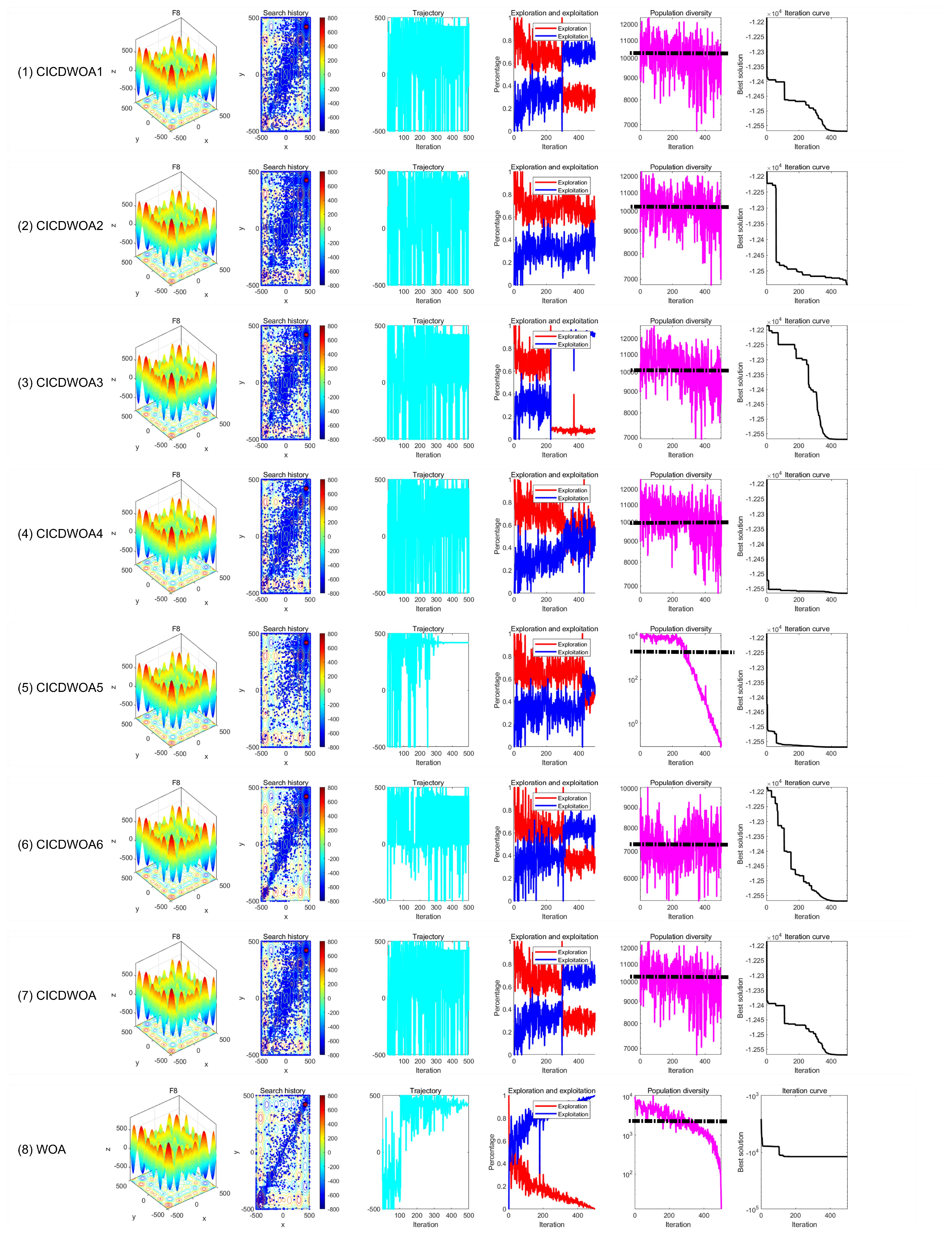}
    \caption{Results of qualitative analysis experiment (F8).} 
    \label{quaF8}
\end{figure*}

\begin{figure*}[htbp]
    \centering
    \includegraphics[width=\textwidth]{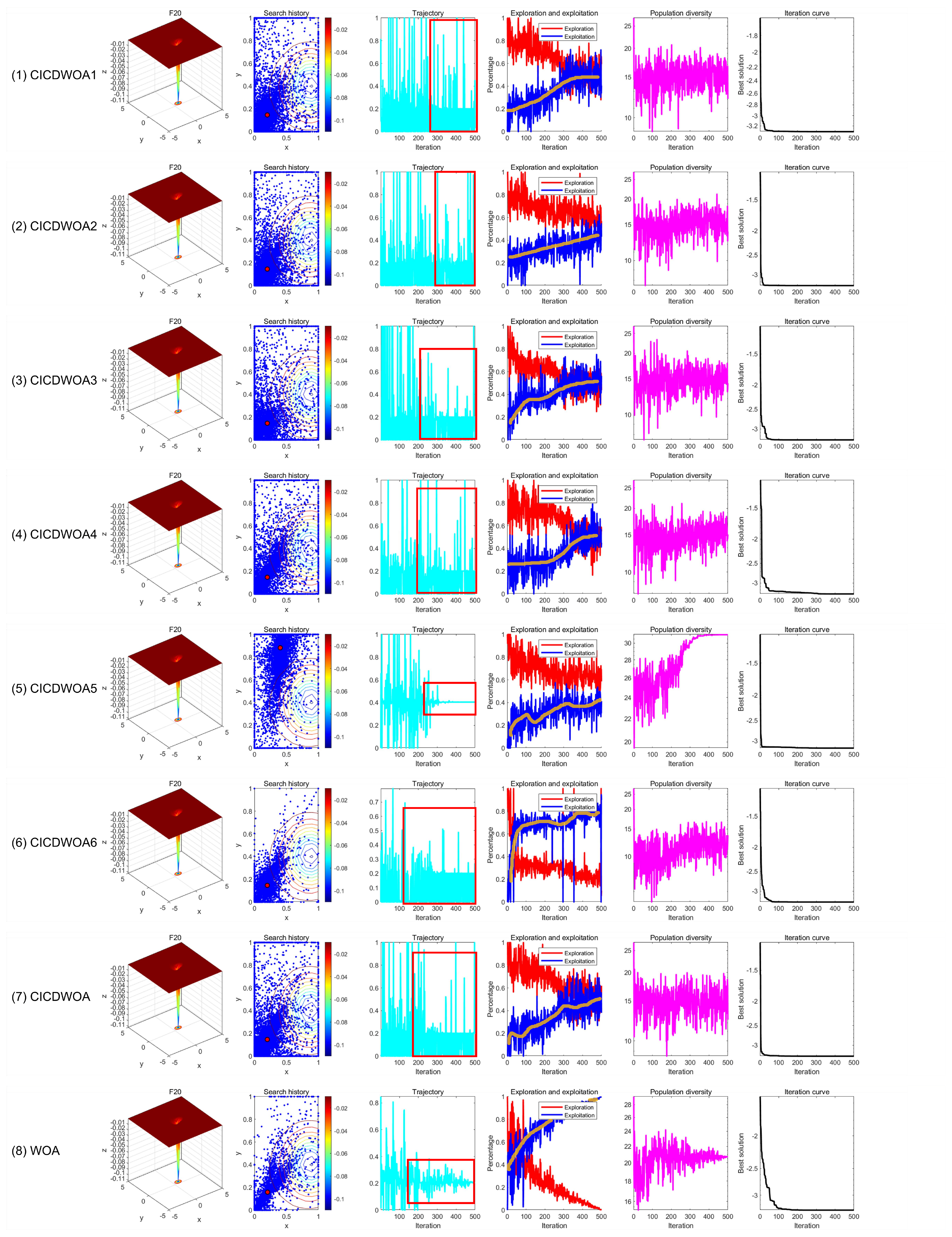}
    \caption{Results of qualitative analysis experiment (F20).} 
    \label{quaF20}
\end{figure*}

Meanwhile we selected three functions from different types and performed a qualitative analysis experiment on the CICDWOAs (CICDWOA1, CICDWOA2, CICDWOA3,..., CICDWOA6, CICDWOA) and WOA to comprehensively evaluate the effects of each component. The results of this qualitative analysis are presented in Figure\ref{quaF5}, Figure\ref{quaF8}, and Figure\ref{quaF20}, which primarily include the following components:\par
\begin{itemize}
  \item Landscapes of the benchmark functions;
  \item Search history of the whale population;
  \item the trajectory of a search agent;
  \item Exploration and exploitation ratio curves;
  \item Population diversity curves;
  \item Iterative convergence curves of CICDWOA.
\end{itemize}\par

From the results in Figure~\ref{abla_fig}, Figure\ref{quaF5}, Figure\ref{quaF8}, and Figure\ref{quaF20}, it is clear that each improvement module has played a significant role in enhancing the performance of CICDWOA. First, Good Nodes Set Initialization helps to generate a more evenly distributed whale population during the initial stage, expanding the coverage of the search space and avoiding search blind spots caused by individual clustering. This mechanism demonstrates strong global exploration ability, especially when dealing with complex multi-modal functions (such as F12-F14 and F20-F23), enabling the algorithm to fully explore potential optimal solution regions. Second, The light orange curve labeled 'Exploration and exploitation' in Figure \ref{quaF20} represents the overall trend of 'Exploitation'. From these curves, it is evident that the nonlinear convergence factor $a$ based on the Sigmoid function introduced in this study provides CICDWOA with a dynamic search balance adjustment mechanism. This design maintains strong global exploration capability during the early iterations while gradually enhancing local exploitation ability in the later stages, achieving a good trade-off between solution accuracy and convergence speed. The control of nonlinear variations smooths the convergence curves for functions such as F5-F8 and F12-F14, allowing the CICDWOA to continuously approach the global optimum during later iterations. However, the curve in CICDWOA2, which replaces the convergence factor $a$ with a linear updating approach, exhibits $a$ steadily increasing linear trend rather than a distinct nonlinear behavior.\par
Moreover, the Collective Cognitive Sharing (CCS) mechanism effectively shares and collaboratively updates global information by introducing a combined feedback from the population's average position and leader distance. This mechanism guides individual movement based on the collective cognitive information of the population and adaptive decay of the leader's attraction, thus achieving a dynamic balance between exploration and exploitation. In the 'Search history' graphs of Figure \ref{quaF5}, it is evident that the Collective Cognitive Sharing (CCS) mechanism effectively enables other search agents to appropriately converge towards the current optimal solution rather than engaging in blind searches. The red box delineates the primary area where the entire whale population is concentrated. In contrast, the search area of the population in CICDWOA3, which excludes the CCS mechanism, is more dispersed compared to the other variants of CICDWOA. This illustrates that the Collective Cognitive Sharing (CCS) mechanism significantly enhances the information utilization among individuals in the population, resulting in a more efficient search process. The results from F1-F6 and F12-F14 demonstrate that this mechanism significantly enhances the global coherence and convergence efficiency of the algorithm. The Adaptive Exponential Spiral (AES) strategy introduces a nonlinear variation in the spiral step size, which results in periodic and random updates to individual positions. This improvement enhances the diversity of the search path, allowing the algorithm to escape local optima and avoid premature convergence in complex multi-modal environments (e.g., F5-F6 and F12-F13). The Enhanced Spiral Updating Strategy further enriches the whale's trajectory. The Trajectory plots in Figure \ref{quaF20} demonstrate that the Cauchy Inverse Cumulative Distribution within the Enhanced Spiral Updating Strategy provides search agents with a substantial ability to escape local optima. This can be observed from the Trajectory plot of CICDWOA5, which omits the Enhanced Spiral Updating Strategy. The search agents exhibit limited 'step sizes' during each movement, resulting in minimal movement towards the later iterations. In contrast, the search agents in the other CICDWOA variants exhibit a combination of very large, large and small step sizes, maintaining significant search capacity and population diversity from the beginning to the end of the iterations. This component accelerates global convergence speed while maintaining search diversity, leading to higher convergence efficiency and solution accuracy when dealing with functions such as F1-F4 and F9-F11. Finally, from the convergence curves of F13-F15, it can be observed that Hybrid Gaussian-Cauchy Mutation based on DE introduces a multi-scale perturbation mechanism. By applying compound mutations from Gaussian and Cauchy distributions, this mechanism effectively enhances population diversity and local jump capability. The 'Population diversity' curve in Figure \ref{quaF8} clearly indicates that the population diversity of CICDWOA6, which excludes the Hybrid Gaussian-Cauchy Mutation based on DE, is significantly lower than that of the other CICDWOA variants (with an average population diversity of over 7000 for CICDWOA6, while the average population diversity for the other CICDWOA variants is consistently above 10000). This mechanism helps individuals break out of local optima traps, ensuring that CICDWOA maintains strong global search activity throughout the optimization process and avoids converging to sub-optimal solutions.\par
In addition to the convergence curves, the results presented in Table~\ref{tab:friedman_cicdwoa} provide further quantitative evidence of the effectiveness of the proposed improvement strategies. The Friedman test results clearly indicate that the complete version of CICDWOA achieves the best overall rank across the 23 benchmark functions, outperforming all its simplified variants (CICDWOA1-CICDWOA6). This demonstrates that the combined integration of the proposed modules yields superior global optimization capability and stability. Specifically, CICDWOA5 and CICDWOA6, which partially include the nonlinear convergence and collective sharing mechanisms, also perform relatively well, confirming that these modules play a pivotal role in enhancing convergence efficiency. Overall, the results in both the figure and the table consistently verify that the hybridized architecture of CICDWOA significantly improves both convergence accuracy and robustness across diverse optimization landscapes.

\begin{table*}[htbp]
\centering
\caption{Friedman test results of the CICDWOAs on 23 benchmark functions.}
\resizebox{\textwidth}{!}{
\begin{tabular}{cccccccc}
\toprule
Function & CICDWOA1 & CICDWOA2 & CICDWOA3 & CICDWOA4 & CICDWOA5 & CICDWOA6 & CICDWOA \\ 
\midrule
F1  & 3.5000 & 3.5000 & 3.5000 & 3.5000 & 7.0000 & 3.5000 & 3.5000 \\ 
F2  & 3.5000 & 3.5000 & 3.5000 & 3.5000 & 7.0000 & 3.5000 & 3.5000 \\ 
F3  & 3.5000 & 3.5000 & 3.5000 & 3.5000 & 7.0000 & 3.5000 & 3.5000 \\ 
F4  & 3.5000 & 3.5000 & 3.5000 & 3.5000 & 7.0000 & 3.5000 & 3.5000 \\ 
F5  & 7.0000 & 5.9000 & 4.9667 & 3.5333 & 1.0000 & 2.9333 & 2.6667 \\ 
F6  & 2.9667 & 5.1667 & 7.0000 & 5.8333 & 1.0000 & 3.2000 & 2.8333 \\ 
F7  & 3.8000 & 3.7333 & 3.8333 & 3.7667 & 6.0667 & 3.2333 & 3.5667 \\ 
F8  & 6.1667 & 6.3000 & 3.3667 & 5.4000 & 1.0000 & 2.7000 & 3.0667 \\ 
F9  & 3.7333 & 3.7333 & 3.7333 & 3.7333 & 5.6000 & 3.7333 & 3.7333 \\ 
F10 & 3.7500 & 3.7500 & 3.7500 & 3.7500 & 5.5000 & 3.7500 & 3.7500 \\ 
F11 & 4.0000 & 4.0000 & 4.0000 & 4.0000 & 4.0000 & 4.0000 & 4.0000 \\ 
F12 & 2.9333 & 5.5000 & 6.3667 & 6.1333 & 1.0000 & 3.0000 & 3.0667 \\ 
F13 & 3.3000 & 5.2667 & 6.3333 & 5.8000 & 1.0333 & 3.5667 & 2.7000 \\ 
F14 & 3.2500 & 6.0000 & 4.9500 & 5.6000 & 1.7000 & 3.9500 & 2.5500 \\ 
F15 & 6.0000 & 4.1333 & 3.6333 & 5.9000 & 1.4667 & 3.6000 & 3.2667 \\ 
F16 & 3.7667 & 6.9667 & 3.4000 & 6.0333 & 2.8000 & 3.6667 & 1.3667 \\ 
F17 & 3.6500 & 6.9333 & 3.8333 & 6.0333 & 2.6500 & 3.4833 & 1.4167 \\ 
F18 & 4.4000 & 4.5333 & 3.9333 & 4.9667 & 1.1333 & 5.7333 & 3.3000 \\ 
F19 & 2.5667 & 4.8333 & 3.4333 & 6.4667 & 2.0000 & 5.8000 & 2.9000 \\ 
F20 & 4.4667 & 5.2667 & 3.5000 & 6.3333 & 2.9667 & 3.0000 & 2.4667 \\ 
F21 & 5.7000 & 5.9667 & 3.2667 & 5.7667 & 1.0000 & 3.0667 & 3.2333 \\ 
F22 & 5.7333 & 5.8667 & 3.3000 & 5.9333 & 1.0000 & 3.1000 & 3.0667 \\ 
F23 & 6.1667 & 6.1000 & 3.3000 & 5.4333 & 1.0000 & 3.0667 & 2.9333 \\ \midrule
Average & 4.2326 & 4.9543 & 4.0826 & 4.9746 & 3.1268 & 3.5906 & \textbf{3.0384} \\ \midrule
Rank & 5 & 6 & 4 & 7 & 2 & 3 & \textbf{1} \\ 
\bottomrule
\end{tabular}
}
\label{tab:friedman_cicdwoa}
\end{table*}

\subsection{Qualitative Analysis Experiment}
\begin{figure*}[htbp]
    \centering
    \includegraphics[width=\textwidth]{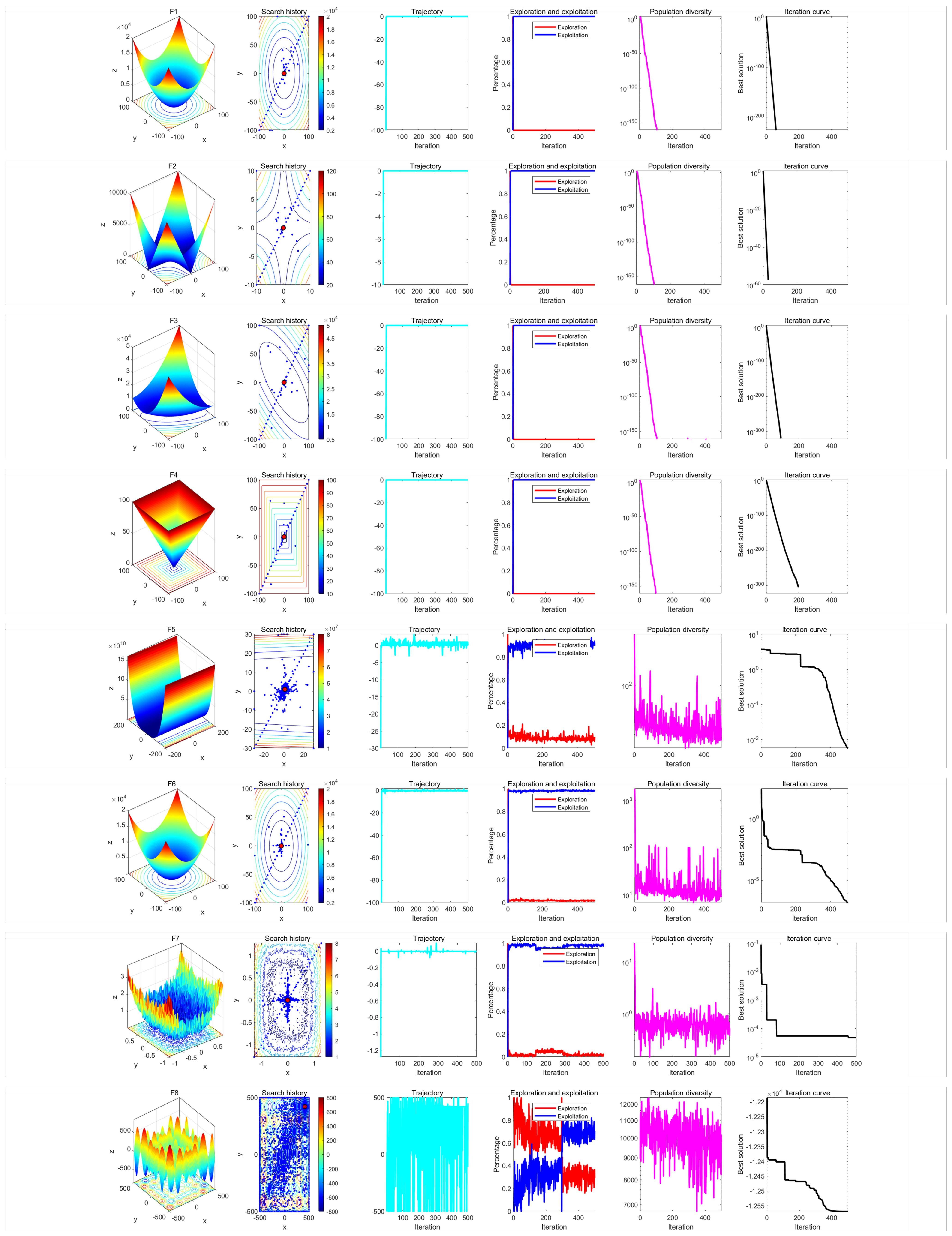}
    \caption{Results of qualitative analysis experiment (F1-F8).} 
    \label{qua1}
\end{figure*}

\begin{figure*}[htbp]
    \centering
    \includegraphics[width=\textwidth]{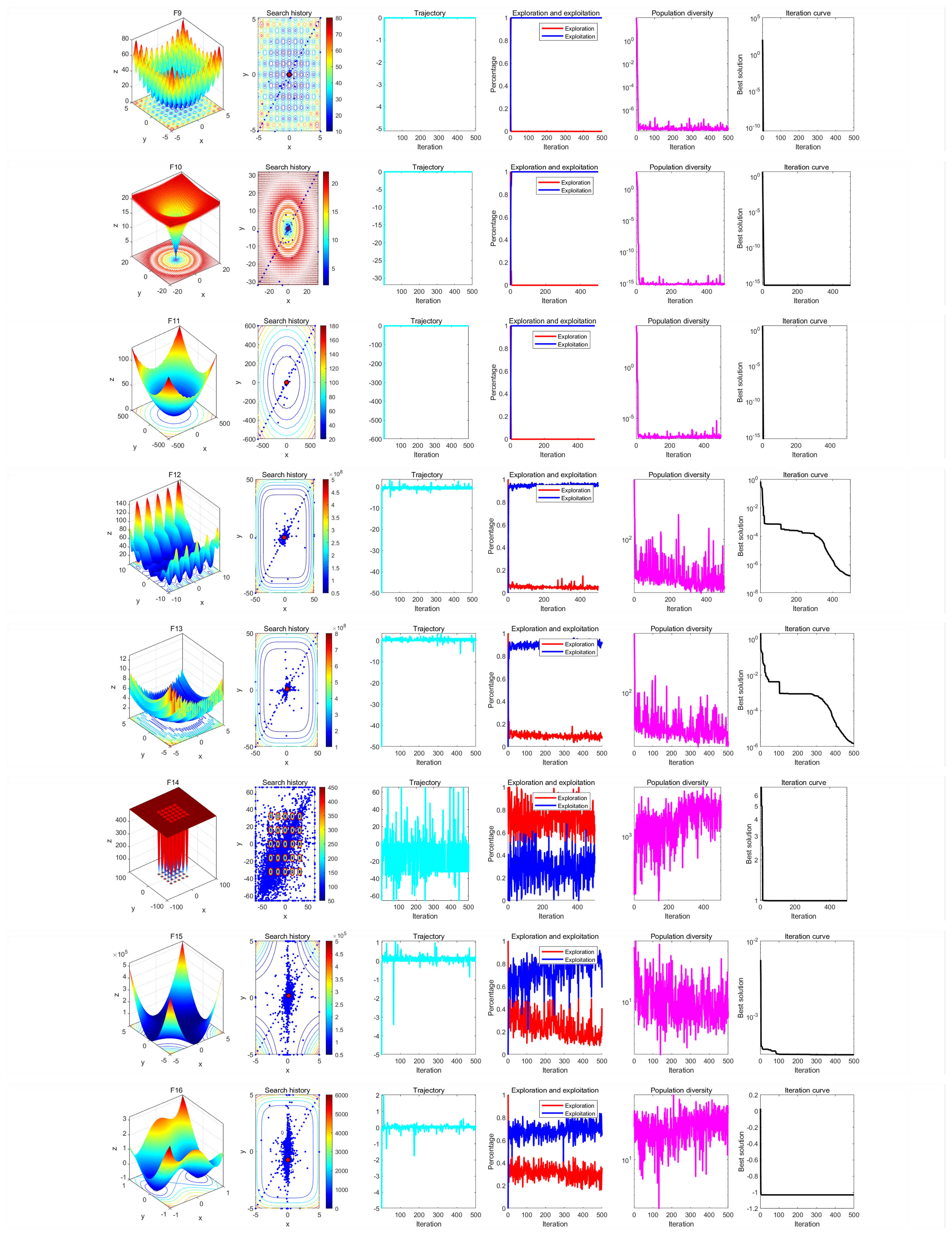}
    \caption{Results of qualitative analysis experiment (F9-F16).} 
    \label{qua2}
\end{figure*}

\begin{figure*}[htbp]
    \centering
    \includegraphics[width=\textwidth]{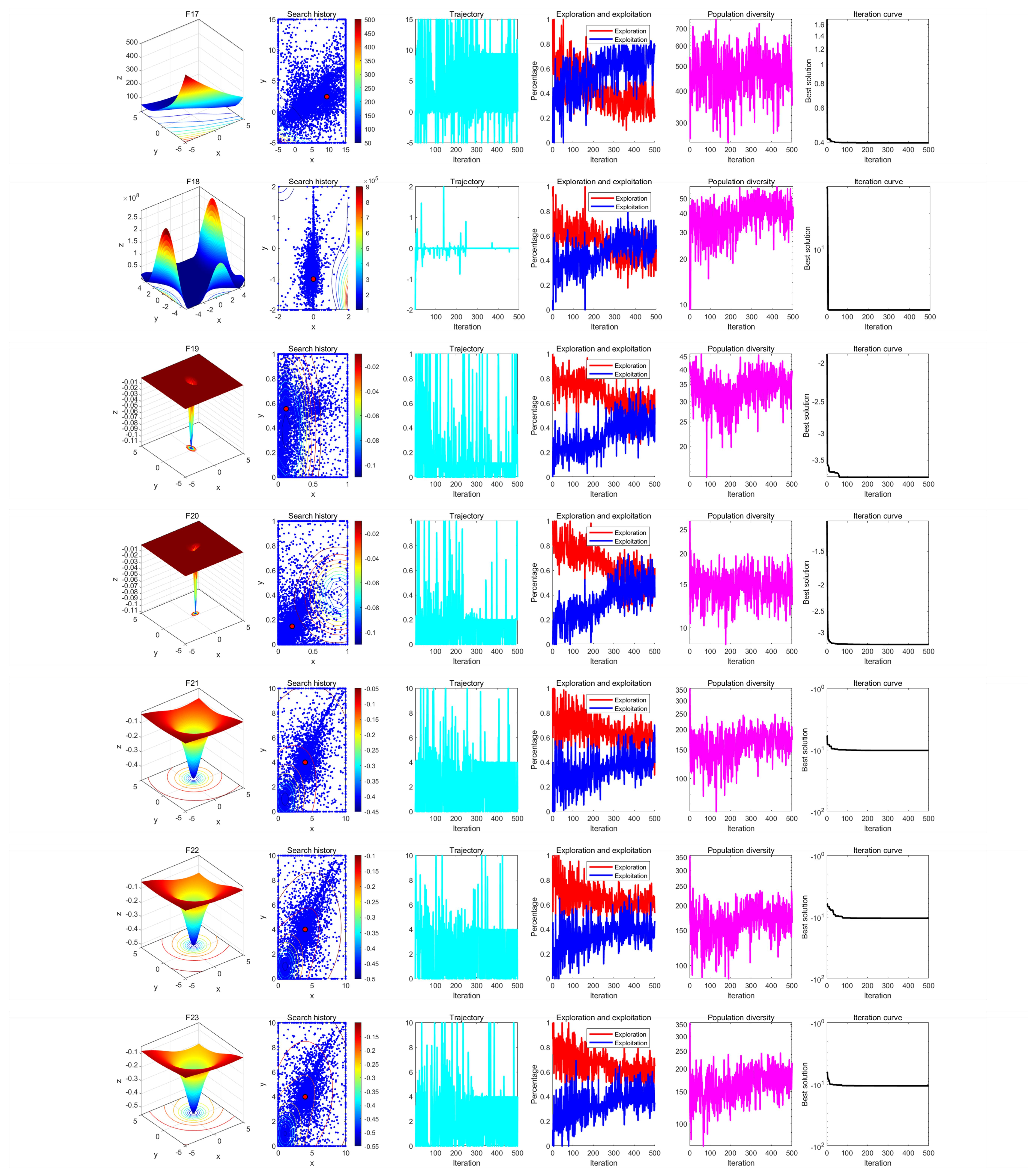}
    \caption{Results of qualitative analysis experiment (F17-F23).} 
    \label{qua3}
\end{figure*}

In the qualitative analysis experiments, we set the number of iterations to $T$=500 and the population size to $N$=30, running CICDWOA independently on 23 benchmark functions, as shown in Table~\ref{funcdetails}, to analyze its search history, exploration and exploitation ratio, and population diversity. Additionally, to facilitate comparison and understanding, we provide the function landscapes and iteration curves for each benchmark function. The results of the qualitative analysis are presented in Figure\ref{qua1}, Figure\ref{qua2}, and Figure\ref{qua3}, which primarily include the following components:\par
\begin{itemize}
  \item Landscapes of the benchmark functions;
  \item Search history of the whale population;
  \item the trajectory of a search agent;
  \item Exploration and exploitation ratio curves;
  \item Population diversity curves;
  \item Iterative convergence curves of CICDWOA.
\end{itemize}\par
The search history graph reflects the distribution of whale individuals' positions during the search process. In the whale population's search history graph, red dots indicate the global optimal position, while blue dots represent the search trajectory of the whale individuals. It is clearly visible that CICDWOA effectively explores the entire search space. For uni-modal functions (e.g., F1-F6), CICDWOA exhibits rapid convergence, with whale individuals finding the optimal solution in fewer iterations, causing the individual distribution in the solution space to quickly concentrate. For complex functions (e.g., F7-F8, F14, and F17-F23), due to the presence of many local optima, CICDWOA performs a rapid global exploration in the early stages and then focuses on refining potential high-quality regions later. The results indicate that whale individuals traverse most of the solution space, with their search trajectories mainly concentrated around the optimal solution. Regarding the balance between exploration and exploitation, CICDWOA demonstrates excellent dynamic control capability, effectively coordinating the relationship between the two. When handling functions such as F1-F13, CICDWOA shows a high exploration ratio in the early stages of iteration, then quickly enhances the exploitation intensity, showcasing its powerful global search ability. On functions such as F14-F23, CICDWOA exhibits a higher exploitation ratio in the early stages, which gradually decreases steadily with iterations, indicating that the algorithm excels in both global exploration and local exploitation. Furthermore, for multi-modal complex functions such as F5-F8 and F12-F23, the population diversity curve of CICDWOA remains highly volatile and maintains a high level. This demonstrates that CICDWOA preserves a high level of population diversity throughout the search process, effectively avoiding premature convergence caused by individual clustering. Overall, CICDWOA exhibits outstanding performance in terms of exploration depth, convergence speed, and stability, achieving a balanced trade-off between global optimization and local refinement.

\subsection{Comparative Experiment}
\begin{figure*}[htbp]
    \centering
    \includegraphics[width=\textwidth]{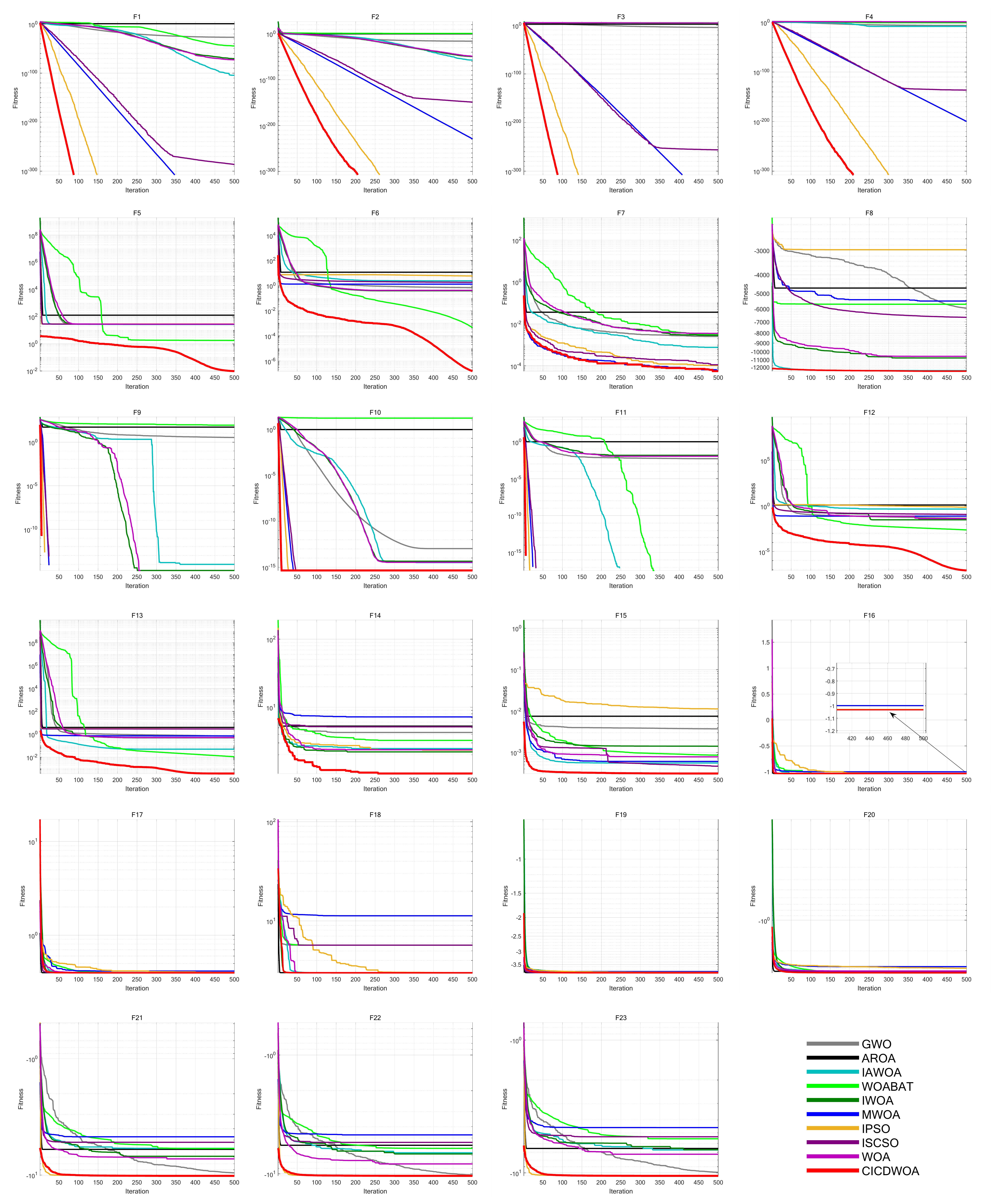}
    \caption{Comparative results of each algorithm in Comparative experiment.} 
    \label{differ1}
\end{figure*}

To validate the superiority of CICDWOA, we selected classic metaheuristic algorithms, outstanding competitive variants, and the latest algorithms from the past two years for comparison with CICDWOA on the benchmark functions listed in Table~\ref{funcdetails}. These competitors include Grey Wolf Optimizer (GWO) \cite{GWO}, Attraction-Repulsion Optimization Algorithm (AROA) \cite{AROA}, IAWOA \cite{IAWOA}, WOABAT \cite{WOABAT}, IWOA \cite{IWOA}, MWOA \cite{MWOA}, IPSO \cite{IPSO}, ISCSO \cite{ISCSO} and WOA \cite{WOA}. The parameter settings for each algorithm are shown in Table~\ref{setting} (To ensure the fairness of the experiments, these parameters were directly taken from the original papers.). The number of iterations was uniformly set to $T$=500, and the population size was set to $N$=30. Each algorithm was independently run 30 times on the 23 benchmark functions for parametric testing and non-parametric testing. The average fitness ($Ave$), standard deviation ($Std$), $p$-values from the Wilcoxon rank-sum test, and Friedman values were recorded for performance analysis. The experimental results are shown in Figure\ref{differ1}, Table~\ref{differ2}, and Table~\ref{differ3}.\par

\begin{table}[htbp]
    \centering
    \caption{Parameter settings for the metaheuristic algorithms in this research.}
    \begin{tabular}{@{}ccc@{}}
        \toprule
        \textbf{Algorithm} & \textbf{Parameter(s)} & \textbf{Value} \\ 
        \midrule
        GWO~\cite{GWO} & Convergence Factor $a$ & 2 decreasing to 0 \\ 
        \midrule
        AROA~\cite{AROA} & Attraction factor $c$ & 0.95 \\ 
		                     & Local search scaling factor 1 & 0.15 \\ 
		                     & Local search scaling factor 2 & 0.6 \\ 
		                     & Attraction probability 1 & 0.2 \\ 
		                     & Local search probability & 0.8 \\ 
		                     & Expansion factor & 0.4 \\ 
		                     & Local search threshold 1 & 0.9 \\
		                     & Local search threshold 2 & 0.85 \\ 
		                     & Local search threshold 3 & 0.9 \\ 
        \midrule
        IAWOA~\cite{IAWOA} & Convergence Factor $a$ & 2 decreasing to 0 \\ 
            & Spiral Factor $b$ & 1 \\ \midrule
        WOABAT~\cite{WOABAT} & Convergence Factor $a$ & 2 decreasing to 0 \\ 
            & Spiral Factor $b$ & 1 \\ \midrule
        IWOA~\cite{IWOA} & Convergence Factor $a$ & 2 decreasing to 0 \\ 
            & Spiral Factor $b$ & 1 \\ \midrule
        MWOA~\cite{MWOA} & Convergence Factor $a$ & 2 decreasing to 0 \\ 
        & Spiral Factor $b$ & 1 \\ 
        & ${CF}_1$ & 2.5 \\ 
        & ${CF}_2$ & 2.5 \\
        \midrule
        IPSO~\cite{IPSO} & Inertia Weight $\omega$  & 0.9 decreasing to 0 \\
            &  $C_1$ & 2 \\ 
            &  $C_2$ & 2 \\
        \midrule
        ISCSO~\cite{ISCSO} & Maximum Sensitivity Range $S$ & 2 \\ 
         & $R$ & [-2, 2] \\
        \midrule
        WOA~\cite{WOA} & Convergence Factor $a$ & 2 decreasing to 0 \\ 
            & Spiral Factor $b$ & 1 \\ 
        \midrule
        CICDWOA & Convergence Factor $a$ & 2 decreasing to 0 \\ 
             & Spiral Factor $b$ & 1 \\ 
             & Inertia weight $\omega$ & 0 increase to 1 \\
             & $s_1$ & 20 \\
             & $s_2$ & 25 \\
        \bottomrule
    \end{tabular}
    \label{setting}
\end{table}

\begin{table*}[htbp]
\centering
\small
\caption{Parametric Comparison results of different algorithms on 23 benchmark functions.}
\resizebox{\textwidth}{!}{
\begin{tabular}{cccccccccccc}
\toprule
\textbf{Function} & \textbf{Metrics} & \textbf{GWO} & \textbf{AROA} & \textbf{IAWOA} & \textbf{WOABAT} & \textbf{IWOA} & \textbf{MWOA} & \textbf{IPSO} & \textbf{ISCSO} & \textbf{WOA} & \textbf{CICDWOA} \\
\midrule
F1 & Ave & 1.7204E-27 & 6.4945E+00 & 3.7944E-105 & 2.7110E-45 & 4.7702E-71 & \textbf{0.0000E+00} & \textbf{0.0000E+00} & 3.4290E-287 & 1.0679E-73 & \textbf{0.0000E+00} \\
   & Std & 2.1454E-27 & 7.6741E+00 & 2.0743E-104 & 7.3667E-45 & 2.5230E-70 & \textbf{0.0000E+00} & \textbf{0.0000E+00} & 4.8678E-287 & 4.8990E-73 & \textbf{0.0000E+00} \\
F2 & Ave & 8.9238E-17 & 7.6754E-01 & 3.5968E-58 & 6.3333E+00 & 2.7761E-50 & 6.5276E-230 & \textbf{0.0000E+00} & 1.1848E-149 & 6.9080E-50 & \textbf{0.0000E+00} \\
   & Std & 5.0561E-17 & 3.3498E-01 & 1.4604E-57 & 1.0981E+01 & 1.2816E-49 & 7.5643E-230 & \textbf{0.0000E+00} & 5.1868E-149 & 2.4554E-49 & \textbf{0.0000E+00} \\
F3 & Ave & 2.8468E-05 & 1.5316E+02 & 1.0423E+04 & 7.9026E+04 & 6.2699E+03 & \textbf{0.0000E+00} & \textbf{0.0000E+00} & 6.5339E-257 & 3.8462E+04 & \textbf{0.0000E+00} \\
   & Std & 7.4423E-05 & 1.3460E+02 & 1.7128E+04 & 2.9909E+04 & 1.2735E+04 & \textbf{0.0000E+00} & \textbf{0.0000E+00} & 7.7580E-257 & 9.1627E+03 & \textbf{0.0000E+00} \\
F4 & Ave & 8.9355E-07 & 1.7370E+00 & 1.2878E-08 & 3.4728E-01 & 1.2564E+01 & 1.7038E-200 & \textbf{0.0000E+00} & 3.6013E-137 & 3.7079E+01 & \textbf{0.0000E+00} \\
   & Std & 1.0346E-06 & 8.9614E-01 & 6.8775E-08 & 1.8747E+00 & 1.5893E+01 & 3.4672E-200 & \textbf{0.0000E+00} & 1.6636E-136 & 2.4459E+01 & \textbf{0.0000E+00} \\
F5 & Ave & 2.7030E+01 & 1.2882E+02 & 2.6670E+01 & 1.7732E+00 & 2.7799E+01 & 2.8676E+01 & 2.8912E+01 & 2.7885E+01 & 2.8018E+01 & \textbf{9.7192E-03} \\
   & Std & 7.9757E-01 & 1.2196E+02 & 4.9212E+00 & 6.7423E+00 & 4.7189E-01 & 1.4465E-01 & 1.4102E-01 & 8.1344E-01 & 5.1526E-01 & \textbf{2.6365E-02} \\
F6 & Ave & 6.7972E-01 & 1.1444E+01 & 2.3666E+00 & 4.3863E-04 & 3.8277E-01 & 1.2880E+00 & 5.7531E+00 & 1.7943E+00 & 4.0871E-01 & \textbf{1.5374E-07} \\
   & Std & 2.3013E-01 & 3.5108E+00 & 1.0543E+00 & 1.1526E-03 & 2.1448E-01 & 3.7236E-01 & 4.0669E-01 & 5.5908E-01 & 2.6185E-01 & \textbf{1.8760E-07} \\
F7 & Ave & 2.5112E-03 & 3.6182E-02 & 7.4509E-04 & 3.0470E-03 & 2.8338E-03 & 6.0267E-05 & 1.0357E-04 & 1.1061E-04 & 3.4982E-03 & \textbf{5.5936E-05} \\
   & Std & 1.2983E-03 & 2.5882E-02 & 5.8719E-04 & 5.2195E-03 & 3.0407E-03 & 4.9004E-05 & 8.4594E-05 & 1.8251E-04 & 4.9858E-03 & \textbf{4.8757E-05} \\
F8 & Ave & -5.9294E+03 & -4.6757E+03 & -1.2482E+04 & -5.6712E+03 & -1.0745E+04 & -5.4580E+03 & -2.9628E+03 & -6.6259E+03 & -1.0541E+04 & \textbf{-1.2569E+04} \\
   & Std & 1.1066E+03 & 7.0449E+02 & 1.4068E+02 & 6.7245E+02 & 1.6436E+03 & 2.1501E+03 & 9.5502E+02 & 7.5120E+02 & 1.7902E+03 & \textbf{1.0649E-02} \\
F9 & Ave & 3.5247E+00 & 5.4279E+01 & 9.4739E-15 & 9.0674E+01 & 1.8948E-15 & \textbf{0.0000E+00} & \textbf{0.0000E+00} & \textbf{0.0000E+00} & \textbf{0.0000E+00} & \textbf{0.0000E+00} \\
   & Std & 4.1913E+00 & 6.7952E+01 & 3.0165E-14 & 9.1154E+01 & 1.0378E-14 & \textbf{0.0000E+00} & \textbf{0.0000E+00} & \textbf{0.0000E+00} & \textbf{0.0000E+00} & \textbf{0.0000E+0}0 \\
F10 & Ave & 1.0454E-13 & 9.1013E-01 & 3.1678E-15 & 1.5766E+01 & 4.4705E-15 & \textbf{4.4409E-16} & \textbf{4.4409E-16} & \textbf{4.4409E-16} & 3.4047E-15 & \textbf{4.4409E-16} \\
    & Std & 1.6952E-14 & 4.2085E-01 & 2.5861E-15 & 8.0961E+00 & 2.0298E-15 & \textbf{0.0000E+00} & \textbf{0.0000E+00} & \textbf{0.0000E+00} & 1.0993E-02 & \textbf{0.0000E+00} \\
F11 & Ave & 5.0962E-03 & 1.0194E+00 & \textbf{0.0000E+00} & \textbf{0.0000E+00} & 1.4710E-02 & \textbf{0.0000E+00} & \textbf{0.0000E+00} & \textbf{0.0000E+00} & 1.0993E-02 & \textbf{0.0000E+00} \\
    & Std & 9.4374E-03 & 6.7871E-02 & \textbf{0.0000E+00} & \textbf{0.0000E+00} & 6.3009E-02 & \textbf{0.0000E+00} & \textbf{0.0000E+00} & \textbf{0.0000E+00} & 4.1835E-02 & \textbf{0.0000E+00} \\
F12 & Ave & 4.6518E-02 & 1.2950E+00 & 4.6876E-01 & 2.3747E-03 & 3.0467E-02 & 7.7196E-02 & 7.3685E-01 & 1.1967E-01 & 4.1019E-02 & \textbf{8.4478E-08} \\
    & Std & 2.7506E-02 & 3.2232E-01 & 3.7622E-01 & 5.2973E-03 & 4.8528E-02 & 4.1035E-02 & 1.5171E-01 & 5.9923E-02 & 1.1193E-01 & \textbf{9.7477E-08} \\
F13 & Ave & 6.8702E-01 & 3.9241E+00 & 4.9433E-02 & 1.0828E-02 & 4.8950E-01 & 7.1263E-01 & 2.8686E+00 & 2.8785E+00 & 5.0927E-01 & \textbf{3.6760E-04} \\
    & Std & 2.6603E-01 & 6.1623E-01 & 3.1472E-02 & 2.2297E-02 & 2.2625E-01 & 1.6917E-01 & 1.6937E-01 & 1.2771E-01 & 2.7143E-01 & \textbf{2.0065E-03} \\
F14 & Ave & 4.2636E+00 & 5.2880E+00 & 2.4753E+00 & 3.2595E+00 & 2.2106E+00 & 7.2213E+00 & 2.3446E+00 & 5.1714E+00 & 2.3756E+00 & \textbf{9.9800E-01} \\
    & Std & 4.0624E+00 & 4.3258E+00 & 2.3320E+00 & 3.1062E+00 & 2.5506E+00 & 4.1595E+00 & 2.3572E+00 & 4.5239E+00 & 2.5634E+00 & \textbf{3.5427E-16} \\
F15 & Ave & 3.7219E-03 & 7.5163E-03 & 5.5514E-04 & 8.5755E-04 & 1.4171E-03 & 6.0723E-04 & 1.1272E-02 & 4.6455E-04 & 7.8188E-04 & \textbf{3.1007E-04} \\
    & Std & 7.5699E-03 & 8.6662E-03 & 3.4804E-04 & 5.6753E-04 & 3.3840E-03 & 1.3887E-04 & 1.6440E-02 & 3.3065E-04 & 5.3488E-04 & \textbf{1.1884E-05} \\
F16 & Ave & \textbf{-1.0316E+00} & \textbf{-1.0316E+00} & -1.0314E+00 & \textbf{-1.0316E+00} & \textbf{-1.0316E+00} & -9.9854E-01 & -1.0298E+00 & \textbf{-1.0316E+00} & \textbf{-1.0316E+00} & \textbf{-1.0316E+00} \\
    & Std & 2.526E-08 & 7.7876E-05 & 9.8690E-04 & 5.9236E-13 & 2.1342E-09 & 2.6893E-02 & 3.4207E-03 & 1.1354E-08 & 3.2999E-09 & \textbf{1.0192E-15} \\
F17 & Ave & 3.9791E-01 & 3.9850E-01 & \textbf{3.9789E-01} & \textbf{3.9789E-01} & 3.9790E-01 & 4.1529E-01 & 4.0194E-01 & \textbf{3.9789E-01} & \textbf{3.9789E-01} & \textbf{3.9789E-01} \\
    & Std & 1.0323E-04 & 1.5824E-03 & 2.3743E-06 & 2.1987E-10 & 1.8799E-05 & 2.2798E-02 & 5.2837E-03 & 1.1045E-06 & 1.1522E-05 & \textbf{7.3043E-14} \\
F18 & Ave & 5.7000E+00 & 3.0083E+00 & 3.0051E+00 & 5.7000E+00 & 3.0001E+00 & 1.1271E+01 & 3.0053E+00 & 5.7000E+00 & 3.0001E+00 & \textbf{3.0000E+00} \\
    & Std & 1.4789E+01 & 2.9149E-02 & 1.0849E-02 & 8.2385E+00 & 1.2381E-04 & 1.1786E+01 & 8.3324E-03 & 1.4789E+01 & 3.9482E-04 & \textbf{1.8337E-05} \\
F19 & Ave & -3.8612E+00 & -3.8561E+00 & -3.8446E+00 & \textbf{-3.8628E+00} & -3.8506E+00 & -3.8046E+00 & -3.8430E+00 & -3.8602E+00 & -3.8581E+00 & \textbf{-3.8628E+00} \\
    & Std & 2.2669E-03 & 1.1968E-02 & 4.0853E-02 & 1.2524E-06 & 2.8434E-02 & 4.8814E-02 & 6.1681E-02 & 3.6448E-03 & 5.9421E-03 & \textbf{3.8494E-07} \\
F20 & Ave & -3.2410E+00 & -3.2009E+00 & -3.2069E+00 & -3.2930E+00 & -3.1914E+00 & -2.8826E+00 & -2.9929E+00 & -3.2277E+00 & -3.1813E+00 & \textbf{-3.3141E+00} \\
    & Std & 9.4645E-02 & 9.2827E-02 & 5.5802E-02 & 5.3606E-02 & 1.0826E-01 & 3.1418E-01 & 1.6892E-01 & 1.2155E-01 & 1.4796E-01 & \textbf{3.0179E-02} \\
F21 & Ave & -9.5658E+00 & -6.0138E+00 & -5.8849E+00 & -5.9049E+00 & -6.8957E+00 & -4.6899E+00 & -1.0118E+01 & -5.2251E+00 & -7.2554E+00 & \textbf{-1.0153E+01} \\
    & Std & 1.8248E+00 & 3.1373E+00 & 1.8896E+00 & 1.9324E+00 & 2.4614E+00 & 9.9256E-01 & 2.5822E-02 & 9.3049E-01 & 2.7963E+00 & \textbf{1.0869E-10} \\
F22 & Ave & -1.0224E+01 & -5.7512E+00 & -6.6479E+00 & -6.1053E+00 & -6.8036E+00 & -4.7053E+00 & -1.0367E+01 & -5.4420E+00 & -8.2618E+00 & \textbf{-1.0403E+01} \\
    & Std & 9.7005E-01 & 3.1365E+00 & 2.4272E+00 & 2.1996E+00 & 2.5786E+00 & 6.6944E-01 & 2.3729E-02 & 1.3483E+00 & 2.6364E+00 & \textbf{6.2428E-11} \\
F23 & Ave & -9.9040E+00 & -6.5499E+00 & -6.7207E+00 & -5.5359E+00 & -6.7014E+00 & -4.5629E+00 & -1.0498E+01 & -5.3496E+00 & -7.2414E+00 & \textbf{-1.0536E+01} \\
    & Std & 1.9682E+00 & 3.6234E+00 & 2.4785E+00 & 1.7788E+00 & 2.5512E+00 & 5.6267E-01 & 2.5978E-02 & 1.6028E+00 & 2.9978E+00 & \textbf{1.3651E-10} \\
\bottomrule
\end{tabular}
\label{differ2}
}
\end{table*}

\begin{table}[htbp]
    \centering
    \caption{Non-parametric Comparison results of different algorithms on 23 benchmark functions.}
    \begin{tabular}{ccccc}
        \toprule
        \textbf{Algorithm} &  \textbf{Average Friedman Value} & \textbf{Rank}  & \textbf{+/=/-} \\
        \midrule
        GWO      & 5.6732   & 7  & 23/0/0        \\
        AROA      & 8.3449   & 10  & 23/0/0       \\
        IAWOA      & 5.5210   & 4  & 22/1/0       \\
        WOABAT     & 5.0384   & 2 & 22/1/0        \\
        IWOA      & 5.6594   & 6  & 23/0/0        \\
        MWOA     & 6.4094   & 9  & 18/5/0       \\
        IPSO    & 5.9268   & 8  & 16/7/0        \\
        ISCSO     & 5.1239   & 3  & 20/3/0        \\
        WOA      & 5.6239   & 5  & 22/1/0        \\
        CICDWOA  & \textbf{1.6790}  & \textbf{1}  & - \\
        \bottomrule
    \end{tabular}
    \label{differ3}
\end{table}

In order to comprehensively evaluate the practical applicability, versatility, and robustness of the proposed CICDWOA, a series of simulation experiments were conducted across multiple engineering domains. Benchmark tests have demonstrated the algorithm's strong global optimization capability and convergence efficiency. However, it is equally important to assess its performance in solving complex real-world optimization problems that involve nonlinear, high-dimensional, and multimodal characteristics. To this end, the rest part of this study designs four following representative case studies to validate the superiority and effectiveness of CICDWOA in optimization contexts and engineering applications.
\begin{itemize}
  \item 2D Robot Path Planning;
  \item 3D UAV Path Planning;
  \item Engineering Design Optimization;
  \item Feature Selection.
\end{itemize}

\section{2D Robot Path Planning}

\begin{figure}[htbp]
    \centering
    \includegraphics[width=0.9\textwidth]{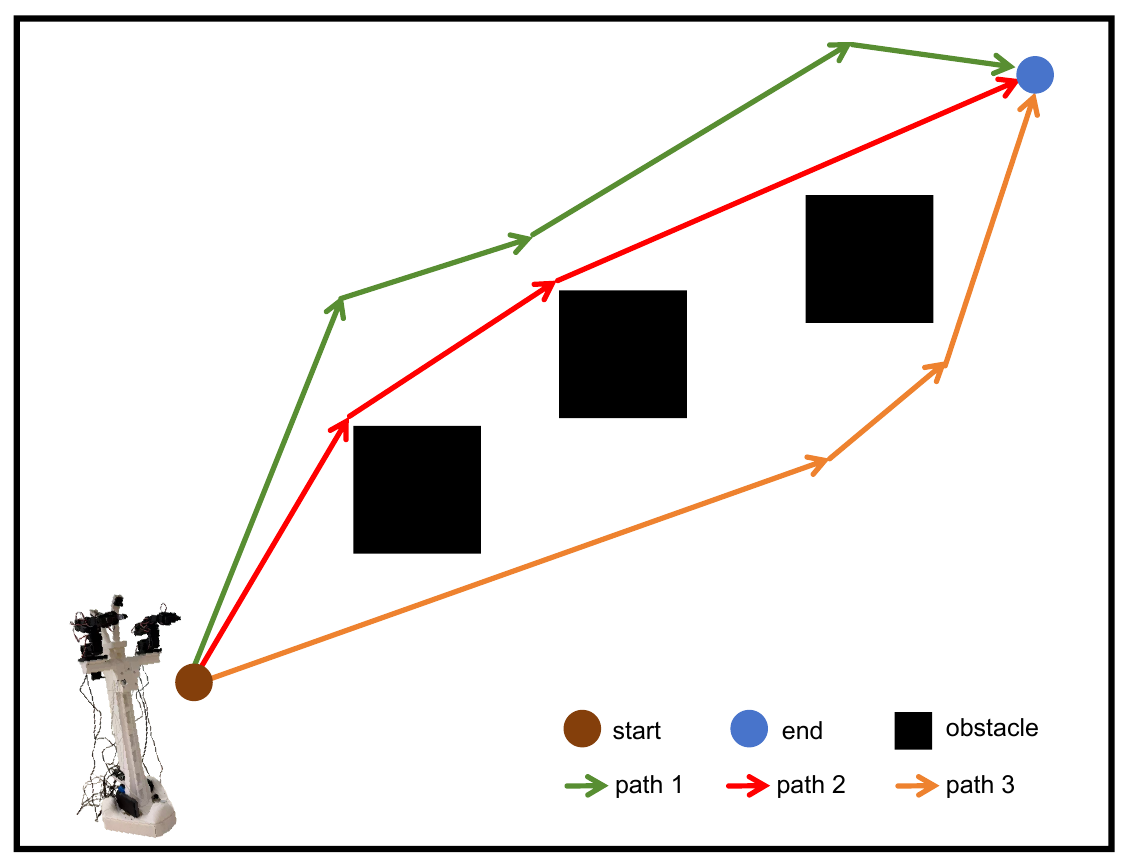}
    \caption{A schematic illustration of 2D robot path planning, showing the start point, target point, obstacles, and the planned paths.} 
    \label{pp1}
\end{figure}

Robot Path Planning is a critical component in the autonomous navigation systems of mobile robots. Its primary objective is to generate a feasible and optimal travel path from the starting point to the target point in a known or partially known environment. The ideal path should ensure obstacle avoidance while minimizing travel distance, maximizing smoothness, and reducing energy consumption. Figure~\ref{pp1} illustrates a typical 2D robot path planning scenario, including the start and target positions, obstacle distribution, and the feasible navigation path.

Due to the nonlinear, high-dimensional, and multi-constrained nature of this problem, traditional deterministic methods often face limitations in complex environments, including high computational costs, susceptibility to local optima, and difficulties in adapting to dynamic changes. As a result, path planning strategies based on metaheuristic algorithms, with their global search capabilities and strong robustness, have become an important research direction for path optimization in complex environments.\par

This study focuses on 2D robot path planning and compares the proposed CICDWOA with nine other metaheuristic algorithms, including Grey Wolf Optimizer (GWO) \cite{GWO}, Attraction-Repulsion Optimization Algorithm (AROA) \cite{AROA}, IAWOA \cite{IAWOA}, WOABAT \cite{WOABAT}, IWOA \cite{IWOA}, MWOA \cite{MWOA}, IPSO \cite{IPSO}, ISCSO \cite{ISCSO} and WOA \cite{WOA}. All algorithms are executed in the same 2D g20$\times$20 grid environment, where the map consists of walkable and obstacle cells. The starting point is set at (1, 1), and the target point is set at ($m$, $n$). The robot is required to search for the optimal path from the starting point to the target while avoiding obstacles.\par
In this task, the path planning problem is formalized as a continuous optimization problem. Each candidate solution (i.e., an individual) is encoded as a sequence of path node coordinates, representing the potential trajectory of the robot. The optimization objective is to minimize the total path cost function, which is determined by both the path length and smoothness metrics, with significant penalties imposed for crossing obstacles or exceeding map boundaries. The fitness function $f(x)$ is calculated as follows: First, the individual position vector $x$ generated by the algorithm is combined with the map information to construct the complete path sequence $[S, x, E]$. Continuous and smoothing operations are then applied to eliminate abrupt nodes in the path, ensuring that the generated trajectory better aligns with the actual movement characteristics of the robot. When the path does not collide with obstacles and remains within the map boundaries, the fitness function computes the sum of the Euclidean distances of the path, which serves as the path cost.\par
\begin{equation}
    f(x)=\sum_{i=1}^{N-1}\sqrt{(p_{i+1}^x-p_i^x)^2+(p_{i+1}^y-p_i^y)^2}
    \label{eq37}
\end{equation}
where $p_i(x,y)$ represents the coordinates of the $i^{th}$ node in the path, and $N$ denotes the number of path nodes. \par
If the path encounters obstacles or exhibits out-of-bounds behavior, a strong penalty term is applied based on the collision count $n_B$, such that:\par
\begin{equation}
    f(x)=C \cdot n_B
    \label{eq38}
\end{equation}
where the parameter $C$ represents a penalty coefficient related to the map size, ensuring that the fitness of invalid paths is significantly worse than that of feasible paths.\par
This mechanism, while ensuring the feasibility of the path, encourages the algorithm to generate shorter, smoother, and completely obstacle-free paths during the search process. All algorithms are executed under the same initial conditions and termination criteria, and iteratively optimize the coordinate distribution of the path nodes. During the search, each algorithm dynamically adjusts the balance between exploration and exploitation according to its specific population update strategy, aiming to find the global optimal path in the fewest iterations. The final results are comprehensively compared in terms of convergence performance, path length, and other factors to validate the global optimization capability and stability of each algorithm in the 2D robot path planning task.\par
\begin{figure}[htbp]
    \centering
    \includegraphics[width=\textwidth]{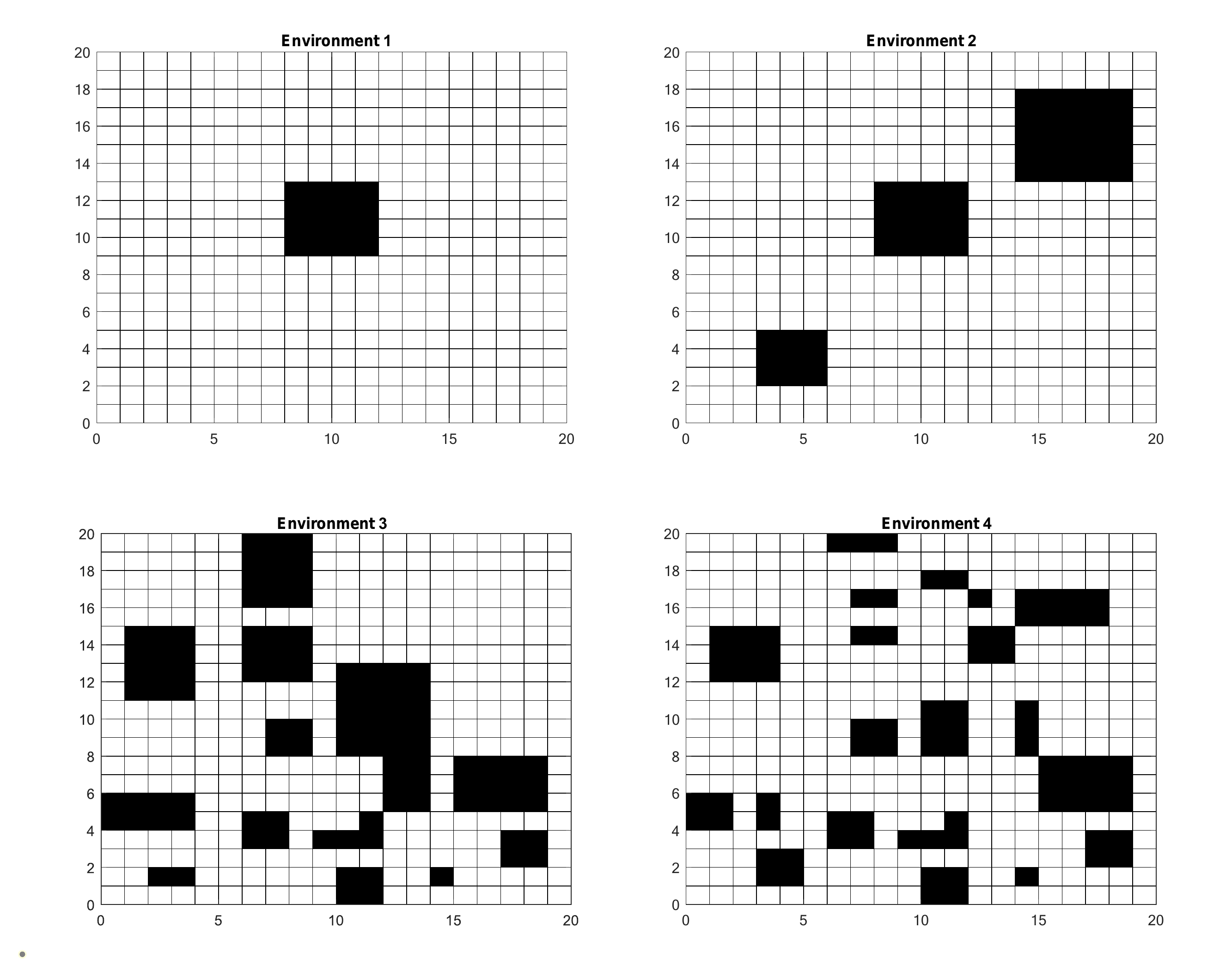}
    \caption{Environments with four different obstacle densities.} 
    \label{maps}
\end{figure}
\begin{figure}[htbp]
    \centering
    \includegraphics[width=\textwidth]{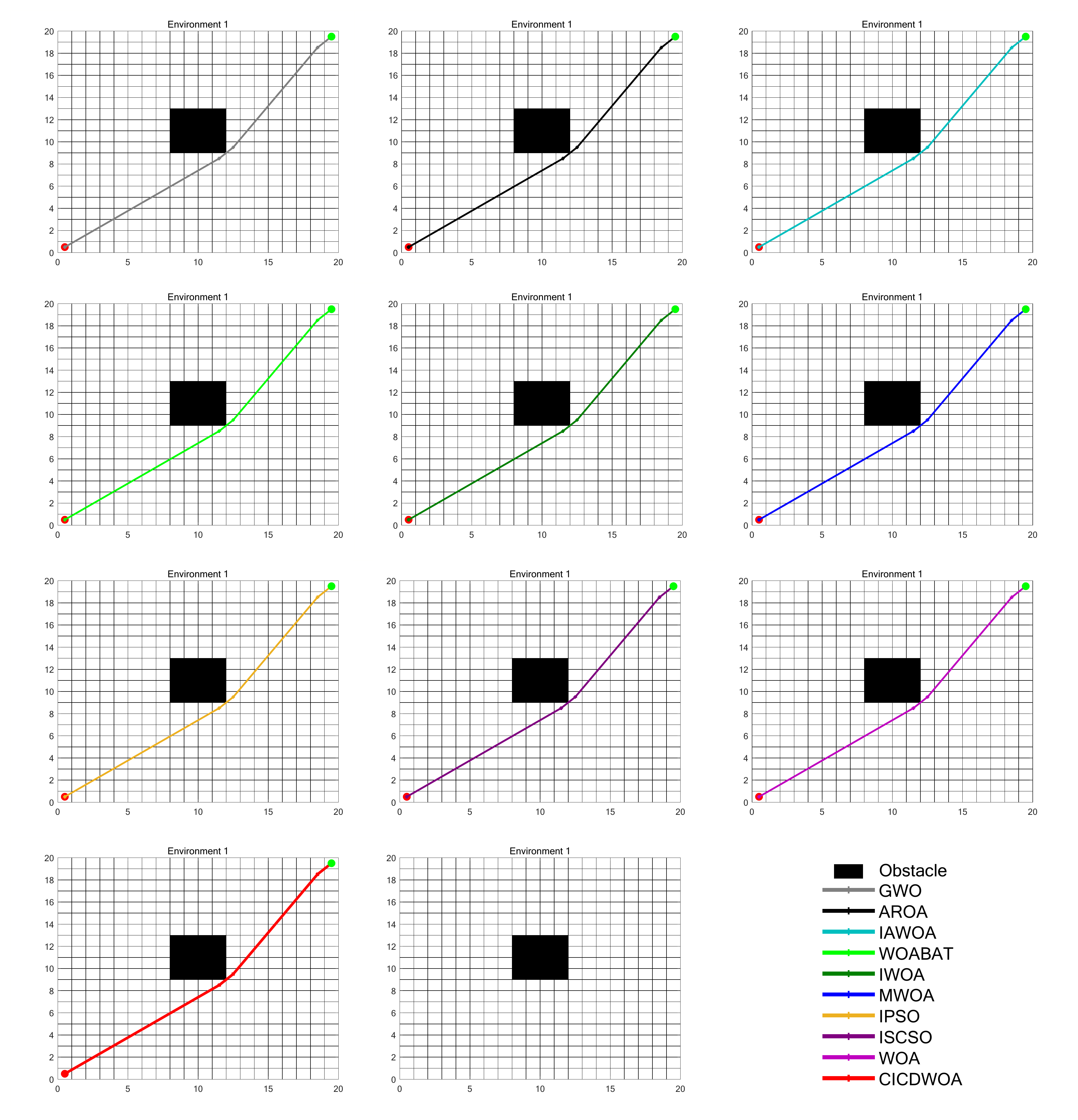}
    \caption{Paths planned by the algorithms in Environment 1.} 
    \label{map11}
\end{figure}
\begin{figure}[htbp]
    \centering
    \includegraphics[width=\textwidth]{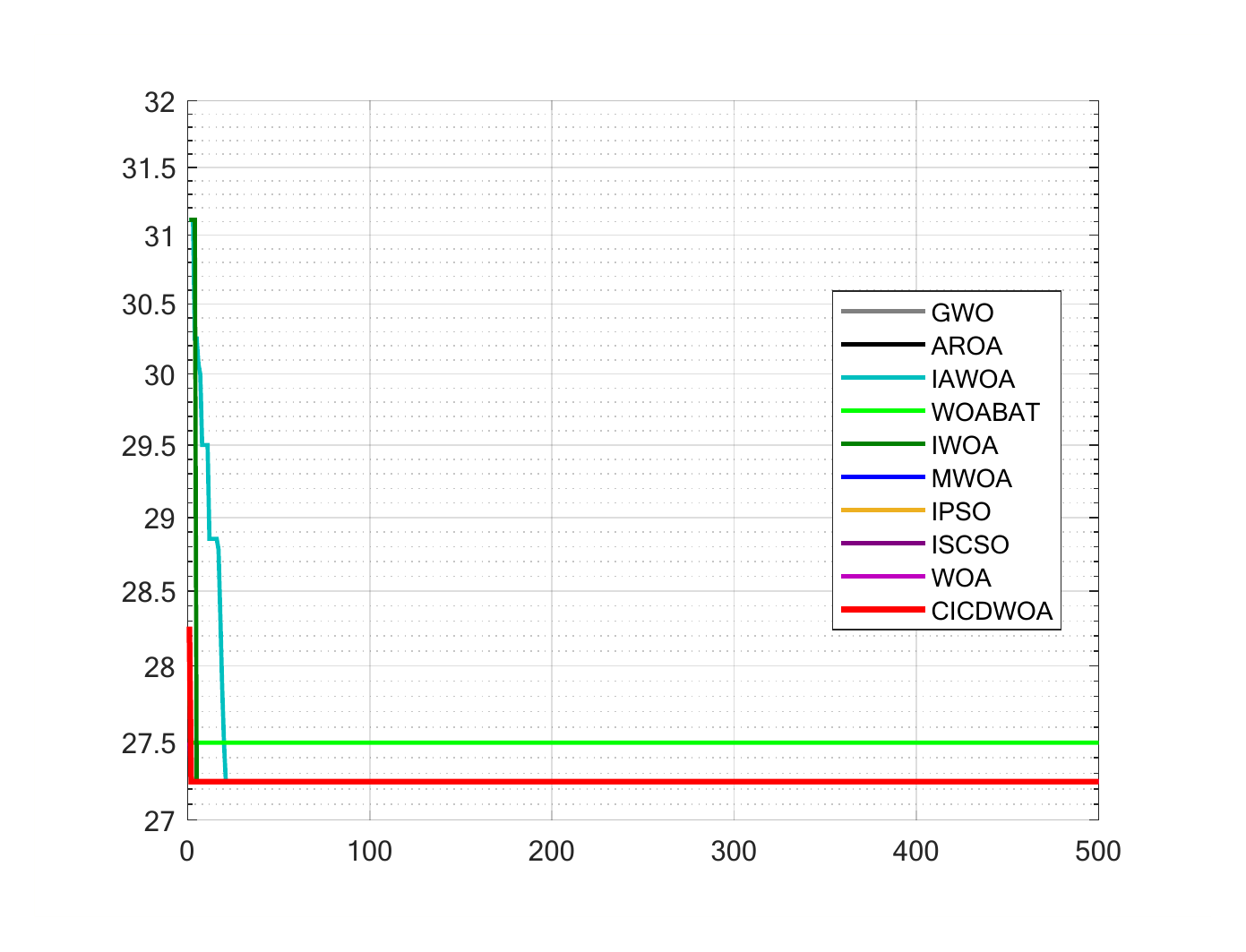}
    \caption{Iteration curves of the algorithms during path planning in Environment 1.} 
    \label{map12}
\end{figure}
\begin{figure}[htbp]
    \centering
    \includegraphics[width=\textwidth]{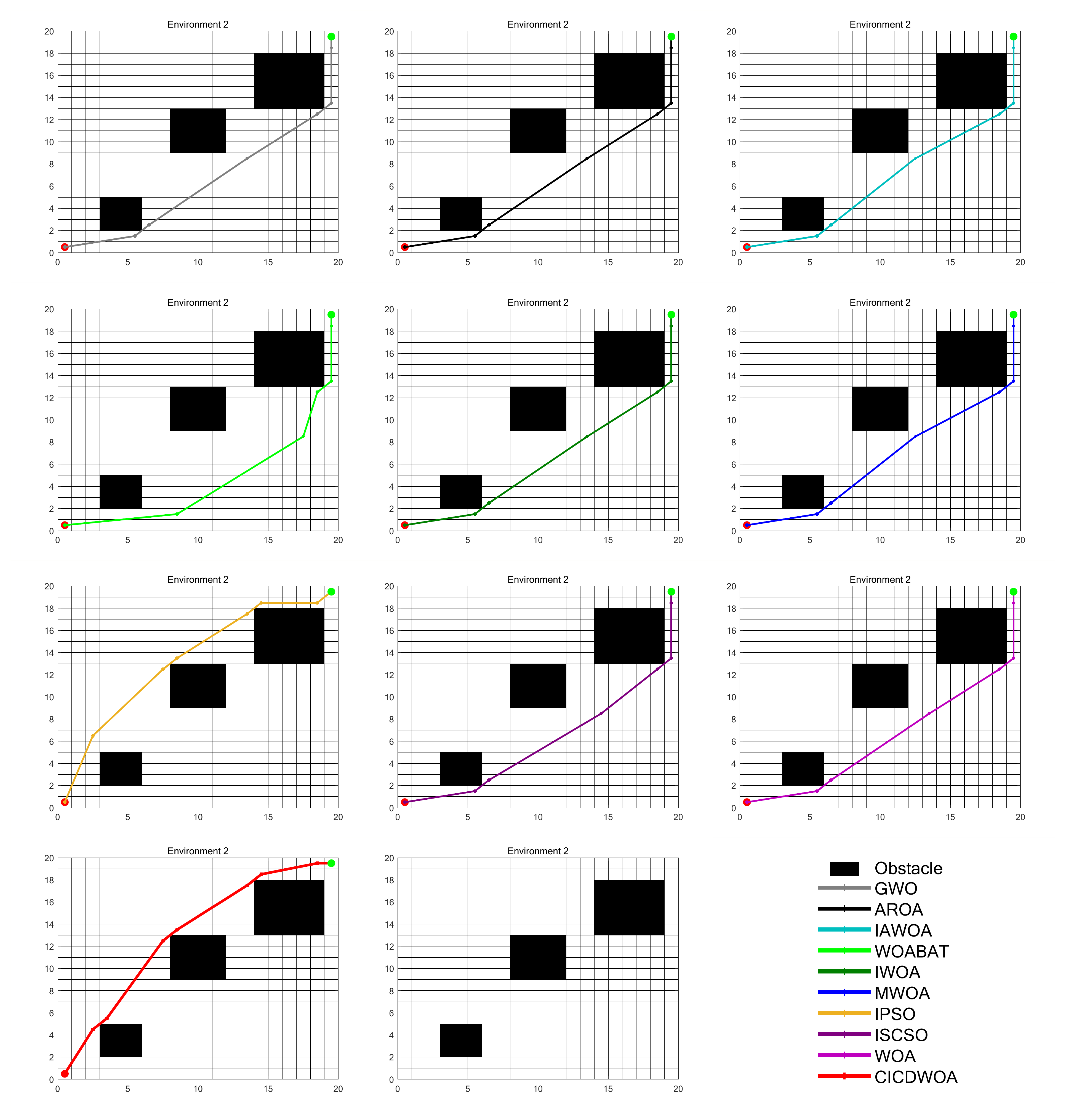}
    \caption{Paths planned by the algorithms in Environment 2.} 
    \label{map21}
\end{figure}
\begin{figure}[htbp]
    \centering
    \includegraphics[width=\textwidth]{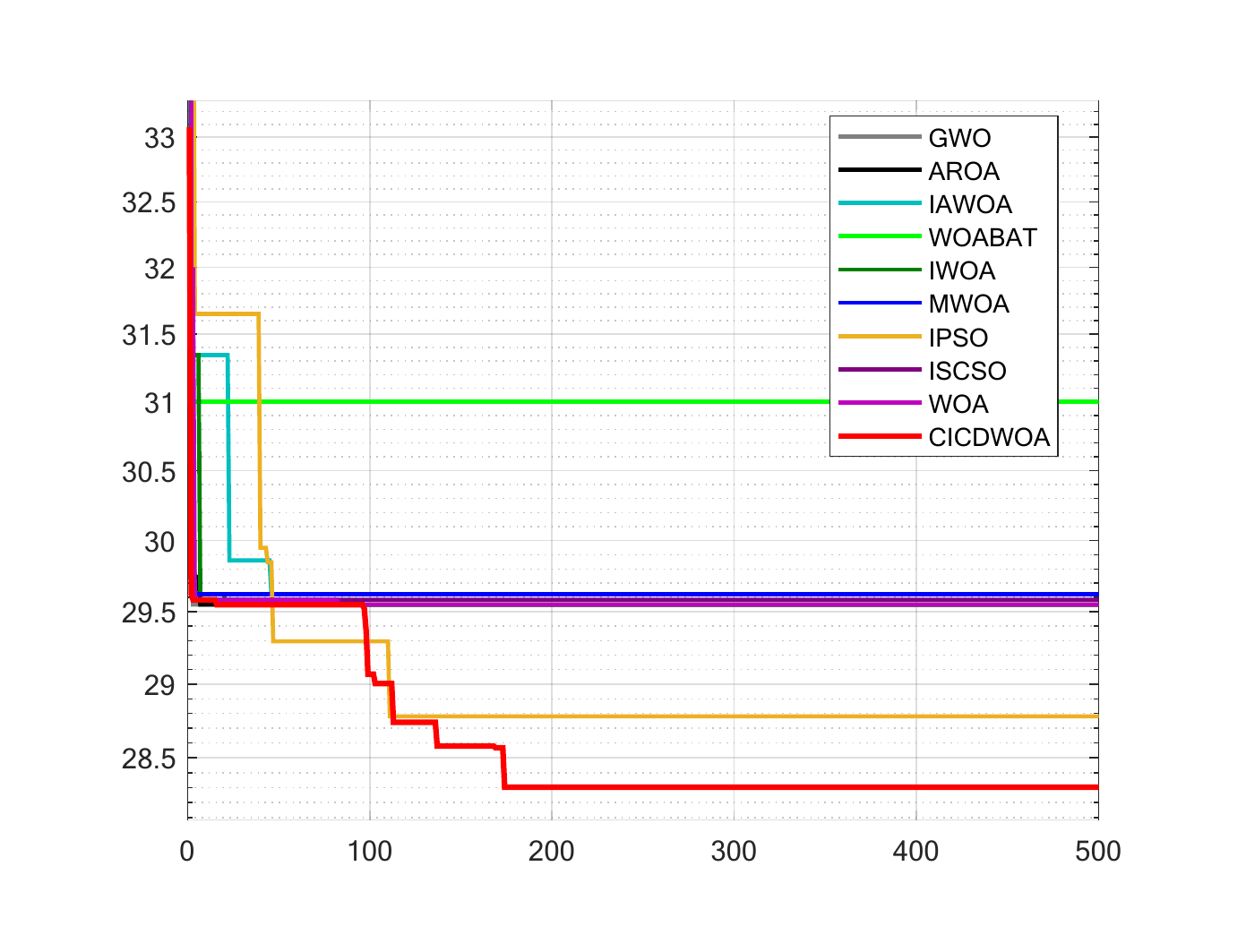}
    \caption{Iteration curves of the algorithms during path planning in Environment 2.} 
    \label{map22}
\end{figure}
\begin{figure}[htbp]
    \centering
    \includegraphics[width=\textwidth]{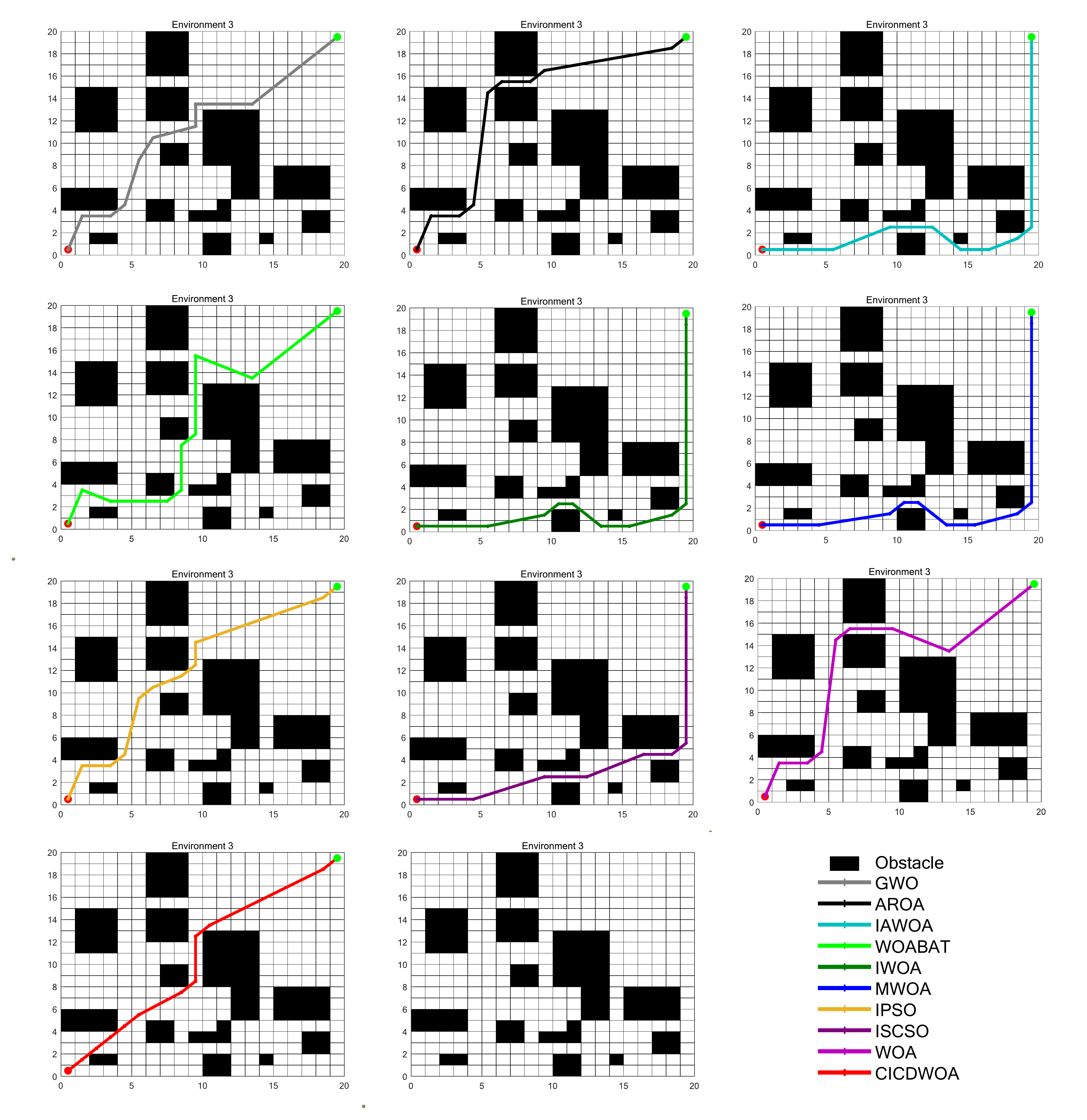}
    \caption{Paths planned by the algorithms in Environment 3.} 
    \label{map31}
\end{figure}
\begin{figure}[htbp]
    \centering
    \includegraphics[width=\textwidth]{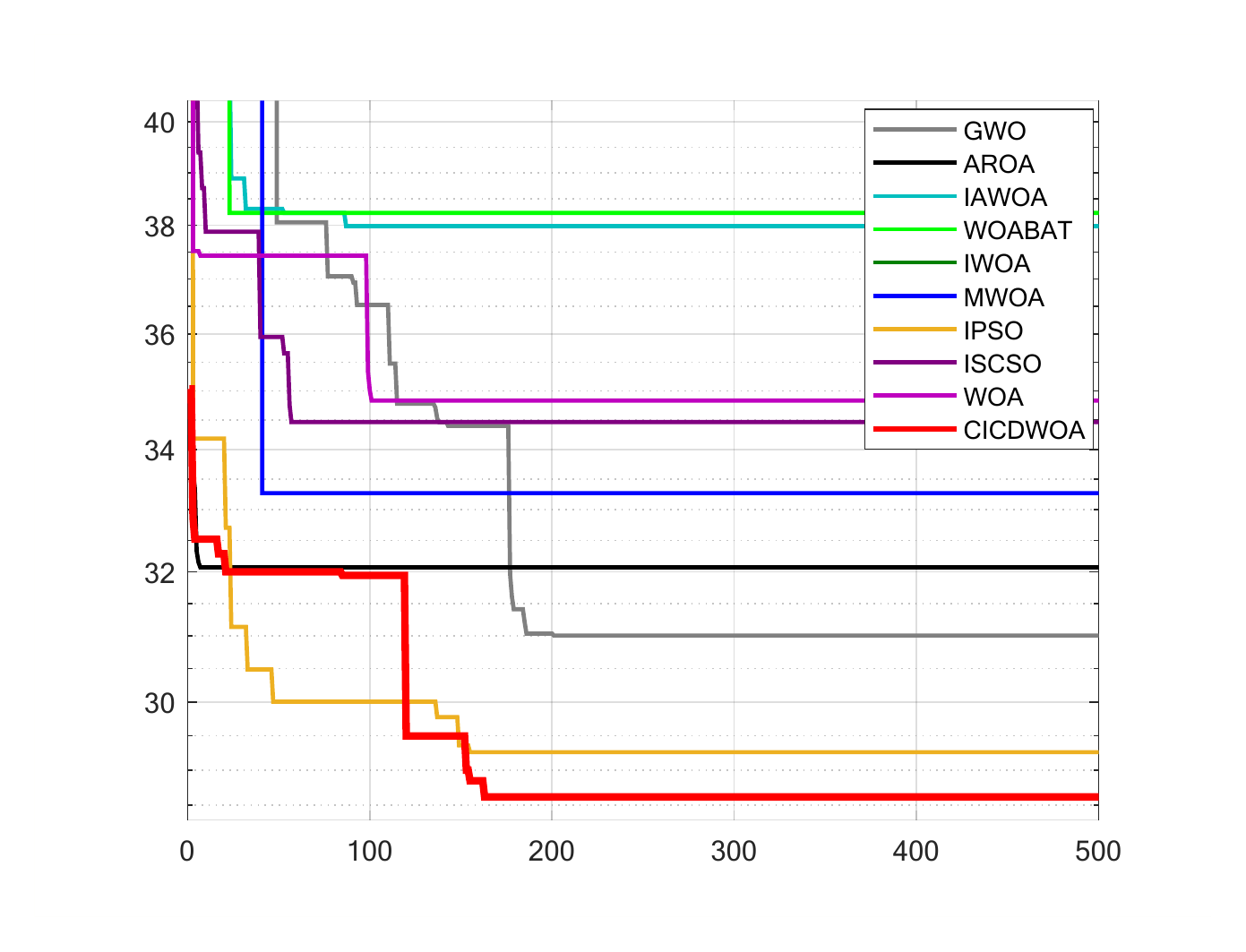}
    \caption{Iteration curves of the algorithms during path planning in Environment 3.} 
    \label{map32}
\end{figure}
\begin{figure}[htbp]
    \centering
    \includegraphics[width=\textwidth]{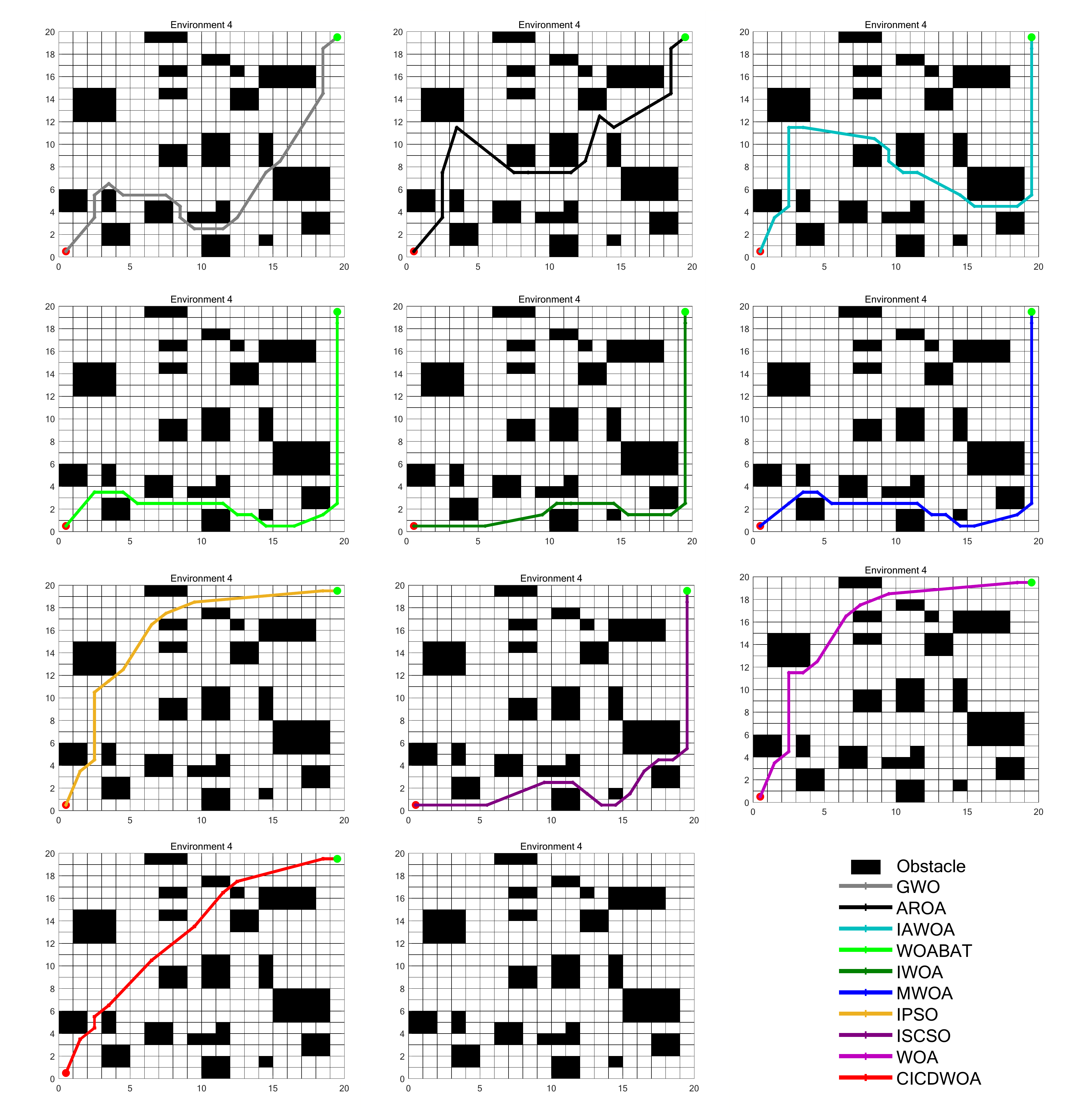}
    \caption{Paths planned by the algorithms in Environment 4.} 
    \label{map41}
\end{figure}
\begin{figure}[htbp]
    \centering
    \includegraphics[width=\textwidth]{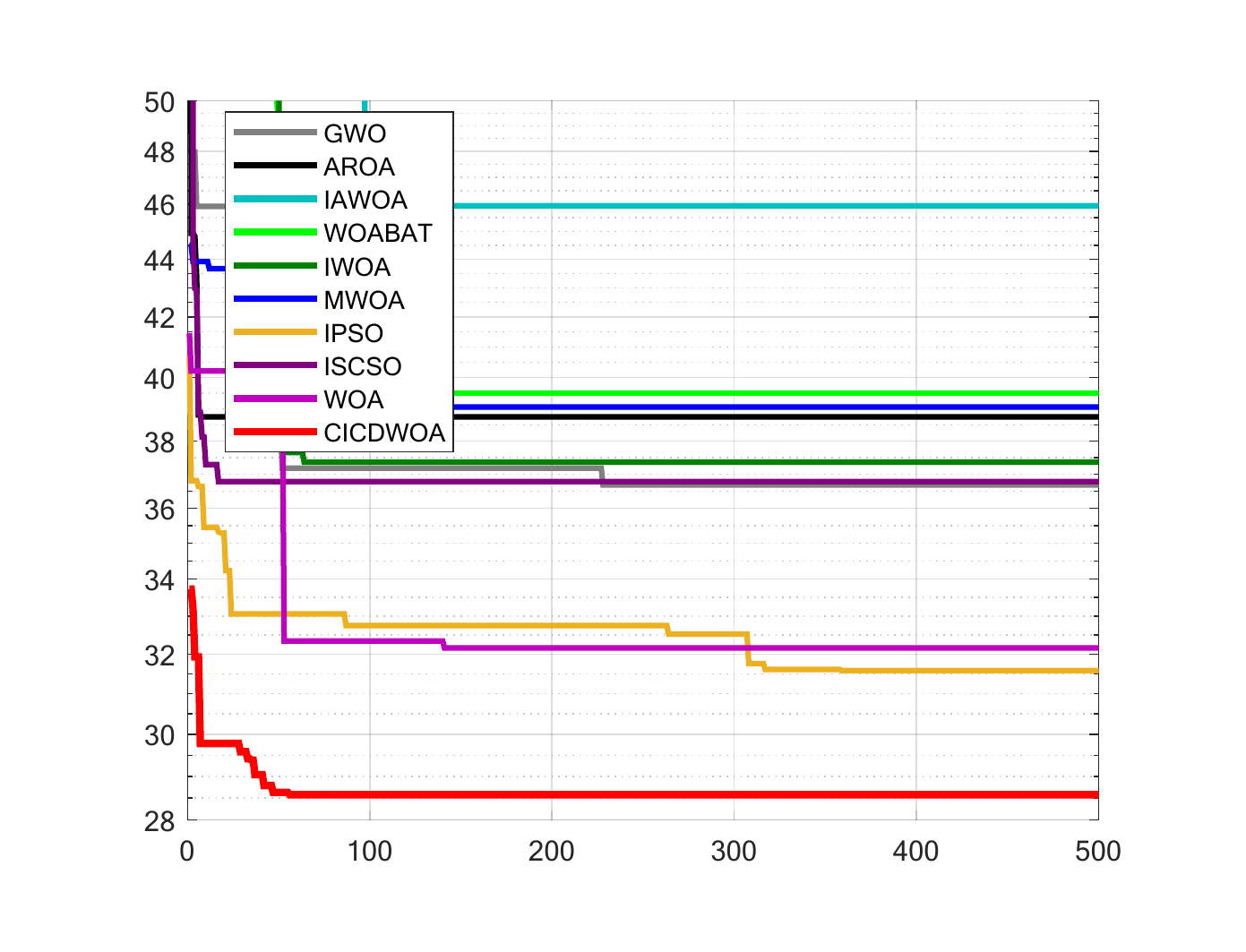}
    \caption{Iteration curves of the algorithms during path planning in Environment 4.} 
    \label{map42}
\end{figure}

The number of iterations is set uniformly as $T$=500, and the population size is $N$=30. Each algorithm runs 30 times independently on the four different environments with varying obstacle densities, as shown in Figure~\ref{maps}. The iteration curves and planned paths of the algorithms are presented in Figure~\ref{map11} through Figure~\ref{map42}, while the path planning metrics of the algorithms in different environments are shown in Table~\ref{2dpp}.\par
The experimental results clearly show that CICDWOA demonstrates optimal or near-optimal path planning performance across all four environments with different obstacle densities, exhibiting strong global optimization capability and stability. In the sparsely obstructed Environment 1, all algorithms converge relatively quickly to the same optimal path length of 27.2466, where the problem is simpler and differences between algorithms are not significant. However, as the obstacle density increases, the advantages of CICDWOA become evident. In Environment 2, some algorithms (e.g., AROA, HHO, IWOA, MWOA) show significant path degradation and result fluctuation, with maximum path lengths reaching up to 400, indicating that these algorithms are prone to getting stuck in local optima or failing in complex environments. In contrast, CICDWOA achieves an average path length of 28.6523 with a standard deviation of only 0.3894, significantly outperforming all comparison algorithms and demonstrating its high stability and reliable convergence under complex obstacle distributions. This trend becomes even more pronounced in Environment 3, which features higher obstacle density. Traditional WOA and its improved versions (e.g., IWOA, MWOA) fail to perform effectively, with path search failures or substantial deviations from the optimal solution, and maximum path lengths often reaching 400 or even 800. In contrast, CICDWOA maintains excellent convergence properties, with an average path length of 28.8979 and a standard deviation of only 0.7287, far lower than other algorithms. This result indicates that CICDWOA is able to effectively escape from local optimum traps under complex constraints and nonlinear obstacle conditions, exhibiting stronger global search ability and ensuring path feasibility. In the most complex environment, Environment 4, CICDWOA again achieves the best results, with an average path length of 28.6140, which is approximately 3.2\% shorter than the second-best algorithm, GWO (29.5550). Additionally, its path standard deviation is 0.3810, reflecting the algorithm's high stability across multiple runs. CICDWOA consistently demonstrates significantly better planning performance than the comparison algorithms across different obstacle densities. Its stable optimal convergence results, smaller variance, and shorter average path lengths strongly highlight its robustness, accuracy, and global optimization advantage in 2D path planning tasks.\par
These findings indicate that the proposed CICDWOA not only converges quickly to high-quality solutions in complex environments but also possesses strong adaptability and generalization potential. This provides a solid algorithmic foundation for drone path planning and the extension to three-dimensional environments.\par

\begin{table}[htbp]
\centering
\caption{Comparison results of different algorithms in 2D Path Planning.}
\resizebox{\textwidth}{!}{
\begin{tabular}{ccccccccccccc}
\toprule
MAP & Metrics & GWO & AROA & IAWOA & WOABAT & IWOA & MWOA & IPSO & ISCSO & WOA & CICDWOA \\
\midrule
Map1 & Ave  & 27.2466 & 27.2466 & 27.2550 & 27.2466 & 27.2909 & 27.2466 & 27.2466 & 27.2466 & 27.2466 & 27.2466 \\
                      & Std  & 0.0000  & 0.0000  & 0.0460  & 0.0000  & 0.2003  & 0.0000  & 0.0000  & 0.0000  & 0.0000  & 0.0000  \\
\midrule
Map2 & Ave  & 30.2740 & 48.1761 & 31.9843 & 63.8829 & 49.0379 & 49.7954 & 31.3592 & 32.5649 & 33.5059 & 28.6523 \\
                      & Std  & 1.6869  & 66.6237 & 2.6763  & 91.4499 & 66.3681 & 66.3530 & 1.9471  & 3.2329  & 2.8281  & 0.3894  \\
\midrule
Map3 & Ave  & 44.0732 & 34.4608 & 32.7825 & 225.2360 & 136.4660 & 138.3511 & 44.8516 & 57.4308 & 45.9602 & 28.8979 \\
                      & Std  & 67.2703 & 5.6168  & 2.6243  & 274.7254 & 243.0678 & 218.5455 & 67.2078 & 93.1685 & 66.9291 & 0.7287  \\
\midrule
Map4 & Ave  & 29.5550 & 29.7132 & 29.5952 & 29.9012 & 29.4837 & 30.2154 & 29.4625 & 29.6210 & 29.7795 & 28.6140 \\
                      & Std  & 0.0124  & 0.2160  & 0.0546  & 0.6300  & 0.3711  & 0.7996  & 0.1969  & 0.1502  & 0.3852  & 0.3810  \\
\bottomrule
\end{tabular}
}
\label{2dpp}
\end{table}

\section{3D UAV Path Planning}
\subsection{Modeling}

\begin{figure}[htbp]
    \centering
    \includegraphics[width=0.9\textwidth]{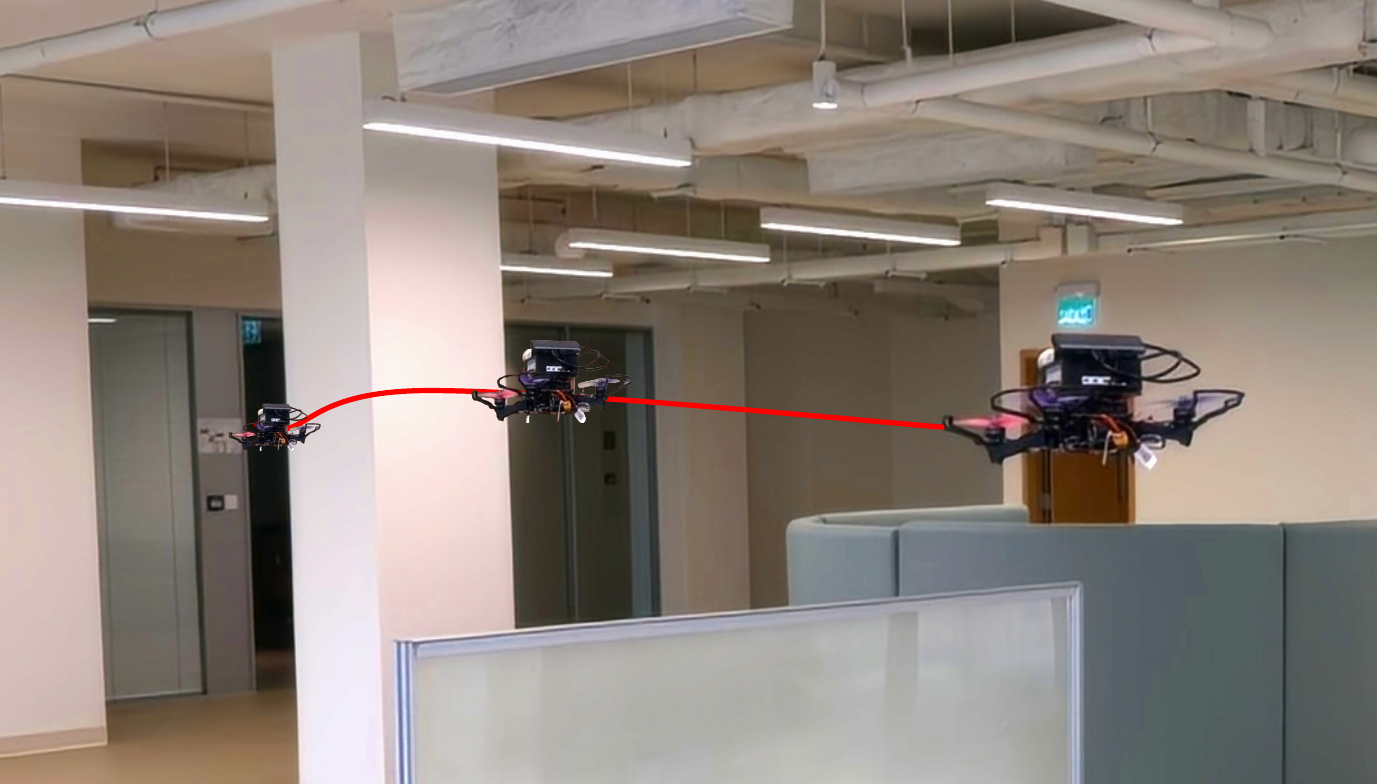}
    \caption{3D UAV Path Planning.} 
    \label{3duav}
\end{figure}

Figure~\ref{3duav} illustrates the schematic diagram of 3D UAV path planning, including the spatial environment, obstacle distribution, and the feasible flight trajectory. The core objective of 3D path planning for drones is typically to minimize the path length while successfully avoiding obstacles and satisfying the drone's dynamic constraints. To facilitate the comparison of different metaheuristic optimizers' search capabilities using a single-objective optimization algorithm, this study employs a weighted composite single-objective function $F_{tc}$, which integrates three key performance metrics into a scalar objective:\par

\begin{equation}
    F_{tc}=w_1F_{pc}+w_2F_{hc}+w_3F_{sc}
    \label{eq39}
\end{equation}

where $F_{pc}$ represents the path length cost, $F_{hc}$ denotes the height cost, and $F_{sc}$ represents the drone's dynamic constraint cost. The weight coefficients satisfy $w_i \geq 0$ and $w_1+w_2+w_3=1$.\par

In all experiments, to maintain a balance between path efficiency and flight stability, the weights are uniformly set to $w_1$=0.5, $w_2$=0.3, $w_3$=0.2, which is a commonly used setting. This weighted strategy allows the constraints on path length, flight height, and trajectory smoothness to be addressed within the single-objective framework, enabling comparisons among all the algorithms based on the same scalar criterion. Mathematically, the discretized path is represented by $g$ waypoints:
\begin{equation}
    P=\{P_1,\dots,P_g\}
    \label{eq40}
\end{equation}
The path length cost term is defined as:\par
\begin{equation}
    F_{pc}=\sum_{m=1}^{g-1}\|P_{m+1}-P_m\|_2
    \label{eq41}
\end{equation}
The height cost term is defined as the standard deviation of the height samples:\par
\begin{equation}
    F_{hc}=\sqrt{\frac{1}{g}\sum_{m=1}^g(z_m-\bar{z})^2}
    \label{eq42}
\end{equation}
The dynamic constraint cost is expressed by the cumulative angle between adjacent segments:\par
\begin{equation}
    F_{sc}=\sum_{m=1}^{g-2}\arccos\left(\frac{(P_{m+1}-P_m)\cdot(P_{m+2}-P_{m+1})}{\|P_{m+1}-P_m\|\|P_{m+2}-P_{m+1}\|}\right)
    \label{eq43}
\end{equation}
The obstacle constraint is incorporated into the objective function through a penalty term (assigning a large penalty value to solutions that conflict with obstacles), leading to the final optimization problem formulation:\par
\begin{equation}
    \min_PF(P)=F_{tc}(P)+\Phi_{{obs}}(P),
    \label{eq44}
\end{equation}
where the penalty term $\Phi_{{obs}}$ uses a combination of distance threshold and squared penalties to prioritize feasible solutions.\par

\subsection{Results}
In this study, the starting point is set as (20,20,20) and the destination as (180,180,20). Several irregular static obstacles are arranged in the 3D scene to simulate complex terrain. The number of iterations is set to $T$=500, and the population size is $N$=30. To evaluate the stability and statistical performance of the algorithms, each algorithm is independently repeated 30 times. The proposed CICDWOA is compared with selected algorithms, including GWO, AROA, IWOA, WOABAT, IWOA, MWOA, IPSO, ISCSO, and WOA. The performance metric is the $F_{tc}$ value obtained from each run, and the average fitness ($Ave$) and standard deviation ($Std$) of each algorithm are recorded to compare their optimization quality and robustness.\par
\begin{table}[htbp]
\centering
\caption{Comparison results of different algorithms in 3D Path Planning.}
\begin{tabular}{ccc}
\toprule
\textbf{Algorithm} & \textbf{Metrics} & \textbf{Value} \\
\midrule
GWO     & Ave  & 80.2345  \\
        & Std  & 12.5821  \\
AROA    & Ave  & 105.1432 \\
        & Std  & 45.29    \\
IAWOA   & Ave  & 77.9589  \\
        & Std  & 10.8559  \\
WOABAT  & Ave  & 156.1647 \\
        & Std  & 152.1973 \\
IWOA    & Ave  & 78.3323  \\
        & Std  & 5.335    \\
MWOA    & Ave  & 116.701  \\
        & Std  & 27.3932  \\
IPSO    & Ave  & 80.3255  \\
        & Std  & 5.5188   \\
ISCSO   & Ave  & 80.0839  \\
        & Std  & 11.775   \\
WOA     & Ave  & 87.1815  \\
        & Std  & 27.5501  \\
CICDWOA & Ave  & \textbf{76.829}   \\
        & Std  & \textbf{1.9231}   \\
\bottomrule
\end{tabular}
\label{3dpp1}
\end{table}
The experimental results demonstrate that CICDWOA achieves the optimal average performance (76.95) while having the smallest standard deviation (1.83), indicating that the algorithm not only finds lower objective values across multiple independent runs but also exhibits high stability and robustness. In comparison, algorithms such as GWO, IWOA, ISCSO, and IAWOA provide reasonably good solutions in most experiments, with average values concentrated in the 77-79 range. However, their standard deviations are generally higher than that of CICDWOA, suggesting some result fluctuation. Furthermore, AROA, WOABAT, and MWOA show significantly higher average values and variances, with WOABAT exhibiting an especially high average value of 140.45 and a very large variance. This indicates that these methods are prone to premature convergence or getting stuck in local optima, resulting in poor or unacceptable path solutions. Lastly, although IPSO does not have the lowest average value, its standard deviation is only 1.99, indicating good performance in maintaining stability.\par
Further analysis through convergence curves and path visualizations reveals that CICDWOA has a faster early convergence rate and stabilizes in the later stages, suggesting a good balance between exploration and exploitation under the weighted objective. The path visualizations also show that the trajectories generated by CICDWOA are shorter, with smoother turns, successfully avoiding obstacles while satisfying dynamic and height constraints. CICDWOA demonstrates the best optimization capability and the highest stability in solving the drone path planning problem.
\begin{figure*}[htbp]
    \centering
    \includegraphics[width=\textwidth]{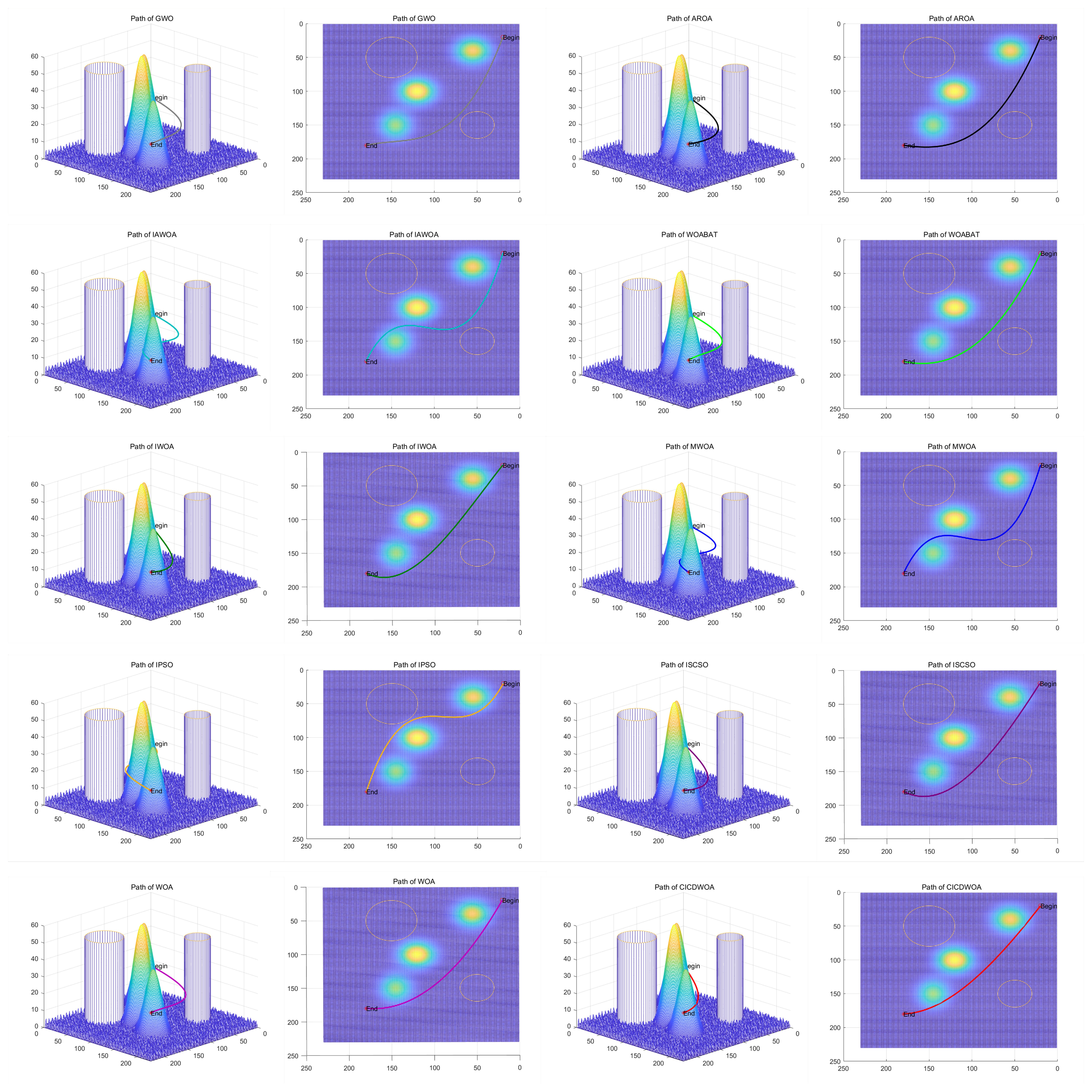}
    \caption{Paths obtained by the algorithms in solving the UAV path planning problem.} 
    \label{3dpp2}
\end{figure*}

\begin{figure}[htbp]
    \centering
    \includegraphics[width=\textwidth]{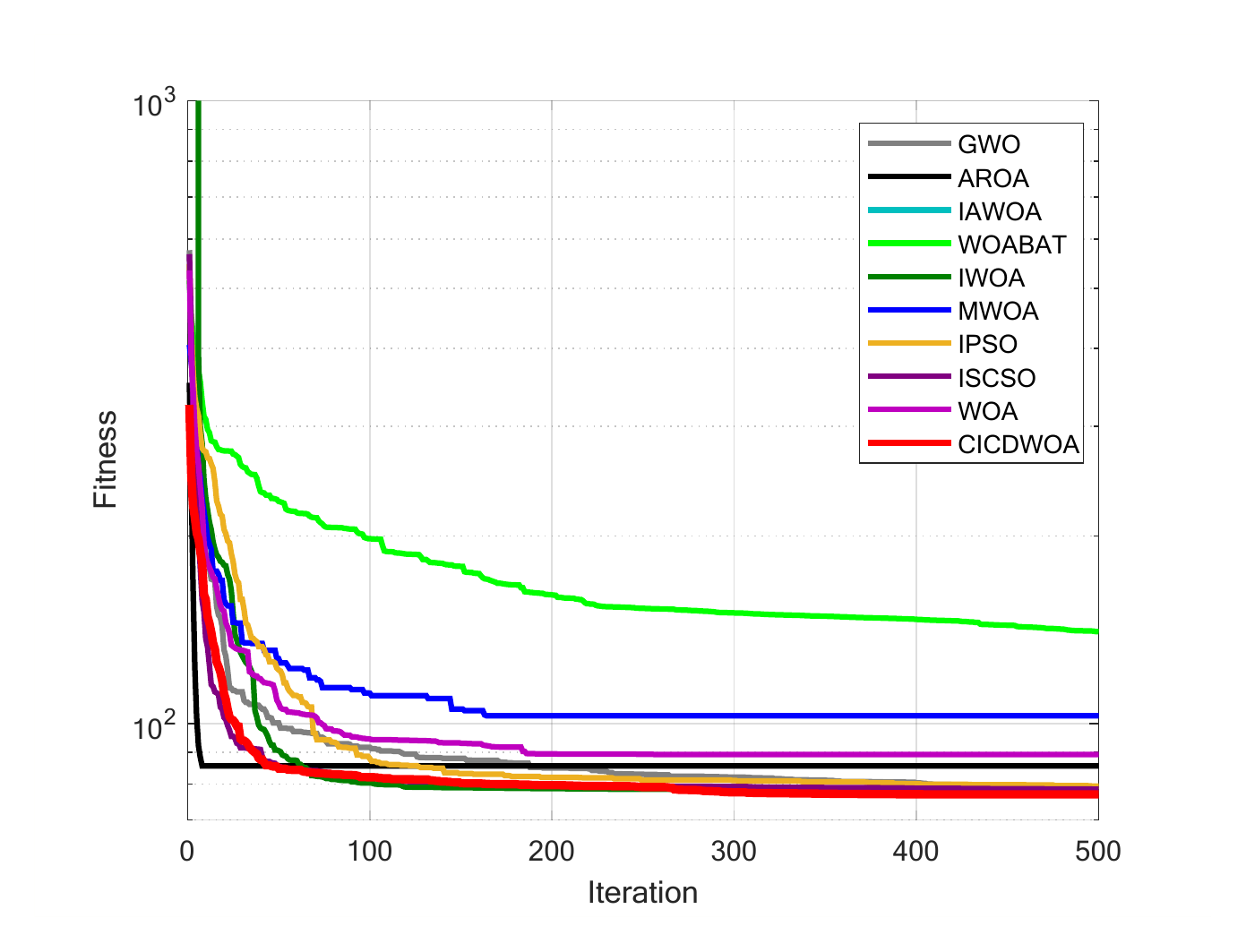}
    \caption{Iteration curves of the algorithms in solving the UAV path planning problem.} 
    \label{3dpp3}
\end{figure}

\section{Engineering Design Optimization}
Engineering Design Optimization aims to find the optimal solution for engineering structures or practical application scenarios under specific constraints, achieving the optimal balance among performance, cost, weight, energy consumption, and other indicators. These problems often exhibit characteristics such as high nonlinearity, strong constraints, multivariable conditions, and non-differentiability, making it difficult for traditional analytical or gradient-based methods to solve them effectively. As a result, researchers widely employ metaheuristic algorithms to address such challenges. Metaheuristic algorithms simulate natural phenomena like predation behavior, group collaboration, energy transfer, and other mechanisms, using random search techniques to iteratively explore and optimize the global solution space. These algorithms do not rely on gradient information, offering strong global search capabilities and robustness, enabling them to effectively escape local optima and thus demonstrating excellent solving performance in complex engineering design optimization problems.

To handle the constraints present in these engineering design problems, a penalty function method is employed in this study. The penalty function transforms the constrained optimization problem into an unconstrained one by adding a penalty term to the objective function whenever a candidate solution violates any constraint. The magnitude of the penalty is proportional to the degree of violation, thereby discouraging infeasible solutions during the search. This approach allows metaheuristic algorithms to efficiently explore the feasible solution space while maintaining constraint satisfaction. In the CICDWOA algorithm, a dynamic penalty function is integrated, ensuring that solutions with smaller constraint violations are preferentially selected, which effectively balances the search for high-performance and feasible solutions.

In this section, ten engineering design problems are used to test the superior performance of the CICDWOA algorithm proposed in this paper for solving various practical applications. The CICDWOA is compared with GWO, AROA, IWOA, WOABAT, IWOA, MWOA, IPSO, ISCSO, and WOA. The parameter settings for each algorithm are shown in Table~\ref{setting}. The iteration number is uniformly set to $T$=500, and the population size to $N$=30. Each algorithm is run 30 times for each engineering design optimization problem, recording the mean values and standard deviations for performance analysis. The experimental results are presented in Figure~\ref{engineering11}-Figure~\ref{engineering102} and Table~\ref{engineering_metrics}.

\subsection{Sawmill Operation Problem}
\begin{figure}[htbp]
    \centering
    \includegraphics[width=0.6\textwidth]{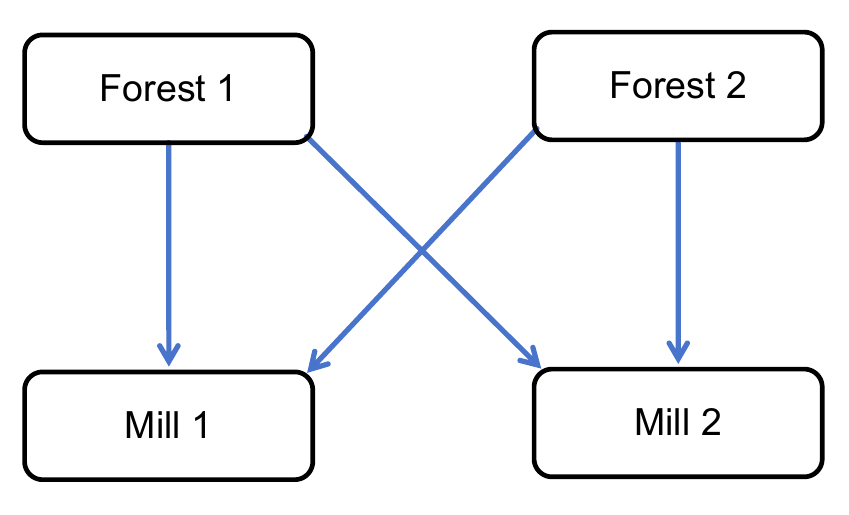}
    \caption{Sawmill Operation Problem.} 
    \label{engineering11}
\end{figure}

The sawmill operation problem is a typical engineering design optimization problem. The objective is to minimize the total cost of transporting logs, subject to the constraints of factory capacity and demand. Assume a company owns two forests and two sawmills. During a production cycle, each forest can provide a maximum of 200 logs per day, while each sawmill requires at least 300 logs per day to maintain normal production. The transportation cost of logs is known to be 10 dollars per kilometer per log, and the distances between the forests and the sawmills, as well as the factories' processing capacities, are provided in Table~\ref{engineering13}. In this problem, the design variables correspond to the number of logs allocated from Forest 1 and Forest 2 to Sawmill A and Sawmill B, denoted as $x_1$, $x_2$, $x_3$, $x_4$ as shown in Figure~\ref{engineering11}. The objective function for optimization is to minimize the total daily transportation cost, which can be expressed as:\par
\begin{flushleft}
    \textit{Variable:}
\end{flushleft}
\vspace{-\baselineskip}
\begin{flushleft}
    \[
        \begin{aligned}
            x= [x_1, x_2, x_3, x_4]
        \end{aligned}
    \]
\end{flushleft}

\begin{flushleft}
    \textit{Minimize:}
\end{flushleft}
\begin{equation}
            y=10(24x_1+20.5x_2+17.2x_3+10x_4)+punishment
\end{equation}

\begin{flushleft}
    \textit{Subject to:}
\end{flushleft}
\begin{equation}
        g_1=x_1+x_2-240 \leq 0;
\end{equation}
\begin{equation}
        g_2=x_3+x_4-300 \leq 0;
\end{equation}
\begin{equation}
        g_3=x_1+x_3-200 \leq 0;
\end{equation}
\begin{equation}
        g_4=x_2+x_4-200 \leq 0;
\end{equation}

\begin{flushleft}
    \textit{Variable range:}
\end{flushleft}
\vspace{-\baselineskip}
\begin{flushleft}
    \[
        0 \leq x_1 \leq 200; \quad 0 \leq x_2 \leq 200; \quad 0 \leq x_3 \leq 200; \quad 0\leq x_4 \leq 200;
    \]
\end{flushleft}

\begin{flushleft}
    \textit{Where:}
\end{flushleft}

\begin{flushleft}
    \[
        \begin{aligned}
        punishment=10^3 \cdot \sum_{i=1}^{4} \max{(0, g_i)^2}
        \end{aligned}
\]
\end{flushleft}

\begin{table}[htbp]
\centering
\caption{Details of Sawmill operation problem.}
\begin{tabular}{cccc}
\toprule
\textbf{Mill} & \textbf{Distance from } & \textbf{Distance from} & \textbf{Mill capacity per }\\
 & \textbf{forest one/km} & \textbf{forest two/km} & \textbf{day/logs}\\
\midrule
A     & 24.0  & 20.5 & 240 \\
B     & 17.2  & 18.0 & 300 \\
\bottomrule
\end{tabular}
\label{engineering13}
\end{table}

The first four constraints limit the production capacity and transportation capacity between each forest and sawmill, while the fifth constraint ensures that the total supply meets the minimum demand. This model allows for a trade-off between cost and capacity utilization across different transportation routes and allocation strategies.\par
\begin{figure}[htbp]
    \centering
    \includegraphics[width=0.8\textwidth]{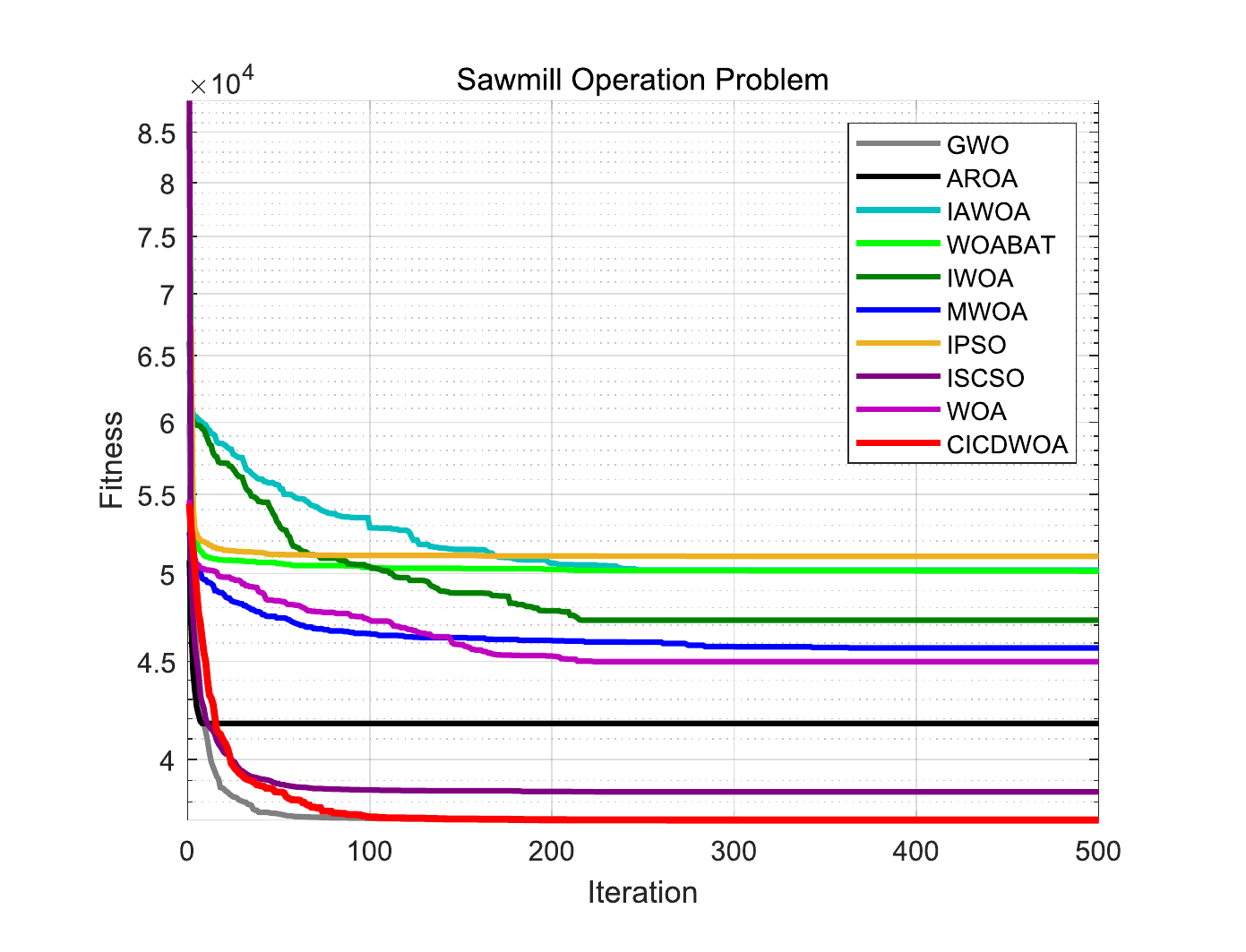}
    \caption{Iteration curves of the algorithms in solving the Sawmill Operation Problem.} 
    \label{engineering12}
\end{figure}

The experimental results are shown in Figure~\ref{engineering12}, and Table~\ref{engineering_metrics}. As indicated in Table~\ref{engineering_metrics}, the stability of CICDWOA in the Sawmill Operation Problem significantly outperforms other algorithms, and its optimization accuracy is the best among all the algorithms. This demonstrates the significant advantage of CICDWOA in handling such optimization problems.

\subsection{Multiple Disk Clutch Brake}
The multiple-disc clutch brake is a critical component commonly used in transmission systems and mechanical equipment, with widespread applications in automobiles, machine tools, construction machinery, and the aerospace industry. Its primary function is to transmit and disconnect power through friction, thereby effectively improving the control accuracy and energy conversion efficiency of mechanical systems. Since this device is subjected to complex thermal, force, and frictional effects during operation, a reasonable structural design is essential for enhancing its performance, prolonging its lifespan, and reducing manufacturing costs. The core objective of optimizing the multiple-disc clutch brake is to minimize the total system cost while satisfying structural and performance constraints. To achieve this, key design parameters of the system need to be optimized, including the thickness of the friction discs, inner and outer radii, actuating force, and the number of friction surfaces. By designing these parameters appropriately, a balance can be achieved between cost, torque transmission capacity, thermal load, and durability.\par
\begin{figure}[htbp]
    \centering
    \includegraphics[width=0.6\textwidth]{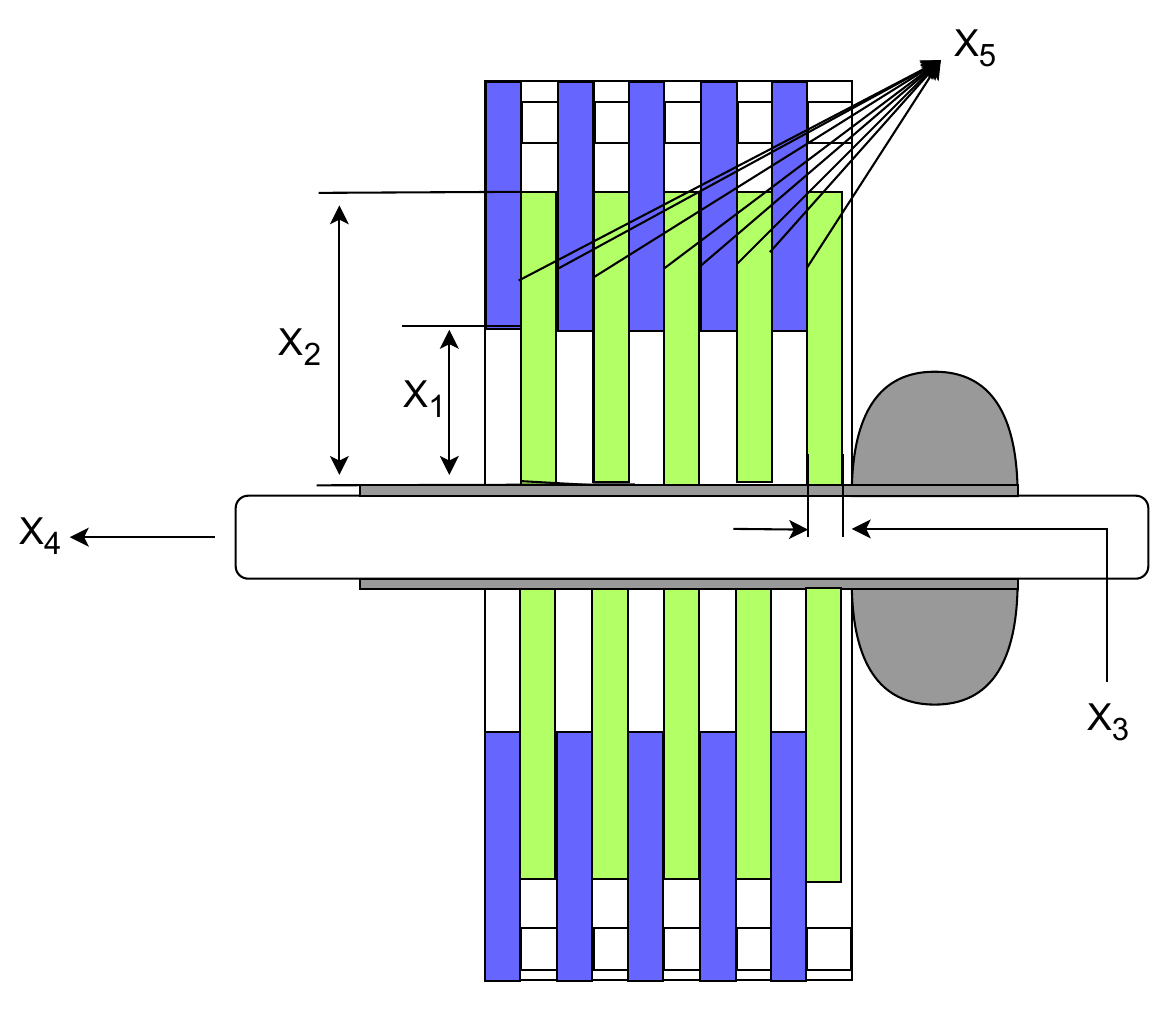}
    \caption{The structure of a multiple-disc clutch brake.} 
    \label{engineering21}
\end{figure}

As shown in Figure~\ref{engineering21}, this problem involves five optimization variables, which are:\par
\begin{itemize}
  \item $x_1$: inner radius
  \item $x_2$: outer radius
  \item $x_3$: thickness of discs
  \item $x_4$: actuating force
  \item $x_5$: the number of friction surfaces
\end{itemize}
The multiple-disc clutch brake optimization problem typically involves several physical and structural constraints, including torque transmission capacity, temperature rise limitations, friction material strength, and geometric structure requirements, with a total of eight constraints. These constraints ensure that the system meets transmission performance requirements while remaining within safety and manufacturing limits. The objective function for the Multi-disc Clutch Brake design problem can be described as:\par
\begin{flushleft}
    \textit{Variable:}
\end{flushleft}
\begin{flushleft}
    \[
        \begin{aligned}
            x = [x_1, x_2, x_3, x_4, x_5]
        \end{aligned}
    \]
\end{flushleft}

\begin{flushleft}
    \textit{Minimize:}
\end{flushleft}
\begin{equation}
        y = \pi x_3\rho\left(x_2^2-x_1^2\right)(x_5+1).+punishment
\end{equation}

\begin{flushleft}
    \textit{Subject to:}
\end{flushleft}
\begin{equation}
	g_1=\Delta r+x_1-x_2\leq0;
\end{equation}

\begin{equation}
	g_2=(x_5+1)(x_3+\delta)-l_{\max}\leq0;
\end{equation}

\begin{equation}
    g_3=P_{rz}-P_{max}\leq0; 
\end{equation}

\begin{equation}
	g_4= P_{rz}\cdot V_{sr}-P_{max}\cdot V_{srmax}\leq0;
\end{equation}

\begin{equation}
    g_5=V_{sr}-V_{srmax}\leq0;
\end{equation}

\begin{equation}
    g_6=T-T_{max}\leq0;
\end{equation}

\begin{equation}
    g_7=s\cdot M_s-M_h\leq0;
\end{equation}

\begin{equation}
    g_8=-T\leq0;
\end{equation}

\begin{flushleft}
    \textit{Where:}
\end{flushleft}

\begin{flushleft}
    \[
        \begin{aligned}
        punishment=10^3 \cdot \sum_{i=1}^{8} \max{(0, g_i)^2}
        \end{aligned}
\]
\end{flushleft}
\begin{flushleft}
    \[
        \begin{aligned}
			M_h=\frac{2}{3}\mu Fx_5\frac{x_2^3-x_1^3}{x_2^2-x_1^2};
        \end{aligned}
    \]
\end{flushleft}
\begin{flushleft}
    \[
        \begin{aligned}
			P_{rz}=\frac{F}{\pi\left(x_2^2-x_1^2\right)};
        \end{aligned}
    \]
\end{flushleft}
\begin{flushleft}
    \[
        \begin{aligned}
			V_{sr}=\frac{2\pi n\left(x_2^3-x_1^3\right)}{90\left(x_2^2-x_1^2\right)};
        \end{aligned}
    \]
\end{flushleft}
\begin{flushleft}
    \[
        \begin{aligned}
			T=\frac{I_z\pi n}{30\left(M_h+M_f\right)};
        \end{aligned}
    \]
\end{flushleft}
\begin{flushleft}
    \[
        \begin{aligned}
			P_{rz}=\frac{x_4}{\pi\cdot (x_2^2-x_1^2)};
        \end{aligned}
    \]
\end{flushleft}
\begin{flushleft}
    \[
        \begin{aligned}
			V_{sr} = \pi\cdot R_{sr}\cdot \frac{n}{30};
        \end{aligned}
    \]
\end{flushleft}
\begin{flushleft}
    \[
        \begin{aligned}
			R_{sr} = \frac{2}{3}\cdot \frac{x_2^3-x_1^3}{x_2^2*x_1^2};
        \end{aligned}
    \]
\end{flushleft}

\vspace{-\baselineskip}
\begin{flushleft}
    \[
        \begin{aligned}
			\Delta r=20; t_{max}=3; t_{min}=1.5; l_{max}=30; Z_{max}=10; V_{max}=10; \mu=0.6; \delta=0.5; M_s=40; 
        \end{aligned}
    \]
\end{flushleft}
\vspace{-\baselineskip}
\begin{flushleft}
    \[
        \begin{aligned}
			M_f=3; n=250; P_{max}=1; I_z=55; T_{max}=15;
			F{max}=1000; r_{imin}=55; 
        \end{aligned}
    \]
\end{flushleft}
\begin{flushleft}
    \[
        \begin{aligned}
			r_{omax}=110; \rho=0.0000078;
        \end{aligned}
    \]
\end{flushleft}

\begin{flushleft}
    \textit{Variable range:}
\end{flushleft}

\begin{flushleft}
    \[
    60\leq x_1 \leq80; \quad 90\leq x_2 \leq 110; 
    \quad 1\leq x_3 \leq 3; \quad 0\leq x_4 \leq 1000; 2\leq x_5 \leq 9
    \]
\end{flushleft}

\begin{figure}[htbp]
    \centering
    \includegraphics[width=0.8\textwidth]{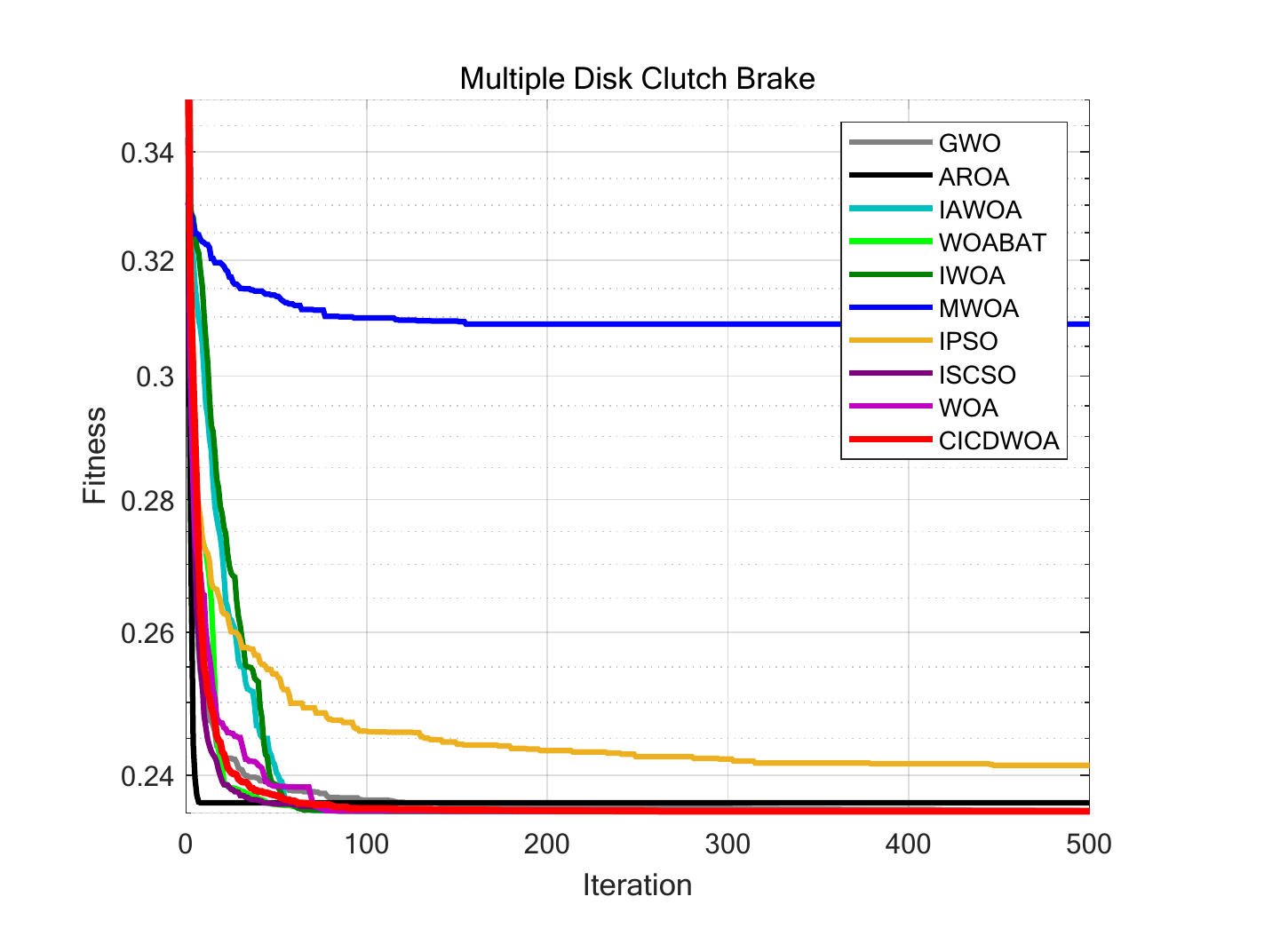}
    \caption{Iteration curves of the algorithms in solving the multiple-disc clutch brake design problem.} 
    \label{engineering22}
\end{figure}

The experimental results are shown in Figure~\ref{engineering22}, and Table~\ref{engineering_metrics}. As indicated in Table~\ref{engineering_metrics}, CICDWOA demonstrates significantly better stability than other algorithms in the Multiple-disc clutch brake design problem, and its optimization accuracy is the best among all the algorithms. This highlights the considerable advantage of CICDWOA in solving such optimization problems.

\subsection{Reactor Network}
\begin{figure}[htbp]
    \centering
    \includegraphics[width=0.8\textwidth]{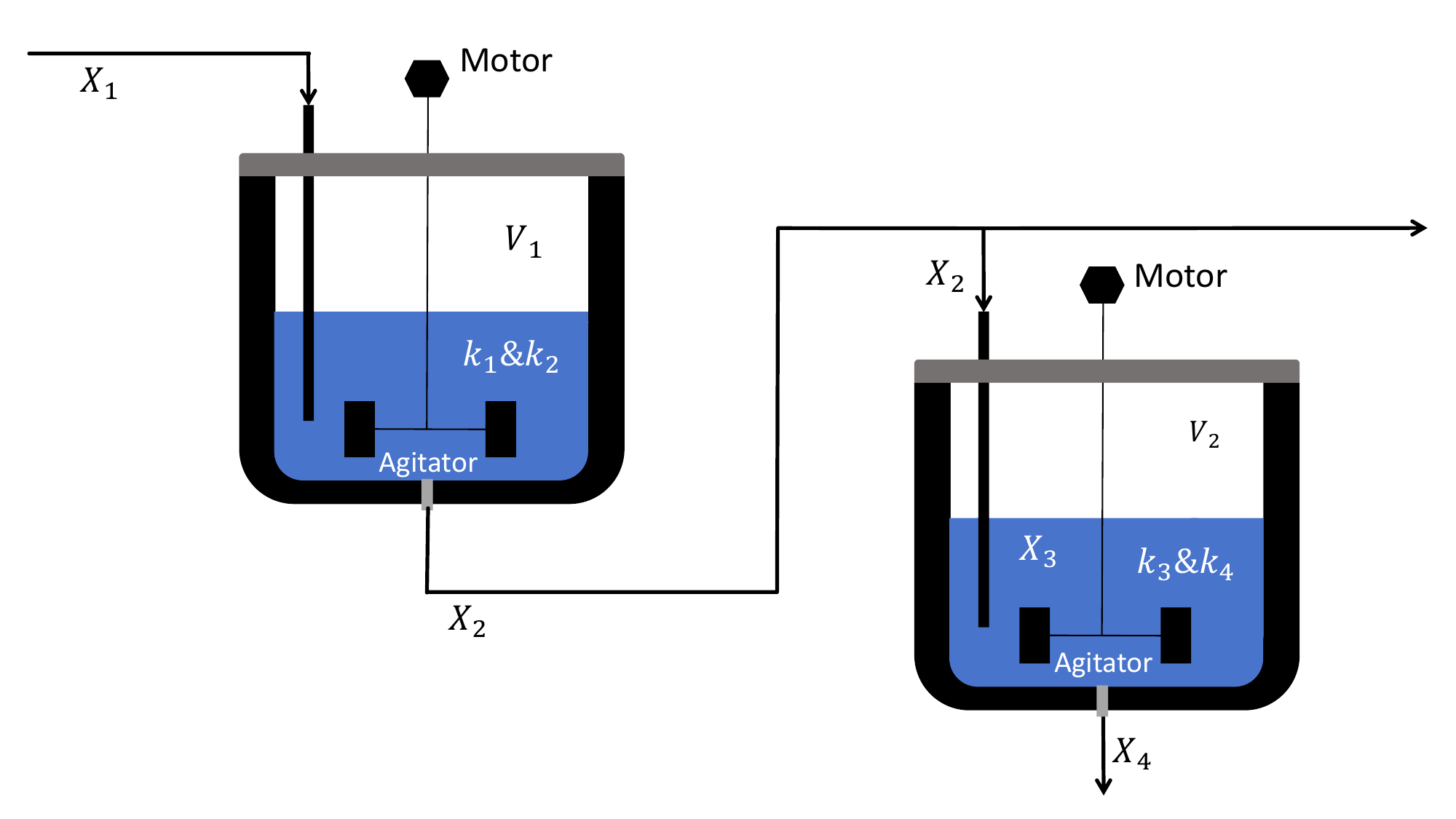}
    \caption{The structure of a reactor network.} 
    \label{engineering31}
\end{figure}

The Reactor Network, as shown in Figure~\ref{engineering31}, is one of the core components in chemical process design. Its optimization is crucial for improving chemical reaction efficiency, reducing energy consumption, and enhancing product quality. The design problem of reactor networks aims to achieve a more efficient, economical, and controllable chemical reaction process by rationally configuring the connection and operating conditions of multiple chemical reactors. In chemical production, different types of reactors, such as Continuous Stirred Tank Reactors (CSTR), Tubular Reactors, and others, are often used in combination to maximize the target product under specific temperature, pressure, and flow conditions. The optimization objective of this problem is to adjust the configuration of the reactors and operating parameters to maximize the final product concentration, while satisfying mass conservation and balance constraints. The model consists of four design variables, which represent the concentration of substances at different reaction stages:\par
\begin{itemize}
  \item $x_1$: Reactant concentration in the first reactor;
  \item $x_2$: Product concentration in the first reactor;
  \item $x_3$: Reactant concentration in the second reactor;
  \item $x_4$: Final product concentration.
\end{itemize}
The constraints $g_1$ to $g_4$ describe the key balance relationships in the reaction process:\par
\begin{itemize}
  \item $g_1$: Ensures the concentration balance between reactants and products in the first reactor;
  \item $g_2$: Constrains mass conservation between the first and second reactors;
  \item $g_3$: Maintains the balance of reactant concentration between the two reactors;
  \item $g_4$: Ensures mass conservation between intermediate and final products.
\end{itemize}

The objective function for the Reactor Network design problem can be described as:\par
\begin{flushleft}
    \textit{Variable:}
\end{flushleft}
\begin{flushleft}
    \[
        \begin{aligned}
            x &= [x_1, x_2, x_3, x_4, x_5, x_6]
        \end{aligned}
    \]
\end{flushleft}

\begin{flushleft}
    \textit{Minimize:}
\end{flushleft}
\begin{equation}
        y = x_4 + punishment
\end{equation}

\begin{flushleft}
    \textit{Subject to:}
\end{flushleft}
\begin{equation}
        g = \sqrt{x_5} + \sqrt{x_6} - 4 \leq 0
\end{equation}
\begin{equation}
        h_1 = x_1 + k_1 \cdot x_2 \cdot x_5 - 1=0; 
\end{equation}
\begin{equation}
        h_2 = x_2 - x_1 + k_2 \cdot x_2 \cdot x_6=0;
\end{equation}
\begin{equation}
        h_3 = x_3 + x_1 + k_3 \cdot x_3 \cdot x_5 - 1=0; 
\end{equation}
\begin{equation}
        h_4 = x_4 - x_3 + x_2 - x_1 + k_4 \cdot x_4 \cdot x_6=0; 
\end{equation}

\begin{flushleft}
    \textit{Where:}
\end{flushleft}
\begin{flushleft}
    \[
        k_1 = 0.09755988; \quad k_2 = 0.99 \cdot k_1; \quad  k_3 = 0.0391908; \quad k_4 = 0.9 \cdot k_3;
    \]
\end{flushleft}
\begin{flushleft}
    \[
        \begin{aligned}
        punishment=10^3 \cdot\max{(0, g)^2} + 10^3 \cdot \sum_{i=1}^{4} h_i^2
        \end{aligned}
\]
\end{flushleft}

\begin{flushleft}
    \textit{Variable range:}
\end{flushleft}
\begin{flushleft}
    \[
        0.00001 \leq x_1 \leq 1; \quad 0.00001 \leq x_2 \leq 1; \quad 0.00001 \leq x_3 \leq 1; \quad 0.00001 \leq x_4 \leq 1;
    \]
\end{flushleft}
\begin{flushleft}
    \[
         0.00001 \leq x_5 \leq 16; \quad 0.00001 \leq x_6 \leq 16;
    \]
\end{flushleft}

\begin{figure}[htbp]
    \centering
    \includegraphics[width=0.8\textwidth]{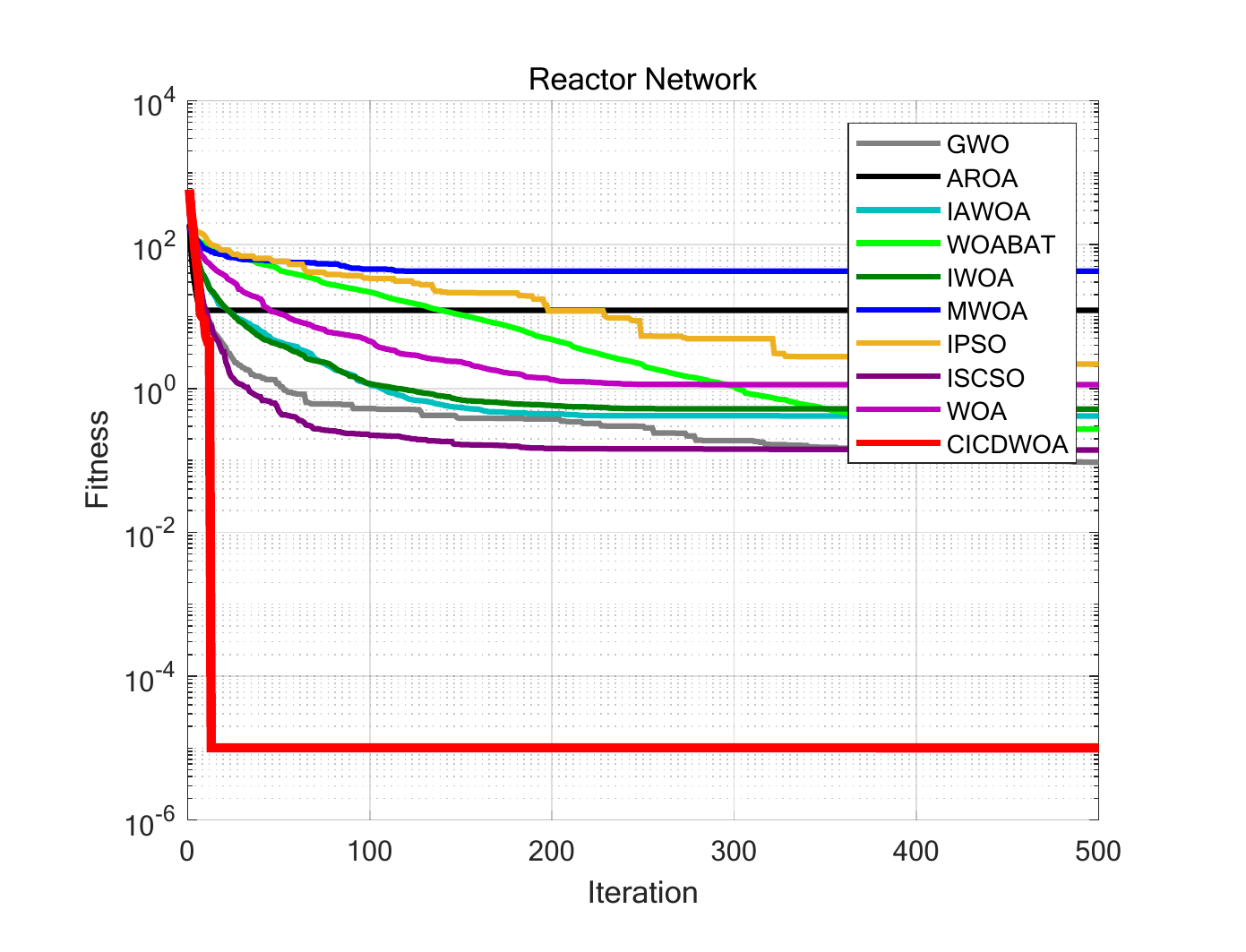}
    \caption{Iteration curves of the algorithms in solving the Reactor Network design problem.} 
    \label{engineering32}
\end{figure}

The experimental results are shown in Figure\ref{engineering32} and Table~\ref{engineering_metrics}. As indicated in Table~\ref{engineering_metrics}, CICDWOA significantly outperforms other algorithms in terms of stability in the Reactor Network design problem, and its optimization accuracy is the best among all the algorithms. This demonstrates the substantial advantage of CICDWOA in solving such optimization problems.

\subsection{Three Bar Truss}
\begin{figure}[htbp]
    \centering
    \includegraphics[width=0.6\textwidth]{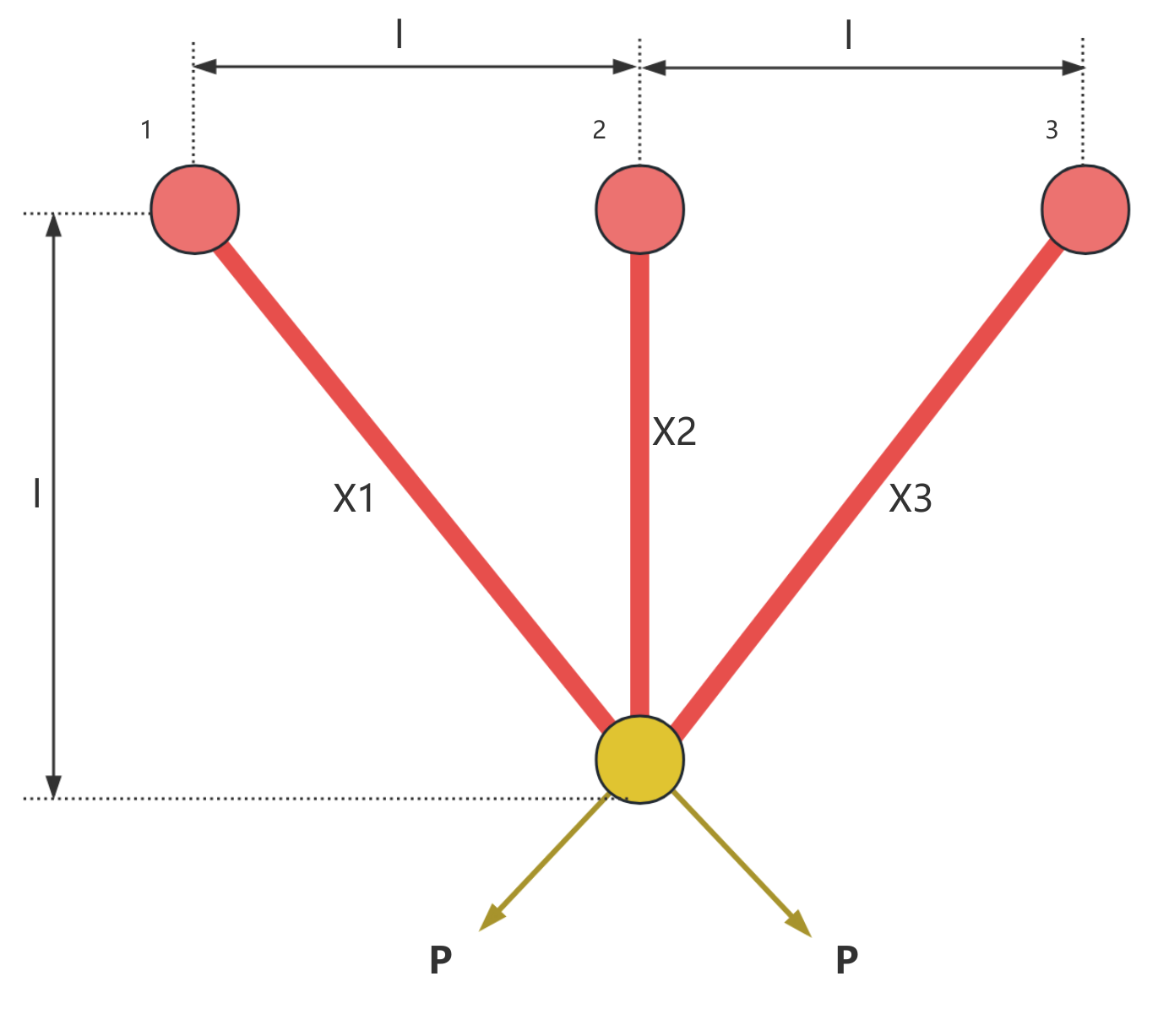}
    \caption{The structure of a three-bar truss.} 
    \label{engineering41}
\end{figure}

The three-bar truss is a basic truss structure consisting of three members, as shown in Figure~\ref{engineering41}. It is commonly used in scenarios where concentrated or distributed loads need to be supported, due to its simple structure, clear load distribution, and ease of calculation. Due to its simplicity and typical nature, the Three-bar truss design problem is often used as a benchmark test model for validating the performance of structural optimization algorithms. It is widely applied in engineering fields such as bridges, building structures, mechanical systems, and aerospace. The primary objective of the Three-bar truss design is to minimize the overall weight or material volume of the truss while ensuring the structural safety and stiffness requirements are met. By adjusting the cross-sectional areas of the truss members, a balance can be achieved between load-bearing capacity and material utilization, thus reducing manufacturing costs and improving structural performance. In this optimization problem, the design variables are two continuous decision variables, $x_1$ and $x_2$, which represent the cross-sectional areas of different members. The objective function is a nonlinear function that reflects the relationship between the total structural weight and the design variables. Additionally, the problem includes three constraint conditions, $g_1$, $g_2$, and $g_3$, which limit the maximum stress, displacement, and structural stability under external forces, ensuring that the design meets the mechanical safety requirements. The objective function for the three-bar truss design problem can be described as follows:\par
\begin{flushleft}
    \textit{Variable:}
\end{flushleft}
\begin{flushleft}
    \[
        \begin{aligned}
            x= [x_1, x_2]
        \end{aligned}
    \]
\end{flushleft}

\begin{flushleft}
    \textit{Minimize:}
\end{flushleft}

\begin{flushleft}
    \[
        \begin{aligned}
            y=(2\sqrt{2}x_1+x_2)\cdot l +punishment
        \end{aligned}
    \]
\end{flushleft}

\begin{flushleft}
    \textit{Subject to:}
\end{flushleft}
\begin{flushleft}
    \[
        g_1=\frac{\sqrt{2}x_1+x_2}{\sqrt{2}x_1^2+2x_1 \cdot x_2}\cdot P-\sigma\leq0;
    \]
\end{flushleft}
\begin{flushleft}
    \[
       g_2=\frac{x_2}{\sqrt{2}x_1^2+2x_1\cdot x_2}\cdot P-\sigma\leq0;
    \]
\end{flushleft}
\begin{flushleft}
    \[
        g_3=\frac{1}{x_1+\sqrt{2}x_2}\cdot P-\sigma\leq0
    \]
\end{flushleft}

\begin{flushleft}
    \textit{Where:}
\end{flushleft}

\begin{flushleft}
    \[
        \begin{aligned}
        punishment=10^3\cdotp\sum_{i=1}^3\max{(0,g_i)^2}
        \end{aligned}
\]
\end{flushleft}
\begin{flushleft}
    \[
        l=100; \quad P=2; \quad \sigma=2;
    \]
\end{flushleft}

\begin{flushleft}
    \textit{Variable range:}
\end{flushleft}
\begin{flushleft}
    \[
        10 \leq x_1 \leq 80; \quad 10 \leq x_2 \leq 50; \quad 0.9 \leq x_3 \leq 5; \quad 0.9\leq x_4 \leq 5;
    \]
\end{flushleft}

\begin{figure}[htbp]
    \centering
    \includegraphics[width=0.8\textwidth]{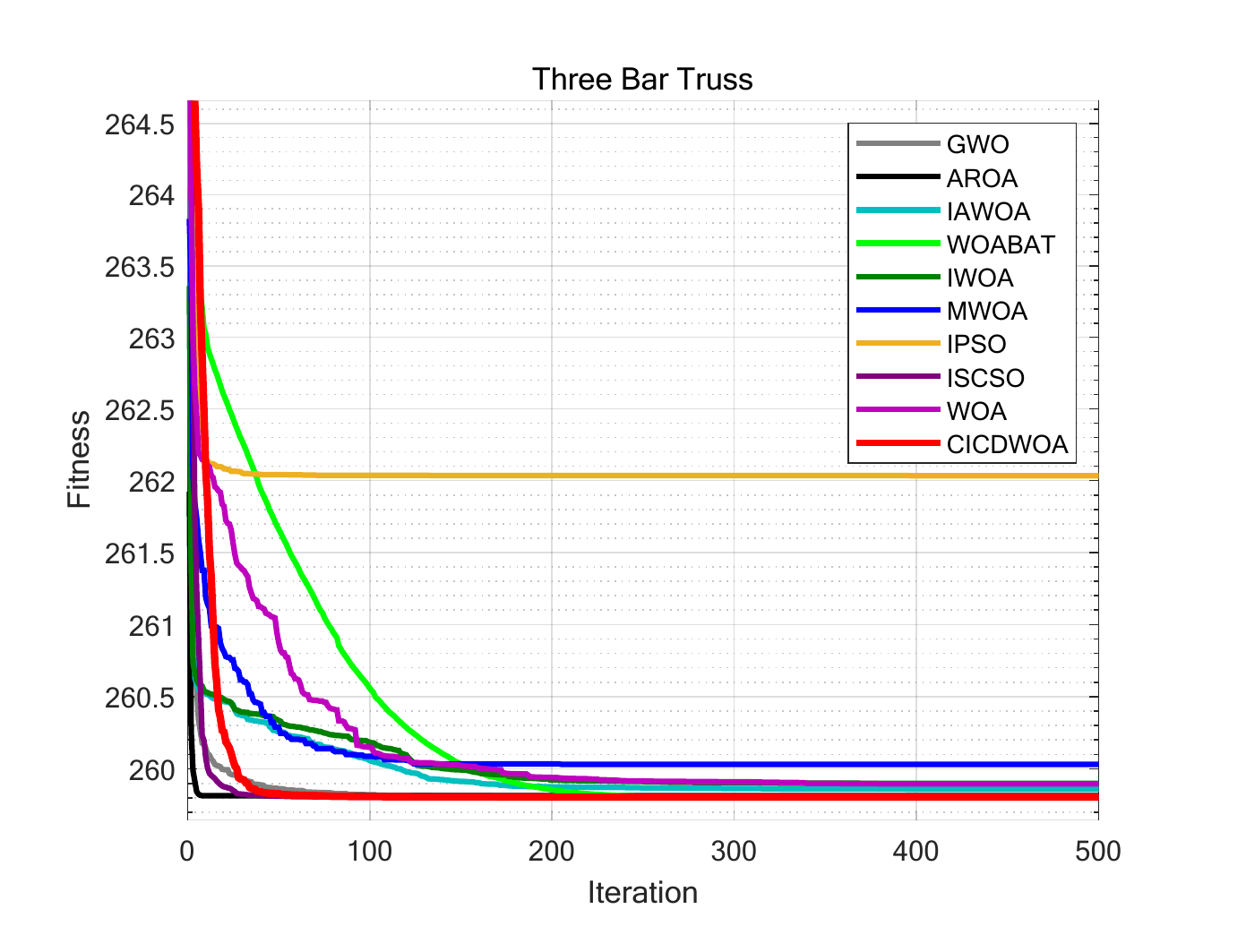}
    \caption{Iteration curves of the algorithms in solving the Three-Bar Truss design problem.} 
    \label{engineering42}
\end{figure}

The experimental results are shown in Figure~\ref{engineering42} and Table~\ref{engineering_metrics}. As indicated in Table~\ref{engineering_metrics}, CICDWOA exhibits far superior stability compared to other algorithms in the Three-bar truss design problem, and its optimization accuracy is the best among all algorithms. This demonstrates the significant advantage of CICDWOA in solving such optimization problems.

\subsection{Tension/Compression Spring}
The tension/compression spring is a common energy storage and vibration damping component, playing a crucial role in industries such as machinery, automotive, electronics, and home appliances. It can store mechanical energy when subjected to external forces and release energy once the external force is removed, thereby performing functions such as buffering, controlling vibrations, and maintaining force balance. To ensure that the spring maintains excellent mechanical performance and stability under long-term cyclic loading, its structural parameters must be designed to achieve lightweight and efficient performance while meeting strength, stiffness, and geometric constraints.\par
\begin{figure}[htbp]
    \centering
    \includegraphics[width=0.8\textwidth]{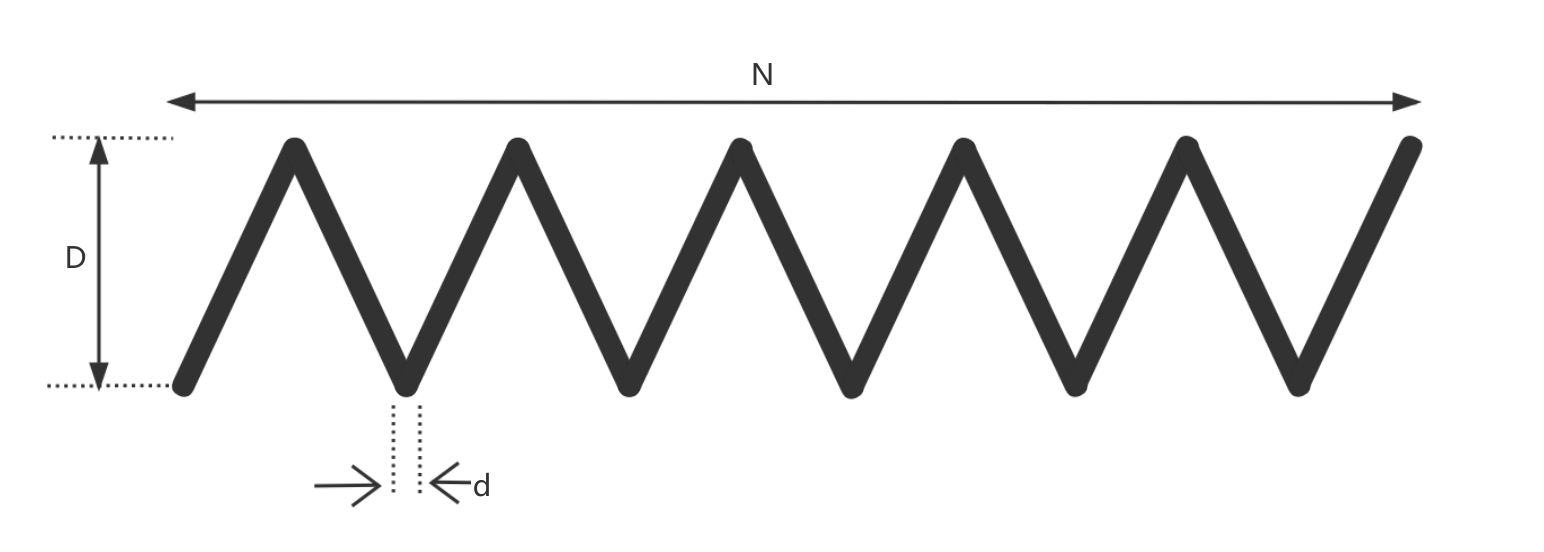}
    \caption{The structure of a tension/compression spring.} 
    \label{engineering51}
\end{figure}
\begin{figure}[htbp]
    \centering
    \includegraphics[width=0.8\textwidth]{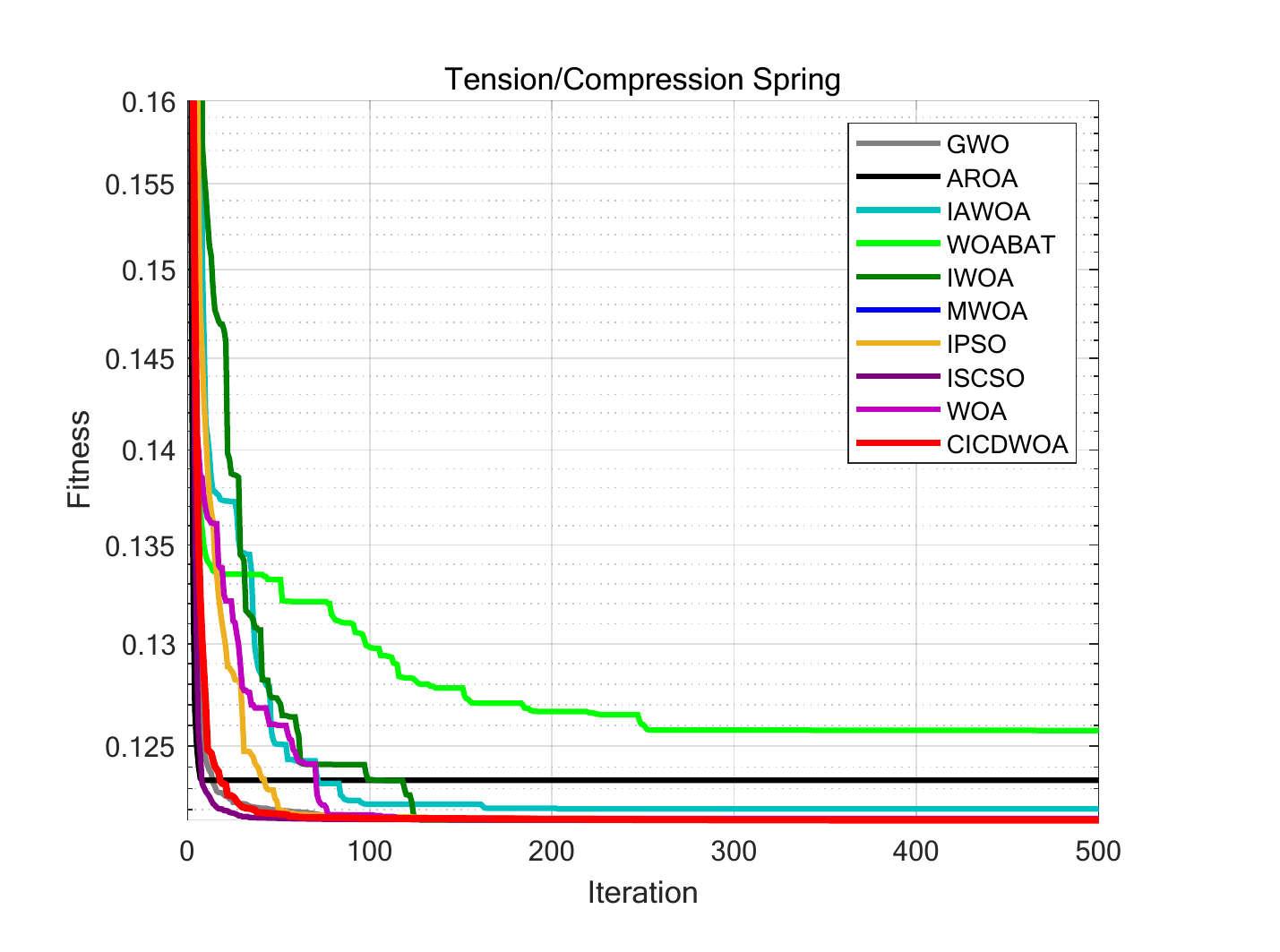}
    \caption{Iteration curves of the algorithms in solving the Tension/Compression Spring design problem.} 
    \label{engineering52}
\end{figure}

The objective of this design optimization problem is to minimize the total mass of the spring while satisfying the operational performance and structural constraints. By rationally adjusting the spring's geometric and material parameters, it is possible to reduce material consumption and manufacturing costs while ensuring the spring's strength and fatigue life. As shown in Figure~\ref{engineering51}, the problem includes three main design variables:\par
\begin{itemize}
  \item $x_1$: wire diameter ($d$)
  \item $x_2$: mean coil diameter ($D$) 
  \item $x_3$: number of active coils ($N$) 
\end{itemize}
During the design process, the spring must simultaneously satisfy several nonlinear constraint conditions $g_1$,$g_2$,$g_3$,$g_4$, which together ensure that the spring has sufficient elastic deformation capability under working loads while preventing yielding or instability. The mathematical model of the problem is as follows,\par
\begin{flushleft}
	\textit{Variable:}
	\end{flushleft}
	\begin{flushleft}
		\[
			\begin{aligned}
			x=[d, D, N]=[x_{1},x_{2},x_{3}]
		\end{aligned}
		\]
	\end{flushleft}
	
	\begin{flushleft}
	\textit{Minimize:}
	\end{flushleft}
	\begin{equation}
		y=(x_3+2)\cdot x_2 \cdot x_1^2+punishment
	\end{equation}
	\begin{flushleft}
		
	\textit{Subject to:}
	\end{flushleft}
	\begin{equation}
		g_1=1-\frac{x_2^3 \cdot x_3}{71785x_1^4}{\leq}0
	\end{equation}
	
	\begin{equation}
		g_2=\frac{4x_2^2-x_1\cdot x_2}{12566(x_2\cdot x_1^3-x_4)}+\frac{1}{5108x_1^2}-1\leq 0
	\end{equation}
	
	\begin{equation}
		g_3=1-\frac{140.45x_1}{x_2^2\cdot x_3}{\leq}0
	\end{equation}
	
	\begin{equation}
		g_4=\frac{x_1+x_2}{1.5}-1\leq0
	\end{equation}
	
\begin{flushleft}
    \textit{Where:}
\end{flushleft}

\begin{flushleft}
    \[
        \begin{aligned}
        punishment=10^3\cdotp\sum_{i=1}^4\max{(0,g_i)^2}
        \end{aligned}
\]
\end{flushleft}

\begin{flushleft}
	\textit{Variable range:}
	\end{flushleft}
	\begin{flushleft}
		\[
			\begin{aligned}
		0.05\leq x_1\leq 2,\quad 0.25\leq x_2\leq 1.3,\quad 2.0\leq x_3\leq 15
	\end{aligned}
	\]
	\end{flushleft}

The experimental results are shown in Figure~\ref{engineering51} and Table~\ref{engineering_metrics}. As indicated in Table~\ref{engineering_metrics}, CICDWOA exhibits significantly superior stability in the Tension/Compression Spring design problem compared to other algorithms, and its optimization accuracy is the best among all the algorithms. This demonstrates the substantial advantage of CICDWOA in solving such optimization problems.

\subsection{Speed Reducer}
\begin{figure}[htbp]
    \centering
    \includegraphics[width=0.6\textwidth]{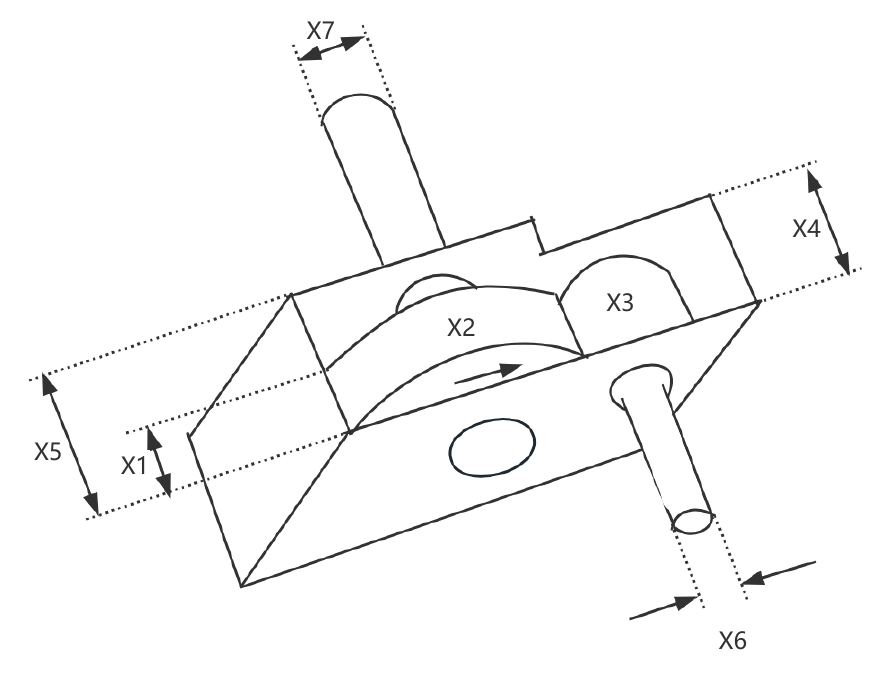}
    \caption{The structure of a speed reducer.} 
    \label{engineering61}
\end{figure}

The speed reducer is a mechanical transmission device used to reduce rotational speed and increase torque, and is widely applied in fields such as automation equipment, robotics, machine tools, transportation machinery, and aerospace. It typically consists of components such as gears, shafts, bearings, and housings, and achieves power transmission and speed regulation through the meshing of gears with different sizes. A well-designed speed reducer can not only improve mechanical transmission efficiency and operational stability, but also significantly reduce vibration, noise, and energy loss, thereby enhancing the overall system's reliability and lifespan. As shown in Figure~\ref{engineering61}, this problem includes seven design variables:\par
\begin{itemize}
  \item $x_1$: width of the gear teeth;
  \item $x_2$: gear module;
  \item $x_3$: number of teeth on the small gear;
  \item $x_4$: length of the first shaft between the bearings;
  \item $x_5$: length of the second shaft between the bearings;
  \item $x_6$: diameter of the first shaft;
  \item $x_7$: diameter of the second shaft.
\end{itemize}
In addition, the problem includes 11 constraint conditions that limit gear strength, shaft bending stress, shear stress, contact stress, and assembly space, among other design requirements. These constraints ensure the structural safety and transmission reliability of the speed reducer during long-term operation. The mathematical model of this problem is as follows,\par
\begin{flushleft}
    \textit{Variable:}
\end{flushleft}
\begin{flushleft}
    \[
        \begin{aligned}
            x = [x_1, x_2, x_3, x_4, x_5, x_6, x_7]
        \end{aligned}
    \]
\end{flushleft}

\begin{flushleft}
    \textit{Minimize:}
\end{flushleft}
\begin{equation}
        y = f(x)+punishment
\end{equation}

\begin{flushleft}
    \textit{Subject to:}
\end{flushleft}
\begin{equation}
    g_1=\frac{27}{x_1\cdot x_2^2\cdot x_3}-1\leq0; 
\end{equation}

\begin{equation}
    g_2=\frac{397.5}{x_1\cdot x_2^2\cdot x_3^2}-1\leq0; 
\end{equation}

\begin{equation}
    g_3=\frac{1.93x_4^3}{x_2\cdot x_6^4\cdot x_3}-1\leq0; 
\end{equation}

\begin{equation}
    g_4=\frac{1.93x_5^3}{x_2\cdot x_7^4\cdot x_3}-1\leq0; 
\end{equation}

\begin{equation}
    g_5=\frac{\sqrt{16.91\cdot 10^6+\left(\frac{745x_4}{x_2\cdot x_3}\right)^2}}{110x_6^3}-1\leq0; 
\end{equation}

\begin{equation}
    g_6=\frac{\sqrt{157.5\cdot 10^6+\left(\frac{745x_4}{x_2\cdot x_3}\right)^2}}{85x_7^3}-1\leq0; 
\end{equation}

\begin{equation}
    g_7=\frac{x_2\cdot x_3}{40}-1\leq0; 
\end{equation}

\begin{equation}
    g_8=\frac{5x_2}{x_1}-1\leq0; 
\end{equation}

\begin{equation}
    g_{9}=\frac{x_{1}}{12x_{2}}-1\leq0; 
\end{equation}

\begin{equation}
    g_{10}=\frac{1.5x_6+1.9}{x_4}-1\leq0; 
\end{equation}

\begin{equation}
    g_{11}=\frac{1.1x_7+1.9}{x_5}-1 \leq0; 
\end{equation}

\begin{flushleft}
    \textit{Where:}
\end{flushleft}
\begin{flushleft}
    \[
        \begin{aligned}
        punishment=10^3\cdotp\sum_{i=1}^{11}\max{(0,g_i)^2}
        \end{aligned}
\]
\end{flushleft}

\begin{equation}
        \begin{aligned}
            f(x) =\;& 0.7854\,x_1 x_2^2 \big( 3.3333\,x_3^2 + 14.9334\,x_3 - 43.0934 \big)\\
 & - 1.508\,x_1 \big( x_6^2 + x_7^2 \big)  + 7.4777 \big( x_6^3 + x_7^3 \big)\\
 & + 0.7854 \big( x_4 x_6^2 + x_5 x_7^2 \big)
            \end{aligned}
\end{equation}

\begin{flushleft}
    \textit{Variable range:}
\end{flushleft}
\vspace{-\baselineskip}
\begin{flushleft}
    \[
    2.6\leq x_1 \leq3.6; \quad 0.7\leq x_2 \leq 0.8; 
    \quad 17\leq x_3 \leq 28; 
    \]
\end{flushleft}
\vspace{-\baselineskip}
\begin{flushleft}
    \[
    7.3\leq x_4 \leq 8.3; \quad 7.3\leq x_5 \leq 8.3; \quad 2.9\leq x_6 \leq 3.9;
    \]
\end{flushleft}
\vspace{-\baselineskip}
\begin{flushleft}
    \[
    5\leq x_7 \leq 5.5;
    \]
\end{flushleft}

\begin{figure}[htbp]
    \centering
    \includegraphics[width=0.8\textwidth]{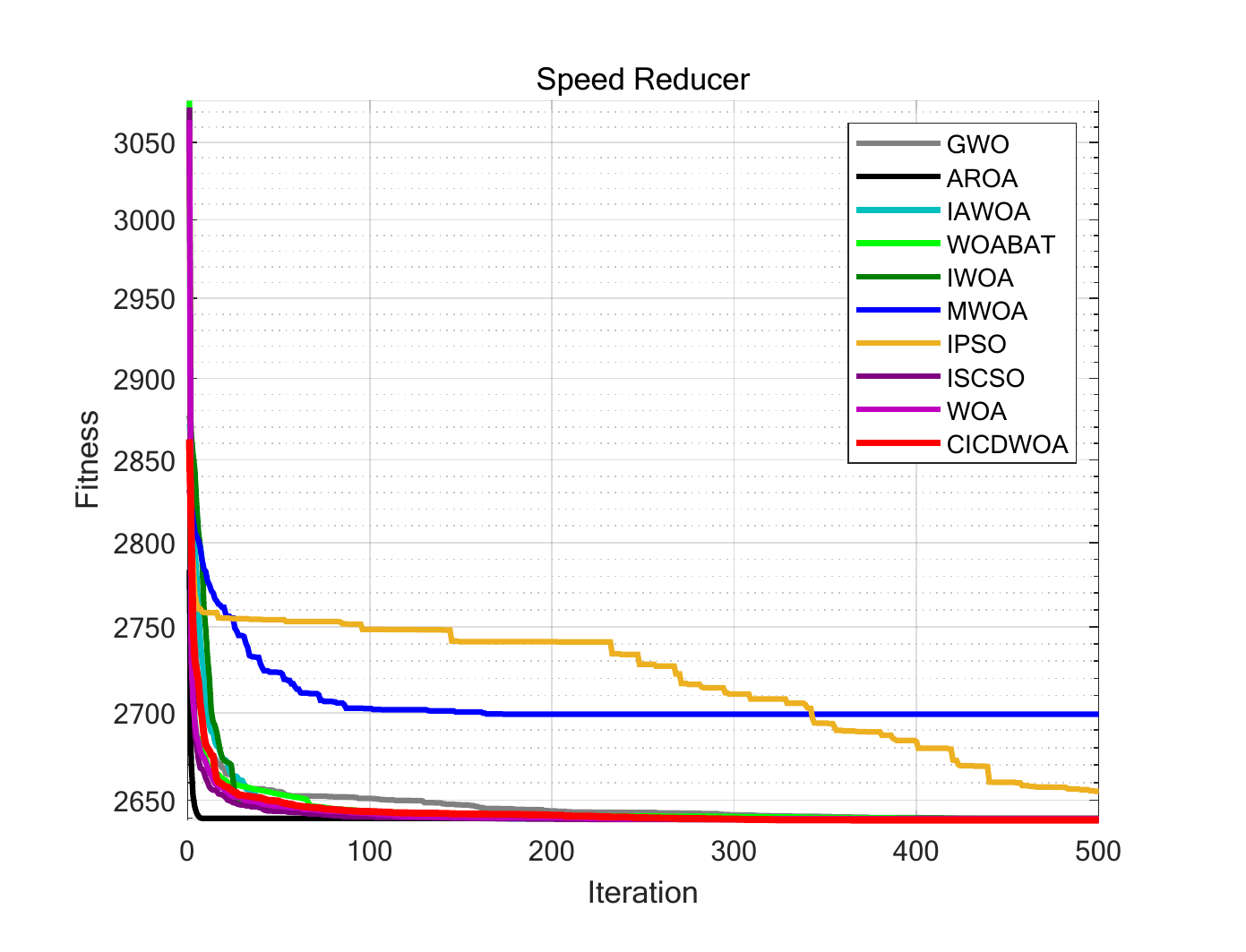}
    \caption{Iteration curves of the algorithms in solving the Speed Reducer design problem.} 
    \label{engineering62}
\end{figure}

The experimental results are shown in Figure~\ref{engineering62} and Table~\ref{engineering_metrics}. As indicated in Table~\ref{engineering_metrics}, CICDWOA exhibits significantly superior stability compared to other algorithms in the Speed Reducer design problem, and its optimization accuracy is the best among all algorithms. This demonstrates the considerable advantage of CICDWOA in handling such optimization problems.

\subsection{I-beam}
\begin{figure}[htbp]
    \centering
    \includegraphics[width=0.5\textwidth]{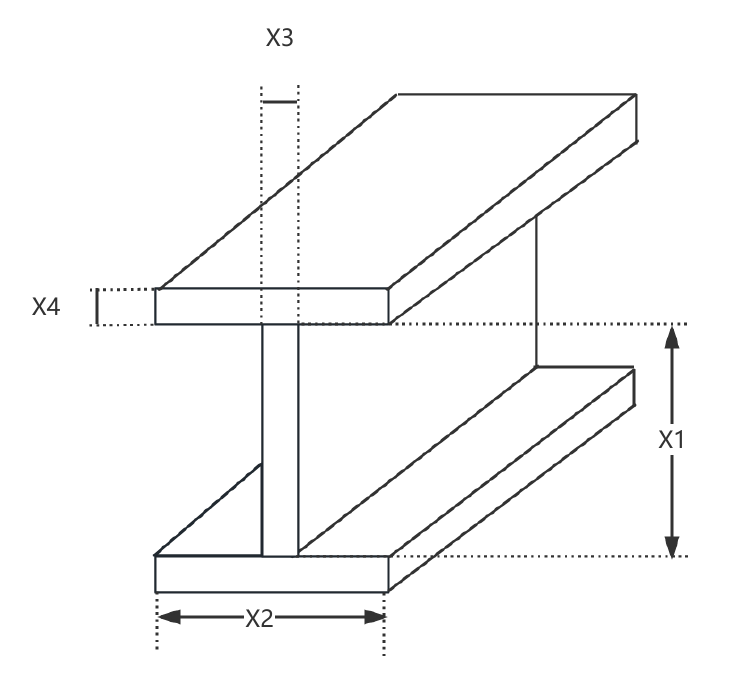}
    \caption{The structure of an I-beam.} 
    \label{engineering71}
\end{figure}

The I-beam, as shown in Figure~\ref{engineering71}, is a load-bearing component with a characteristic 'I'-shaped cross-section. Due to its excellent strength and stiffness properties, it has been widely used in construction engineering, bridge structures, mechanical manufacturing, as well as in the shipbuilding and aerospace industries. The design concept of the I-beam is to achieve high bending resistance under bending loads by distributing materials efficiently, while simultaneously reducing weight and material consumption. The core objective of the I-beam optimization design is to determine the optimal configuration of cross-sectional dimensions while ensuring structural strength and stability, thereby minimizing material volume or mass. This optimization problem typically takes into account the impact of the web and flange geometric parameters on overall performance, and adjusts the variables to balance strength, stiffness, and lightweight design. This problem includes four design variables:\par
\begin{itemize}
  \item $x_1$: height of the web;
  \item $x_2$: width of the flange;
  \item $x_3$: thickness of the web;
  \item $x_4$: thickness of the flange.
\end{itemize}
During the design process, the I-beam must simultaneously satisfy two constraint conditions, $g_1$ and $g_2$, which are used to limit the maximum stress and deflection, ensuring that the beam possesses adequate load-bearing capacity without excessive deformation under applied loads. By implementing these constraints, a reasonable balance between safety, economy, and manufacturability can be achieved. The mathematical model of this problem is as follows,\par
\begin{flushleft}
    \textit{Variable:}
\end{flushleft}
\vspace{-\baselineskip}
\begin{flushleft}
    \[
        \begin{aligned}
            x= [x_1, x_2, x_3, x_4]
        \end{aligned}
    \]
\end{flushleft}

\begin{flushleft}
    \textit{Maximize:}
\end{flushleft}

\begin{equation}
            y=\frac{5000}{\frac{x_3\cdot (x_1-2x_4)^3}{12}+\frac{x_2 \cdot x_4^3}{6}+2x_2 \cdot x_4\left(\frac{x_1-x_4}{2}\right)^2}+punishment
\end{equation}

\begin{flushleft}
    \textit{Subject to:}
\end{flushleft}
\begin{equation}
        g_1=2x_2 \cdot x_3+x_3 \cdot (x_1-2x_4)-300\leq0;
\end{equation}

\begin{equation}
        g_2=\frac{18\cdot 10^4x_1}{x_3\left(x_1-2x_4\right)^3+2x_2\cdot x_3\left(4x_4^2+3x_1\cdot (x_1-2x_4)\right)}\leq0;
\end{equation}

\begin{flushleft}
    \textit{Where:}
\end{flushleft}
\begin{flushleft}
    \[
        \begin{aligned}
        punishment=10^3\cdotp\sum_{i=1}^2\max{(0,g_i)^2}
        \end{aligned}
\]
\end{flushleft}

\begin{flushleft}
    \textit{Variable range:}
\end{flushleft}
\vspace{-\baselineskip}
\begin{flushleft}
    \[
        10 \leq x_1 \leq 80; \quad 10 \leq x_2 \leq 50; \quad 0.9 \leq x_3 \leq 5; \quad 0.9\leq x_4 \leq 5;
    \]
\end{flushleft}

\begin{figure}[htbp]
    \centering
    \includegraphics[width=0.8\textwidth]{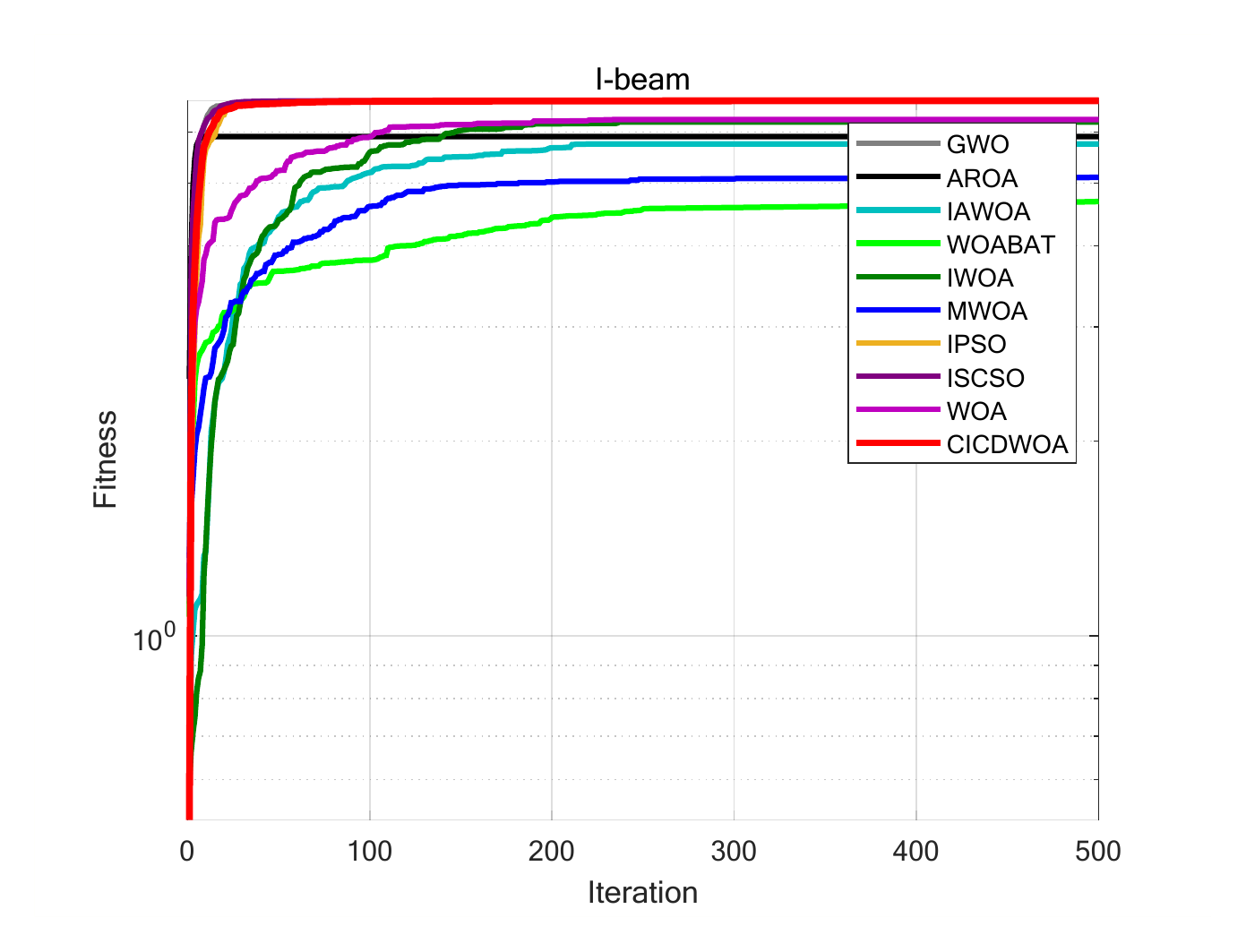}
    \caption{Iteration curves of the algorithms in solving the I-beam design problem.} 
    \label{engineering72}
\end{figure}

\begin{figure}[htbp]
    \centering
    \includegraphics[width=0.6\textwidth]{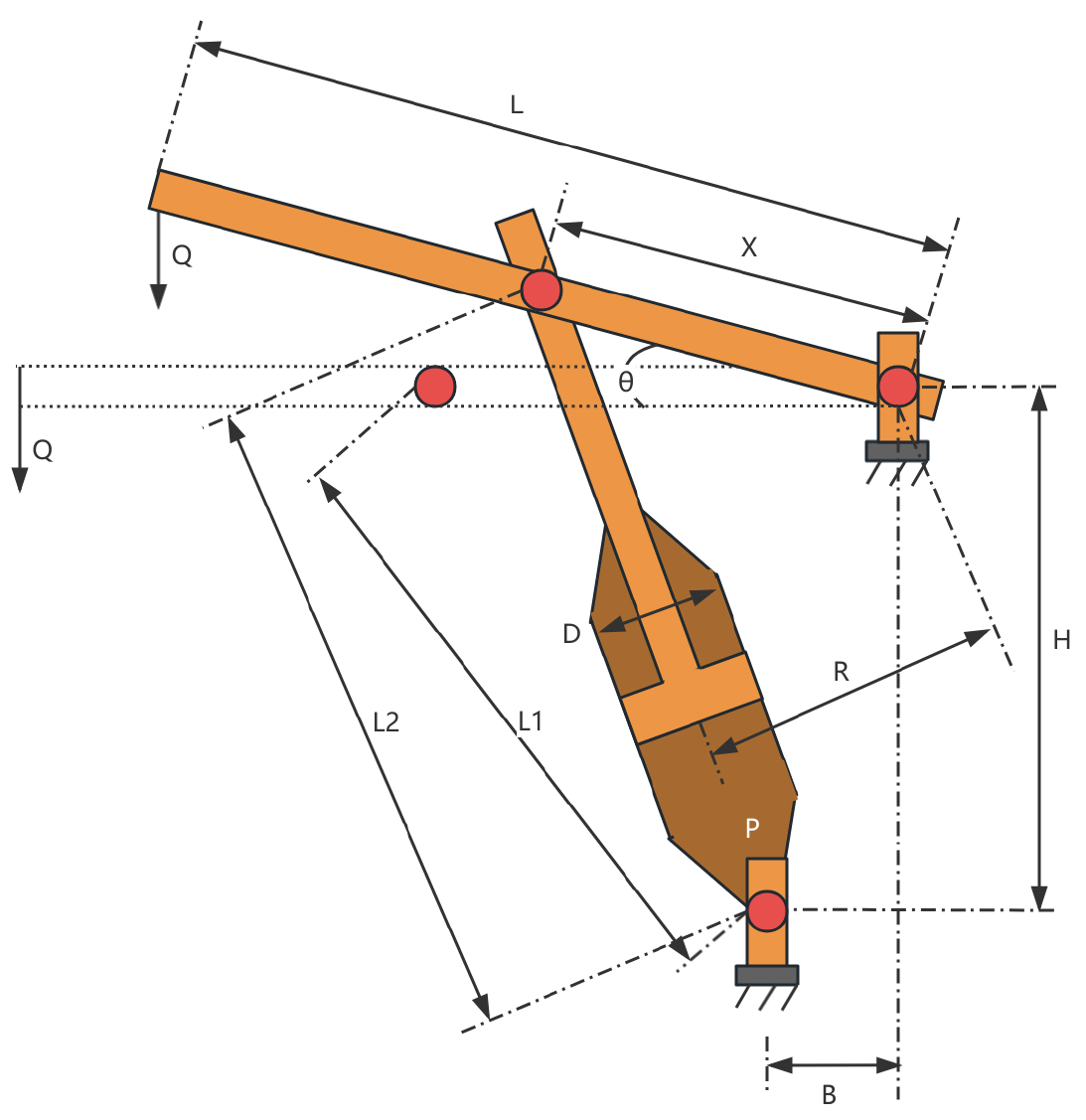}
    \caption{The structure of a piston lever.} 
    \label{engineering81}
\end{figure}

The experimental results are shown in Figure~\ref{engineering72} and Table~\ref{engineering_metrics}. As indicated in Table~\ref{engineering_metrics}, CICDWOA demonstrates significantly superior stability in the I-beam design problem compared to other algorithms, and its optimization accuracy is the best among all algorithms. This highlights the substantial advantage of CICDWOA in addressing such optimization problems.

\subsection{Piston Lever}
\begin{figure}[htbp]
    \centering
    \includegraphics[width=0.8\textwidth]{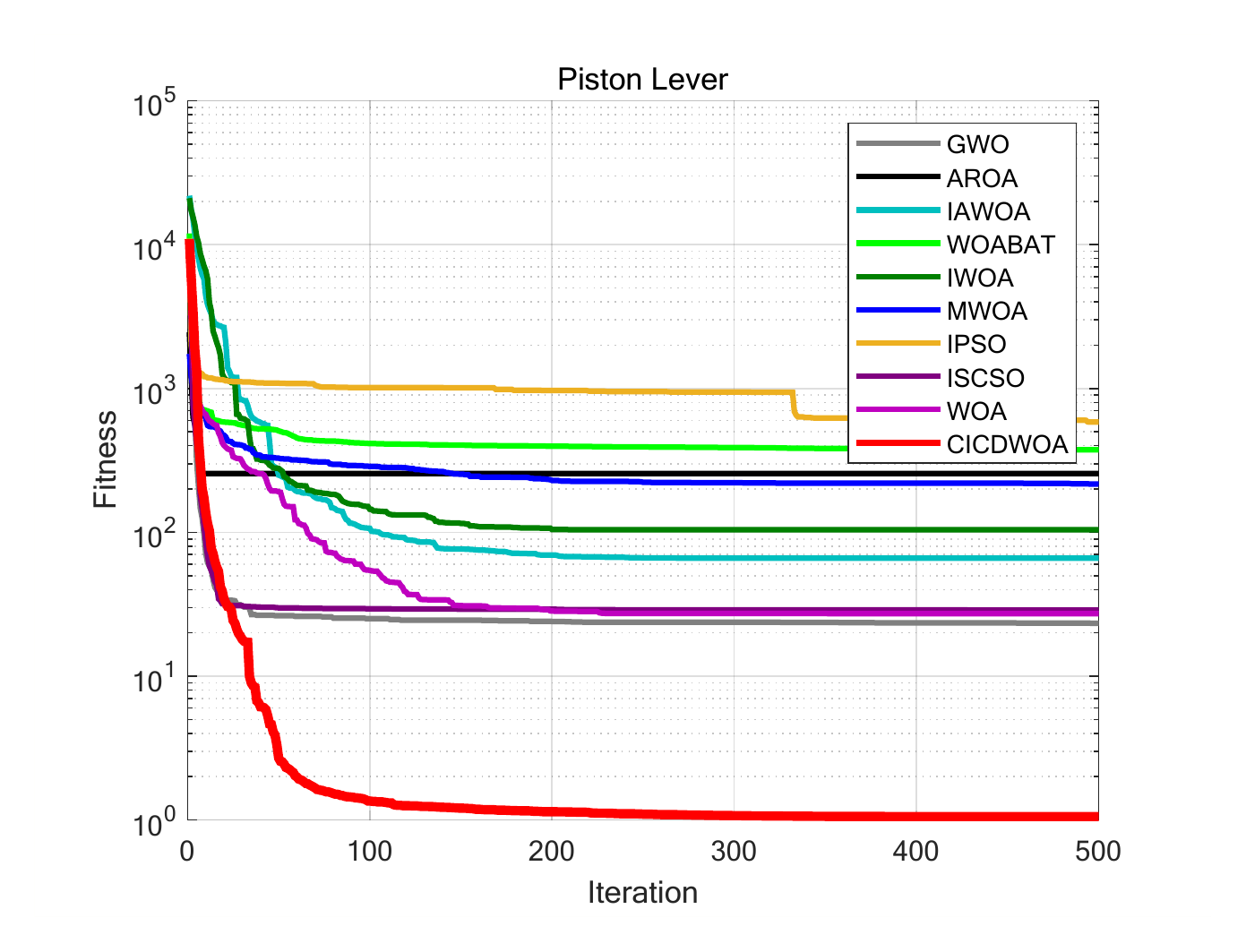}
    \caption{Iteration curves of the algorithms in solving the Piston Lever design problem.} 
    \label{engineering82}
\end{figure}

The Piston Lever is a critical component commonly used in mechanical transmission and power systems, with widespread applications in engines, hydraulic systems, and various actuators. It typically converts linear motion generated by a piston into a specific torque output through the lever principle, thereby driving other mechanical components. Since the piston lever must endure substantial alternating loads and friction during operation, its structural design must balance strength, stiffness, and lightweight properties to ensure efficient, stable, and durable performance. In the optimization of the piston lever, the research goal is to minimize the total material usage while ensuring that the structural strength and geometric constraints are met, i.e., achieving a lightweight design without compromising mechanical performance. This optimization process holds significant engineering importance for reducing manufacturing costs, improving energy efficiency, and extending service life.
As shown in Figure~\ref{engineering82}, the geometry of the piston lever is defined by several key parameters, and its structural characteristics are primarily controlled by four design variables:\par
\begin{itemize}
  \item $x_1$: primary length parameter of the lever body;
  \item $x_2$: primary width parameter of the lever body;
  \item $x_3$: cross-sectional radius at the force point;
  \item $x_4$: geometric dimension related to the support point.
\end{itemize}
The Piston Lever design problem can be described as:\par
\begin{flushleft}
    \textit{Variable:}
\end{flushleft}
\begin{equation}
    x = [x_1, x_2, x_3, x_4]
 \end{equation}

 \begin{flushleft}
    \textit{Minimize:}
\end{flushleft}
\begin{equation}
    y = 0.25\pi x_3^2(L_2-L_1)+punishment
\end{equation}

\begin{flushleft}
    \textit{Subject to:}
\end{flushleft}
\begin{equation}
    g_1=QL\cos\theta-RF\leq0; 
\end{equation}

\begin{equation}
        g_2=Q(L-x_4)-M_{\max}\leq0;
\end{equation}

\begin{equation}
    g_3=1.2(L_2-L_1)-L_1\leq0; 
\end{equation}

\begin{equation}
    g_4=\frac{x_3}2-x_2\leq0;
\end{equation}

\begin{flushleft}
    \textit{Variable range:}
\end{flushleft}
\begin{equation}
    0.05\leq x_1 \leq500; \quad 0.05\leq x_2 \leq 500; 
    \quad 0.05\leq x_4 \leq 500; \quad 0.05\leq x_3 \leq 120;
\end{equation}

\begin{flushleft}
    \textit{Where:}
\end{flushleft}
\begin{flushleft}
    \[
        \begin{aligned}
        punishment=10^3\cdotp\sum_{i=1}^4\max{(0,g_i)^2}
        \end{aligned}
\]
\end{flushleft}

\begin{flushleft}
    \[
    Q=10000;\quad P=1500; \quad L=240; \quad M_{\max}=1.8\times10^6;
    \]
\end{flushleft}

\begin{flushleft}
    \[
    L_1=\sqrt{(x_4-x_2)^2+x_1^2};\quad L_2=\sqrt{(x_4\sin\theta+x_1)^2+(x_2-x_4\cos\theta)^2};
\]
\end{flushleft}

\begin{flushleft}
    \[
    R=\frac{|-x_4(x_4\sin\theta+x_1)+x_1(x_2-x_4\cos\theta)|}{\sqrt{(x_4-x_2)^2+x_1^2}};
\]
\end{flushleft}

\begin{flushleft}
    \[
    F=0.25\pi Px_3^2;
\]
\end{flushleft}

The experimental results are shown in Figure~\ref{engineering82} and Table~\ref{engineering_metrics}. As indicated in Table~\ref{engineering_metrics}, CICDWOA demonstrates significantly superior stability in the piston lever design problem compared to other algorithms, and its optimization accuracy is the best among all algorithms. This highlights the substantial advantage of CICDWOA in handling such optimization problems.

\subsection{Gas Transmission Compressor}
\begin{figure*}[htbp]
    \centering
    \includegraphics[width=\textwidth]{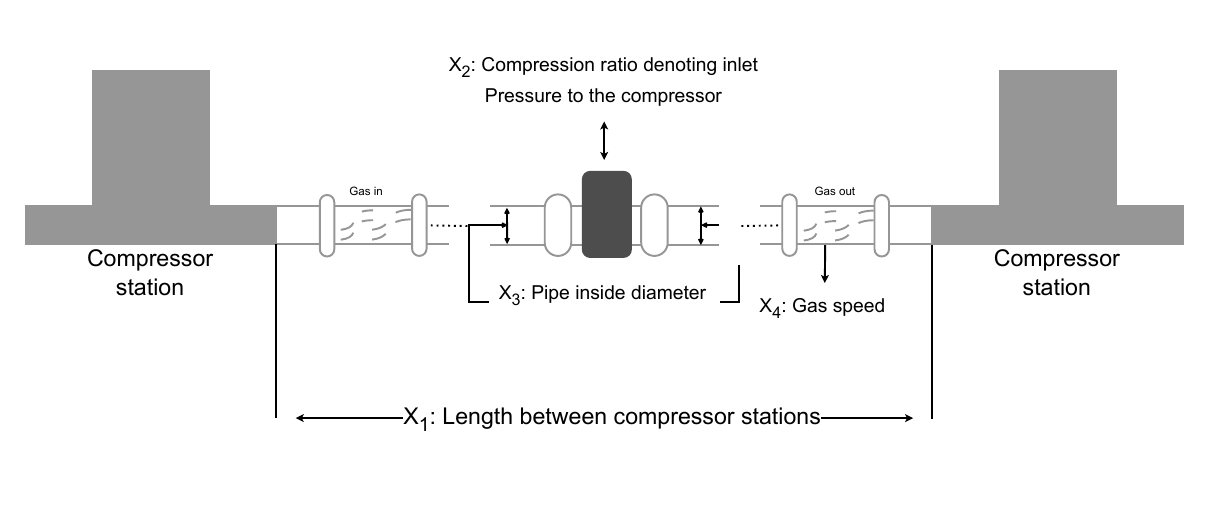}
    \caption{The structure of a gas transmission compressor.} 
    \label{engineering91}
\end{figure*}

The Gas Transmission Compressor, as shown in Figure~\ref{engineering91}, is one of the core components in long-distance natural gas transportation systems. Its primary function is to maintain and increase the pressure of natural gas within the pipeline, overcoming resistive forces and ensuring the continuous flow of gas throughout the network. This equipment is widely used in urban gas supply, industrial energy delivery, as well as in cross-regional and international pipeline systems, making it an essential part of modern energy infrastructure. The operational efficiency of the compressor directly affects the energy consumption, operational costs, and environmental impact of the entire gas transmission system. In long-distance gas transmission networks, compressors are typically arranged at regular intervals, providing stage-by-stage pressurization to maintain a stable gas flow rate over extensive pipeline lengths. Since the compressor's design parameters are closely linked to the pipeline structure, gas flow characteristics, and operational conditions, optimizing the compressor system can significantly reduce energy losses and operational costs while also improving system safety and economic efficiency.\par
The optimization objective for the Gas Transmission Compressor problem is to minimize system energy consumption or operating costs, while satisfying fluid dynamics constraints and operational safety requirements. The design variables include four main parameters:\par
\begin{itemize}
  \item $x_1$: Length between compressor stations, which determines the density of compressor placement in the pipeline network;
  \item $x_2$: Compression ratio (denoting the inlet pressure to the compressor), which controls the ratio of input to output pressure and is a key parameter affecting compression energy consumption;
  \item $x_3$: Pipe inside diameter, influencing gas flow resistance and transmission efficiency;
  \item $x_4$: Gas speed at the output side, closely related to gas flow rate and system stability.
\end{itemize}
The mathematical modeling of the Gas Transmission Compressor optimization problem is as follows:\par
\begin{flushleft}
    \textit{Variable:}
\end{flushleft}
\begin{flushleft}
    \[
        \begin{aligned}
            x &= [x_1, x_2, x_3, x_4]
        \end{aligned}
    \]
\end{flushleft}

\begin{flushleft}
    \textit{Minimize:}
\end{flushleft}
\begin{equation}
        y =f(x) +punishment
\end{equation}

\begin{flushleft}
    \textit{Subject to:}
\end{flushleft}
\begin{equation}
    g =x_{4}x_{2}^{-2}+x_{2}^{-2}-1\leq 0;
\end{equation}

\begin{flushleft}
    \textit{Where:}
\end{flushleft}

\begin{equation}
        \begin{aligned}
        f(x)=8.61\cdot 10^5x_1^{\frac{1}{2}}x_2x_3^{-\frac{2}{3}}x_4^{-\frac{1}{2}}+3.69\cdot 10^4x_3+7.72\cdot 10^8x_1^{-1}x_2^{0.219}-765.43\cdot 10^6x_1^{-1} 
        \end{aligned}
\end{equation}

\begin{flushleft}
    \[
        \begin{aligned}
        punishment=10^3 \cdot\max{(0, g)^2} 
        \end{aligned}
\]
\end{flushleft}

\begin{flushleft}
    \textit{Variable range:}
\end{flushleft}
\begin{flushleft}
    \[
        20 < x_1 < 50; \quad 1 < x_2 < 10; \quad 20 < x_3 < 45; \quad 0.1 < x_4 < 60
    \]
\end{flushleft}

\begin{figure}[htbp]
    \centering
    \includegraphics[width=0.8\textwidth]{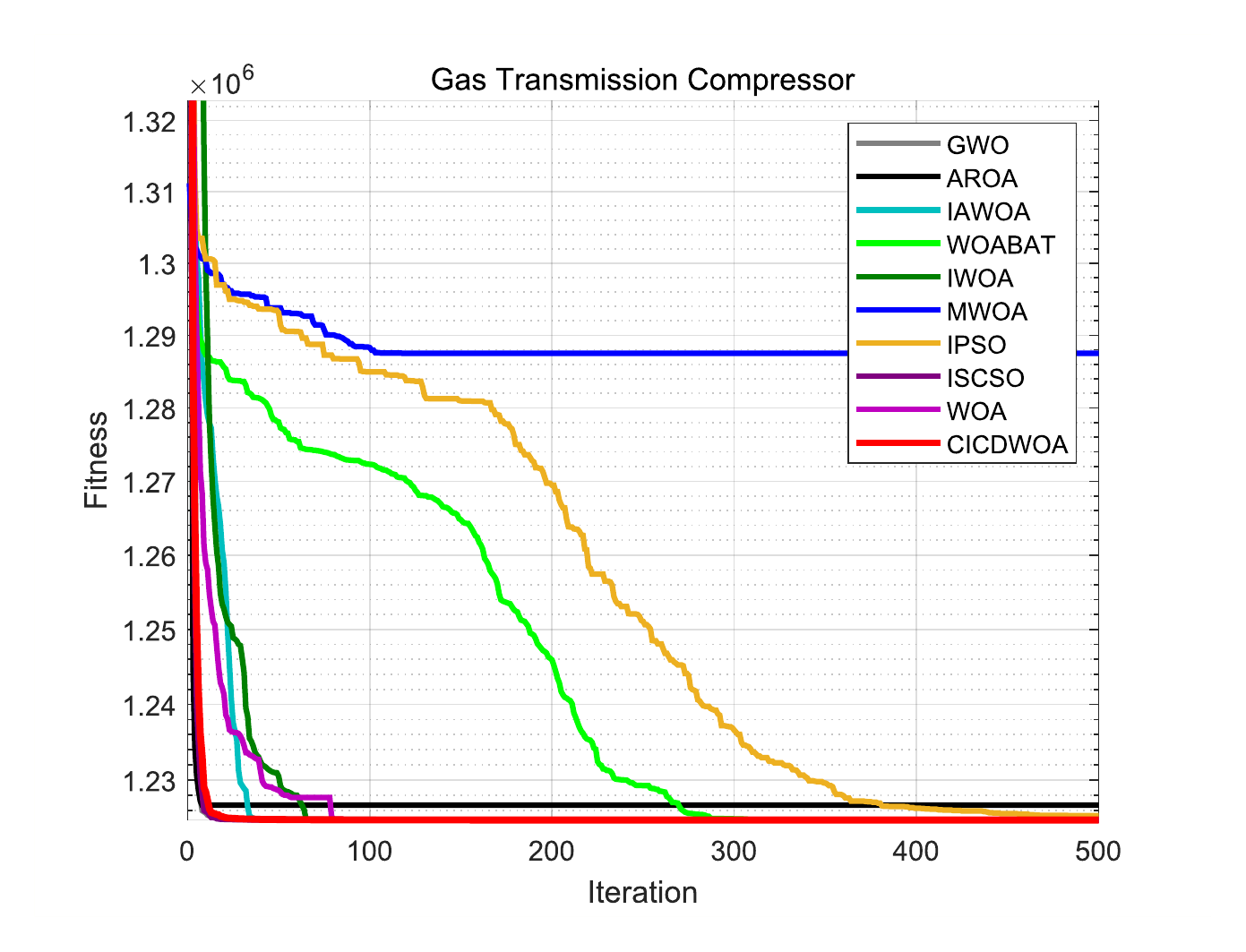}
    \caption{Iteration curves of the algorithms in solving the Gas Transmission Compressor design problem.} 
    \label{engineering92}
\end{figure}

The experimental results are shown in Figure~\ref{engineering92} and Table~\ref{engineering_metrics}. As shown in Table~\ref{engineering_metrics}, CICDWOA demonstrates significantly better stability in solving the Gas Transmission Compressor design problem compared to other algorithms, and its optimization accuracy is the best among all algorithms. This demonstrates the substantial advantage of CICDWOA in handling such optimization problems.

\subsection{Industrial Refrigeration System}
The Industrial Refrigeration System is an indispensable core auxiliary device in industries such as chemicals, pharmaceuticals, food processing, energy, and manufacturing. It is primarily used to maintain temperature stability and thermal balance during industrial processes. The system operates by circulating refrigerants to absorb and release heat, thus providing a constant low-temperature environment for reactors, storage tanks, distillation columns, and various precision equipment. In chemical plants and large-scale production facilities, the refrigeration system not only impacts product quality and reaction safety, but also directly determines energy utilization efficiency and production costs. Therefore, optimizing the design of such systems holds significant economic and engineering importance.\par
\begin{figure*}[htbp]
    \centering
    \includegraphics[width=\textwidth]{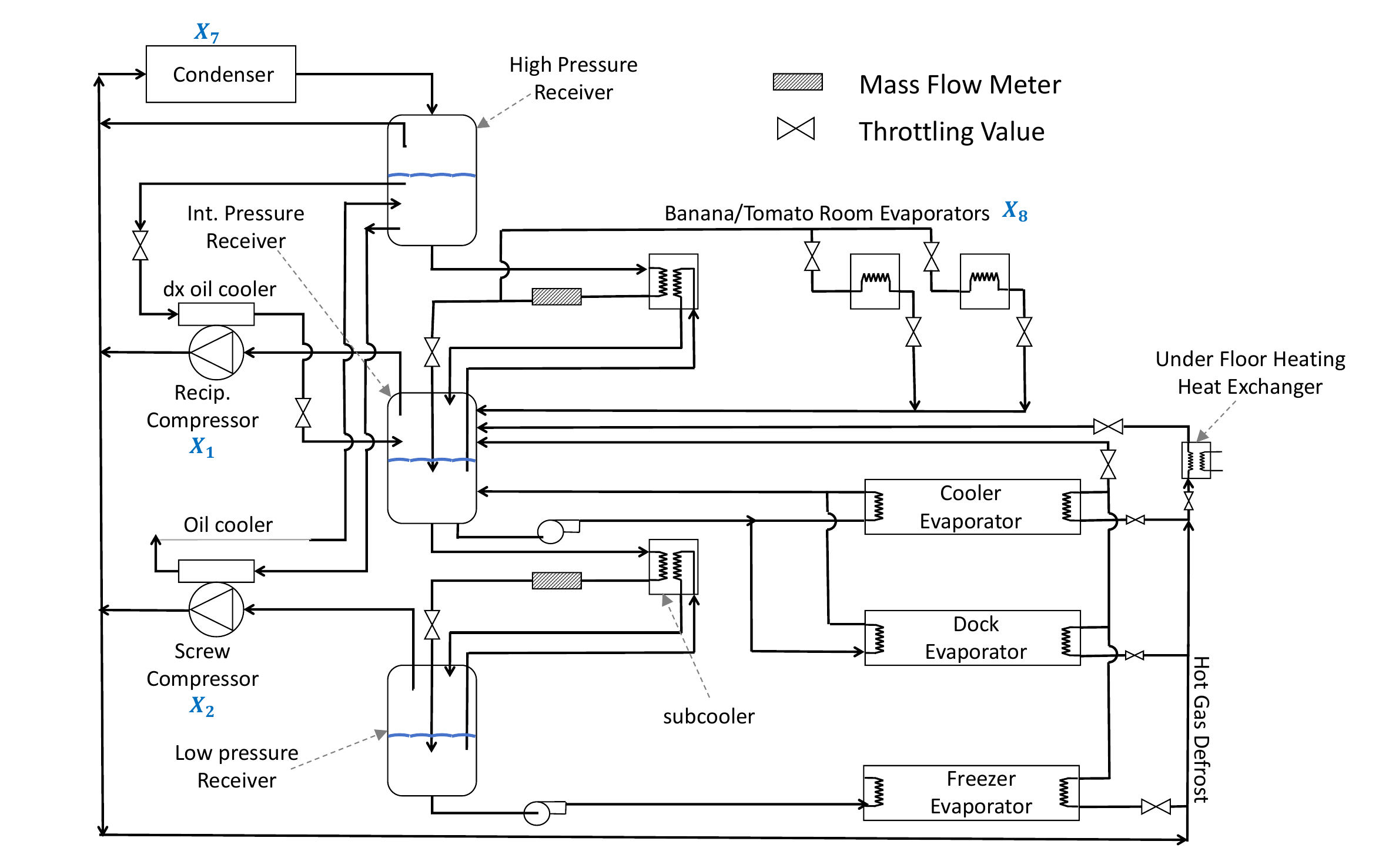}
    \caption{The structure of an industrial refrigeration system.} 
    \label{engineering101}
\end{figure*}

The optimization objective of the Industrial Refrigeration System is to minimize energy consumption and operational costs while ensuring sufficient cooling capacity and heat exchange efficiency. As illustrated in the figure, the system mainly consists of a compressor, condenser, evaporator, expansion valve, and piping network. There is a strong nonlinear coupling relationship among the parameters of these components, making the optimization problem complex and multidimensional. As shown in Figure~\ref{engineering101}, this problem includes fourteen design variables:\par
\begin{itemize}
  \item $x_1$ and $x_2$: Compressor power parameters, which directly determine the system's cooling output capacity and energy consumption level;
  \item $x_3$ to $x_6$: Refrigerant mass and flow rates, which describe the cycling characteristics of the refrigerant in the condenser, evaporator, and accumulator;
  \item $x_7$ and $x_8$: Structural characteristic parameters of the condenser and evaporator, affecting heat transfer efficiency and system stability;
  \item $x_9$ and $x_{10}$: Compression ratio and compressor efficiency, which determine the energy conversion characteristics during the compression process;
  \item $x_{11}$ and $x_{12}$: Temperature parameters, reflecting the temperature differential control during the heat exchange process;
  \item $x_{13}$ and $x_{14}$: Flow control parameters, used to regulate the flow rate of cooling water or refrigerant, balancing cooling performance and energy consumption.
\end{itemize}

The mathematical modeling of this problem is as follows:\par
\begin{flushleft}
    \textit{Variable:}
\end{flushleft}
\vspace{-\baselineskip}
\begin{flushleft}
    \[
        \begin{aligned}
            x &= [x_1, x_2, x_3, x_4, x_5, x_6, x_7, x_8, x_9, x_{10}, x_{11}, x_{12}, x_{13}, x_{14}]
        \end{aligned}
    \]
\end{flushleft}

\begin{flushleft}
    \textit{Minimize:}
\end{flushleft}
\vspace{-\baselineskip}
\begin{equation}
        y = f(x)+punishment
\end{equation}

\begin{flushleft}
    \textit{Subject to:}
\end{flushleft}
\begin{equation}
    g_1 = \frac{1.524}{x_7} - 1 \leq 0;
\end{equation}

\begin{equation}
    g_2 = \frac{1.524}{x_8} - 1 \leq 0;
\end{equation}

\begin{equation}
    g_3 = 0.07789 \cdot x_1 - \frac{2 \cdot x_9}{x_7} - 1 \leq 0;
\end{equation}

\begin{equation}
    g_4 = \frac{7.05305 \cdot x_1^2 \cdot x_{10}}{x_9 \cdot x_8 \cdot x_2 \cdot x_{14}} - 1 \leq 0;
\end{equation}

\begin{equation}
    g_5 = \frac{0.0833 \cdot x_{14}}{x_{13}} - 1 \leq 0;
\end{equation}

\begin{equation}
\begin{aligned}
g_6 =\;& \frac{47.136\,x_2^{0.333}\,x_{12}}{x_{10}}
      - 1.333\,x_8\,x_{13}^{2.1195} \\
     & + \frac{62.08\,x_{13}^{2.1195}\,x_8^{0.2}}{x_{12}\,x_{10}}
      - 1 \leq 0;
\end{aligned}
\end{equation}

\begin{equation}
    g_7 = 0.04771 \cdot x_{10} \cdot x_8^{1.8812} \cdot x_{12}^{0.3424} - 1 \leq 0;
\end{equation}

\begin{equation}
    g_8 = 0.0488 \cdot x_9 \cdot x_7^{1.893} \cdot x_{11}^{0.316} - 1 \leq 0; 
\end{equation}

\begin{equation}
    g_9 = \frac{0.0099 \cdot x_1}{x_3} - 1 \leq 0;
\end{equation}

\begin{equation}
    g_{10} = \frac{0.0193 \cdot x_2}{x_4} - 1 \leq 0;
\end{equation}

\begin{equation}
    g_{11} = \frac{0.0298 \cdot x_1}{x_5} - 1 \leq 0;
\end{equation}

\begin{equation}
    g_{12} = \frac{0.056 \cdot x_2}{x_6} - 1 \leq 0;
\end{equation}

\begin{equation}
    g_{13} = \frac{2}{x_9} - 1 \leq 0;
\end{equation}

\begin{equation}
    g_{14} = \frac{2}{x_{10}} - 1 \leq 0;
\end{equation}

\begin{equation}
    g_{15} = \frac{x_{12}}{x_{11}} - 1 \leq 0;
\end{equation}

\begin{flushleft}
    \textit{Where:}
\end{flushleft}

\begin{flushleft}
    \[
        \begin{aligned}
        punishment=10^3 \cdot \sum_{i=1}^{15} \max{(0, g_i)^2}
        \end{aligned}
\]
\end{flushleft}

\begin{equation}
\begin{aligned}
f(x) =\;& 63098.88\,x_2 x_4 x_{12} 
+ 5441.5\,x_2^2 x_{12} \\
&+ 115055.5\,x_2^{1.664} x_6 
+ 6172.27\,x_2^2 x_6 \\
&+ 63098.88\,x_1 x_3 x_{11} 
+ 5441.5\,x_1^2 x_{11} \\
&+ 115055.5\,x_1^{1.664} x_5 
+ 6172.27\,x_1^2 x_5 \\
& + 140.53\,x_1 x_{11} 
+ 281.29\,x_3 x_{11} \\
&+ 70.26\,x_1^2 
+ 281.29\,x_1 x_3 
+ 281.29\,x_3^2 \\
& + 14437\,x_8^{1.8812} x_{12}^{0.3424} x_{10} x_1^2 
    \frac{x_7}{x_{14} x_9}\\
& + 20470.2\,x_7^{2.893} x_{11}^{0.316} x_{12}
\end{aligned}
\end{equation}

\begin{flushleft}
    \textit{Variable range:}
\end{flushleft}
\vspace{-\baselineskip}
\begin{flushleft}
    \[
        0.001 < x_1 < 5; \quad 0.001 < x_2 < 5; 
    \]
\end{flushleft}
\vspace{-\baselineskip}
\begin{flushleft}
    \[
       0.001 < x_3 < 5; \quad 0.001 < x_4 < 5;
    \]
\end{flushleft}
\vspace{-\baselineskip}
\begin{flushleft}
    \[
        0.001 < x_5 < 5; \quad 0.001 < x_6 < 5;
    \]
\end{flushleft}
\vspace{-\baselineskip}
\begin{flushleft}
    \[
       0.001 < x_7 < 5; \quad 0.001 < x_8 < 5; 
    \]
\end{flushleft}
\vspace{-\baselineskip}
\begin{flushleft}
    \[
        0.001 < x_9 < 5; \quad 0.001 < x_{10} < 5; 
    \]
\end{flushleft}
\vspace{-\baselineskip}
\begin{flushleft}
    \[
       0.001 < x_{11} < 5; \quad 0.001 < x_{12} < 5;
    \]
\end{flushleft}
\vspace{-\baselineskip}
\begin{flushleft}
    \[
         0.001 < x_{13} < 5; \quad 0.001 < x_{14} < 5;
    \]
\end{flushleft}

\begin{figure}[htbp]
    \centering
    \includegraphics[width=0.8\textwidth]{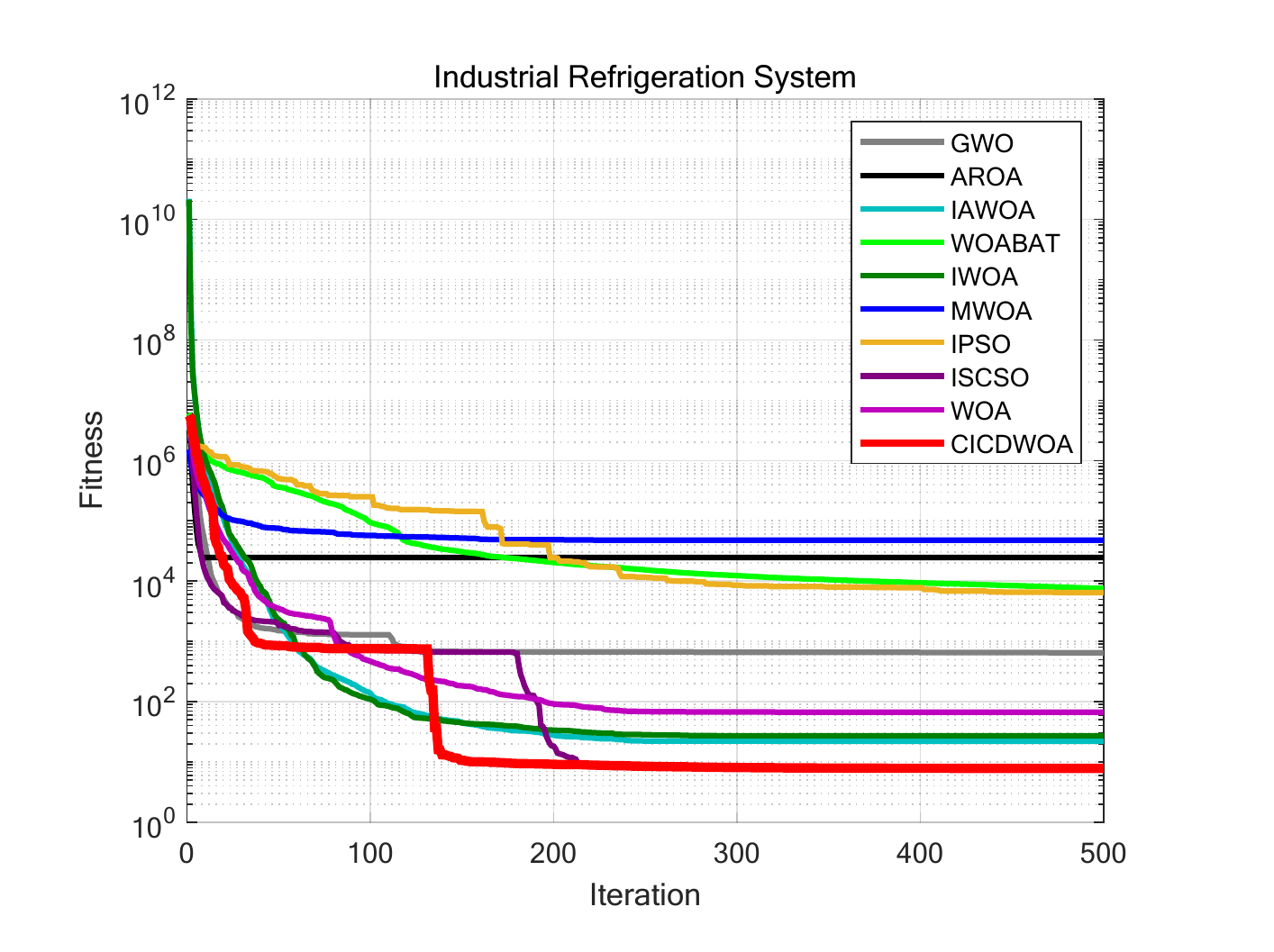}
    \caption{Iteration curves of the algorithms in solving the Industrial Refrigeration System design problem.} 
    \label{engineering102}
\end{figure}

The experimental results are shown in Figure~\ref{engineering102} and Table~\ref{engineering_metrics}. As indicated in Table~\ref{engineering_metrics}, CICDWOA demonstrates significantly better stability in solving the Industrial Refrigeration System design problem compared to other algorithms, and its optimization accuracy is the best among all algorithms. This illustrates the substantial advantage of CICDWOA in handling such optimization problems.\par
From an overall perspective, CICDWOA demonstrates excellent optimization performance and significant stability advantages across all test problems. It achieves global or near-global optimal objective values in all cases, with nearly zero standard deviation, indicating that the algorithm can consistently converge to the optimal solution across multiple independent runs. This reflects the high robustness and search consistency of CICDWOA.\par

\subsection{Results}
In the Sawmill Operation Problem, CICDWOA achieved the lowest average objective value of 37192.612702 and the smallest standard deviation of 0.047364, outperforming nine other algorithms. This suggests that the algorithm exhibits strong convergence precision and global optimality in resource allocation optimization for complex production systems.\par
For the five structural optimization problems, Multiple Disk Clutch Brake, Three-Bar Truss, Tension/Compression Spring, Speed Reducer, and I-beam, CICDWOA's results are nearly identical to the theoretical optimal values, with zero standard deviation. This demonstrates the algorithm's outstanding stability and constraint handling capability in high-constrained, continuous optimization problems.\par
In the Reactor Network problem, CICDWOA significantly outperforms all comparison algorithms, achieving an average objective value of just 0.00001, while the error for the second-best algorithm, AROA, is two orders of magnitude higher. This result effectively validates CICDWOA's strong global exploration ability in nonlinear chemical process optimization.\par
For the Piston Lever problem, traditional algorithms generally suffer from local optima, while CICDWOA achieves a much better optimal result of 1.057179, far outperforming other algorithms (e.g., MWOA with 216.408, IPSO with 583.316), with almost zero standard deviation. This highlights the algorithm's exceptional stability in highly nonlinear mechanical system design.\par
In the Gas Transmission Compressor problem, CICDWOA's standard deviation is zero, clearly outperforming other algorithms and demonstrating its high-precision solving capability in large-scale industrial gas transmission system optimization.\par
Finally, in the Industrial Refrigeration System optimization, CICDWOA achieves the lowest energy consumption objective value of 7.831789, significantly lower than GWO (649.636351) and AROA (24620.83798), with a standard deviation of only 0.062125. This indicates that the algorithm has remarkable robustness and stable convergence characteristics in high-dimensional, non-convex energy consumption optimization problems.\par
CICDWOA achieves theoretical optimal or near-optimal results in all ten engineering design optimization problems. The algorithm yields the lowest average objective values and the smallest standard deviations, fully reflecting its significant advantages in terms of accuracy, stability, and convergence speed. Therefore, the experimental results demonstrate that CICDWOA not only maintains strong global search capabilities on standard benchmark functions but also exhibits outstanding stability, precision, and versatility in engineering design optimization problems. It effectively addresses practical engineering optimization tasks characterized by complex constraints and nonlinear features.

\begin{table*}[htbp]
\centering
\caption{Comparison results of the algorithms in solving engineering design optimization problems.}
\resizebox{\textwidth}{!}{
\begin{tabular}{cccccccccccc}
\toprule
\textbf{Problem} & \textbf{Metrics} & \textbf{GWO} & \textbf{AROA} & \textbf{IAWOA} & \textbf{WOABAT} & \textbf{IWOA} & \textbf{MWOA} & \textbf{IPSO} & \textbf{ISCSO} & \textbf{WOA} & \textbf{CICDWOA} \\
\midrule
Sawmill Operation Problem & Ave & 37208.771992 & 41373.594284 & 49418.119834 & 49054.912341 & 47859.641862 & 44971.139655 & 51053.190612 & 38465.953626 & 43973.576609 & \textbf{37192.612702} \\
 & Std & 12.196434 & 2662.740225 & 5776.401522 & 5548.235581 & 7158.504076 & 1954.456937 & 5842.922027 & 3728.754830 & 5357.644767 & \textbf{0.047364} \\
Multiple Disk Clutch Brake & Ave & 0.235299 & 0.236364 & \textbf{0.235242} & \textbf{0.235242} & \textbf{0.235242} & 0.308785 & 0.241350 & 0.235273 & 0.235246 & \textbf{0.235242} \\
 & Std & 0.000050 & 0.001687 & \textbf{0.000000} & \textbf{0.000000} & \textbf{0.000000} & 0.014214 & 0.017043 & 0.000034 & 0.000018 & \textbf{0.000000} \\
Reactor Network & Ave & 0.093557 & 12.198257 & 0.415237 & 0.270835 & 0.519004 & 42.770049 & 2.179705 & 0.139222 & 1.128079 & \textbf{0.000010} \\
 & Std & 0.131752 & 17.010564 & 0.246738 & 0.111177 & 0.325013 & 46.548289 & 5.860215 & 0.184477 & 1.315195 & \textbf{0.000000} \\
Three Bar Truss & Ave & 259.805061 & 259.815288 & 259.860334 & 259.805047 & 259.900304 & 260.031721 & 262.034823 & 259.805056 & 259.896730 & \textbf{259.805047} \\
 & Std & 0.000014 & 0.023638 & 0.055828 & 0.000000 & 0.118605 & 0.269837 & 2.701473 & 0.000010 & 0.145243 & \textbf{0.000000} \\
Tension/Compression Spring & Ave & 0.121525 & 0.123384 & 0.122038 & 0.125751 & 0.121523 & 7.085268 & 0.121545 & 0.121523 & 0.121594 & \textbf{0.121522} \\
 & Std & 0.000004 & 0.004391 & 0.002529 & 0.010059 & 0.000001 & 31.761084 & 0.000025 & 0.000001 & 0.000392 & \textbf{0.000000} \\
Speed Reducer & Ave & 2638.848555 & 2639.777531 & 2639.509734 & 2639.430147 & 2639.509771 & 2699.232307 & 2655.066568 & 2638.877227 & 2639.509795 & \textbf{2638.819843} \\
 & Std & 0.022406 & 1.155440 & 3.778183 & 1.918549 & 3.778176 & 33.199205 & 6.708155 & 0.054271 & 3.778172 & \textbf{0.000036} \\
I-beam & Ave & 6.702622 & 5.899500 & 5.750464 & 4.686197 & 6.215656 & 5.106279 & 6.701612 & 6.702866 & 6.268948 & \textbf{6.703048} \\
 & Std & 0.000348 & 1.000864 & 1.104984 & 1.998061 & 0.670386 & 1.387003 & 0.001288 & 0.000195 & 0.408080 & \textbf{0.000000} \\
Piston Lever & Ave & 23.303077 & 255.866678 & 66.070314 & 374.878604 & 104.153610 & 216.408246 & 583.316036 & 28.828503 & 27.229055 & \textbf{1.057179} \\
 & Std & 57.674861 & 172.413456 & 105.558816 & 431.000595 & 148.488637 & 129.025307 & 1015.982234 & 63.157492 & 71.048743 & \textbf{0.000004} \\
Gas Transmission Compressor & Ave & 1224745.990 & 1226699.247 & 1224745.937 & 1224745.937 & 1224745.937 & 1287506.115 & 1225304.808 & 1224745.960 & 1224745.937 & \textbf{1224745.937} \\
 & Std & 0.059697 & 5166.16834 & 0.000007 & 0.000031 & 0.000017 & 33136.46341 & 461.973714 & 0.02241 & 0.000011 & \textbf{0.000000} \\
Industrial Refrigeration System & Ave & 649.636351 & 24620.83798 & 22.127301 & 7537.757992 & 27.035287 & 47680.19452 & 6447.80299 & 7.879989 & 66.801153 & \textbf{7.831789} \\
 & Std & 3515.357417 & 19943.89744 & 23.658647 & 6928.20809 & 29.963414 & 73379.81632 & 11805.51147 & 0.154495 & 112.177118 & \textbf{0.062125} \\
\bottomrule
\end{tabular}}
\label{engineering_metrics}
\end{table*}

\section{Discussion}
From both theoretical and experimental perspectives, the outstanding advantages of CICDWOA are primarily manifested in the following three aspects:\par
(a) Significant Enhancement of Global Search Capability: Through the Collective Cognitive Sharing mechanism, CICDWOA establishes dynamic information exchange and global knowledge collaboration among individuals within the population, combining the characteristics of individual exploration and collective cooperation in the search process. This mechanism effectively prevents issues such as premature convergence and loss of population diversity, which are common in traditional WOA, thereby significantly enhancing the algorithm's ability to escape local optima; (b) Improved Local Development Accuracy and Convergence Stability: The introduced Cauchy Inverse Cumulative Distribution disturbance provides nonlinear random perturbations during the local search phase, enabling individuals to maintain search activity even when approaching the optimal solution. This ensures higher convergence accuracy while maintaining global exploration capabilities; (c) Strong Cross-Domain Adaptability and Robustness: Experimental results demonstrate that CICDWOA maintains excellent performance across continuous, multimodal, and high-dimensional optimization problems. It is not only applicable to path planning and engineering optimization but also extendable to complex tasks such as machine learning and feature selection, showcasing its strong generalization potential.\par
However, while CICDWOA offers performance improvements, it also introduces certain computational overhead. Due to the need to perform multiple operations during each iteration (including global information sharing, collaborative updates, and Cauchy perturbation calculations), its computational complexity and runtime are higher compared to the traditional WOA. In standard benchmark tests, CICDWOA's average computational time is approximately 15\%-25\% longer than that of the original WOA. Nonetheless, in most practical engineering and intelligent optimization tasks, this time overhead is acceptable, and when compared to the substantial performance gains in accuracy and stability, it can be considered a negligible cost.\par
In conclusion, CICDWOA significantly surpasses traditional WOA and other advanced algorithms in terms of optimization performance, stability, and adaptability. By achieving a high balance between exploration and exploitation through collective cognitive sharing and nonlinear perturbations, it provides a new research direction and theoretical foundation for solving complex optimization problems such as multi-UAV cooperative path planning, multi-objective optimization, and high-dimensional constrained system optimization.

\section{Conclusion}
This paper proposes a Collective Cognitive Sharing Whale Optimization Algorithm with Cauchy Inverse Cumulative Distribution (CICDWOA), aimed at enhancing the global optimization capability and stability of the traditional Whale Optimization Algorithm (WOA). CICDWOA uses Good Nodes Set method for initialization, incorporates Collective Cognitive Sharing mechanism, Adaptive Exponential Spiral (AES) mechanism, Enhanced Spiral Updating mechanism with CICDWOA and dynamic inertia weight, and Hybrid Gaussian-Cauchy Mutation based on DE. Meanwhile, a nonlinear update method for convergence factor $a$ is designed for balancing the exploration and exploitation. To comprehensively evaluate the performance of CICDWOA, systematic experiments and assessments were conducted from five perspectives: benchmark function testing, two-dimensional robot path planning, three-dimensional UAV path planning and engineering design optimization.\par
In the testing of 23 standard benchmark functions, CICDWOA achieved a leading average Friedman value of 1.6790, significantly outperforming various classical and improved metaheuristic algorithms, including GWO, AROA, IWOA, MWOA, and WOABAT. The algorithm demonstrated faster convergence, higher solution accuracy, and stronger global exploration ability in solving uni-modal, multi-modal, and combinatorial complex functions. The results of the Wilcoxon rank-sum test further indicate that the performance differences between CICDWOA and other algorithms are statistically significant across most test functions.\par
In the two-dimensional robot path planning task, CICDWOA achieved optimal or near-optimal paths under different obstacle densities, with the shortest average path length and the lowest standard deviation, demonstrating exceptional global convergence and interference resistance. Particularly in high-density obstacle environments, traditional algorithms (such as AROA, IWOA, MWOA) tend to get trapped in local optima or fail, whereas CICDWOA maintained smooth and optimal path generation capabilities.\par
In the three-dimensional UAV path planning experiment, CICDWOA achieved the best performance with an average value of 76.95 and a standard deviation of 1.83, producing shorter paths with smoother turns, successfully meeting obstacle avoidance and dynamic constraints. This demonstrates the algorithm's practicality and stability in solving complex spatial optimization problems.\par
Finally, in the engineering design optimization section, CICDWOA obtained global or near-optimal solutions for ten typical continuous and highly constrained engineering problems, with a nearly zero standard deviation. The algorithm exhibited excellent solving ability and robustness across various fields such as mechanical design, chemical process optimization, and energy system design. Especially in nonlinear complex systems like the Reactor Network and Industrial Refrigeration System, CICDWOA's objective values were significantly lower than those of other algorithms, validating its powerful search capability and stable convergence characteristics in high-dimensional non-convex spaces.\par
The experimental results thoroughly validate the superior performance of CICDWOA across a wide range of optimization tasks. The algorithm achieves fast and stable convergence while maintaining strong global exploration capabilities, combining precision, robustness, and broad applicability. It provides an efficient and reliable metaheuristic optimization framework for solving complex real-world optimization problems.\par

\section{Acknowledgements}
The supports provided by Macao Polytechnic University (RP/FCA-01/2025) and Macao Science and Technology Development Fund (FDCT-MOST: 0018/2025/AMJ) enabled us to conduct data collection, analysis, and interpretation, as well as cover expenses related to research materials and participant recruitment. MPU and FDCT investment in our work (MPU submission code: fca.36c2.d726.5) have significantly contributed to the quality and impact of our research findings.

\appendix
\section{Abbreviations}
Abbreviations and their corresponding full names used in this paper are listed in Table \ref{tab:abbreviations}.
\begin{table}[!t]
\centering
\caption{List of abbreviations and their corresponding full names used in this paper.}
\label{tab:abbreviations}
\begin{tabular}{@{}ll@{}}
\toprule
\textbf{Full name} & \textbf{Abbreviation} \\
\midrule
Metaheuristic Algorithm & MA \\
State-of-the-art & SOTA \\
Whale Optimization Algorithm & WOA \\
Differential Evolution & DE \\
Grey Wolf Optimizer & GWO \\
Attraction--Repulsion Optimization Algorithm & AROA \\
Good Nodes Set & GNS \\
Collective Cognitive Sharing & CCS \\
Adaptive Exponential Spiral & AES \\
Average fitness & Ave \\
Standard deviation & Std \\
Overall Effectiveness & OE \\
\bottomrule
\end{tabular}
\end{table}

\section{Code Availability}
\begin{itemize}
    \item The code of CICDWOA are available on \href{URL}{https://github.com/JunhaoWei-mpu/ROBIS-Lab/tree/CICDWOA}.
    \item To facilitate the experiments in this study, we employed a set of Standard Benchmark Functions. The corresponding modeling data has been made publicly available on Figshare:\par \href{URL}{https://figshare.com/articles/dataset/Get\_F\_m/28440863?file=52488527}, allowing readers to access, reference, and analyze it further.
    \item The modeling details for the engineering optimization problems used in this paper are also available on Figshare:\par \href{URL}{https://figshare.com/articles/thesis/engineering\_m/28673777}, provided to support replication and additional investigation by readers.
\end{itemize}

\section{Benchmark Functions}
The details of the benchmark functions used in this research are shown in Table~\ref{funcdetails}.

\begin{table}[htbp]
    \centering
    \caption{Standard Benchmark Functions~\cite{CEC} \cite{ESTGWOA}.}
    \resizebox{\linewidth}{!}{
    \begin{tabular}{l c c c c}
    \hline
    Function & Function's Name &  Best Value \\
    \hline
    $F_1\left(x\right)=\sum_{i=1}^nx_i^2$ & Sphere &  0 \\
    $\begin{aligned}&F_{2}\left(x\right)=\sum_{i=1}^{n}\left|x_{i}\right|+\prod_{i=1}^{n}|x_{i}|\end{aligned}$ & Schwefel's Problem 2.22 & 0 \\
    $F_3\left(x\right)=\sum_{i=1}^n\left(\sum_{j-1}^ix_j\right)^2$ & Schwefel's Problem 1.2 & 0 \\
    $\begin{aligned}
        F_{4}(x) &= \max\limits_{1 \leq i \leq n} \left\{ \left| x_i \right| \right\}
        \end{aligned}$ & Schwefel's Problem 2.21 &  0 \\
    $F_{5}(x)=\sum_{i=1}^{n-1}[100(x_{i+1}-x_{i}^{2})^{2}+(x_{i}-1)^{2}]$ & Generalized Rosenbrock's Function &  0 \\
    $F_{6}(x)=\sum_{i=1}^{n}(\lfloor x_{i}+0.5\rfloor)^{2}$ & Step Function &  0 \\
    $F_{7}(x)=\sum_{i=1}^{n}ix_{i}^{4}+random[0,1)$ & Quartic Function &  0 \\
    $F_{8}(x)=\sum_{i=1}^{n}-x_{i}\sin(\sqrt{|x_{i}|})$ & Generalized Schwefel's Function &  -12569.5 \\
    $F_{9}(x)=\sum_{i=1}^{n}[x_{i}^{2}-10\cos(2\pi x_{i})+10)]$ & Generalized Rastrigin's Function &  0 \\
    $\begin{aligned}F_{10}\left(x\right)&=-20\exp\left(-0.2\sqrt{\frac{1}{n}\sum_{i=1}^{n}x_{i}^{2}}\right)-\exp\left(\frac{1}{n}\sum_{i=1}^{n}\cos2\pi x_{i}\right)\\&+20+e\end{aligned}$ & Ackley's Function &  0 \\
    $F_{11}\left(x\right)=\frac{1}{4000}\sum_{i=1}^{n}x_{i}^{2}-\prod_{i=1}^{n}\cos\left(\frac{x_{i}}{\sqrt{i}}\right)+1$ & Generalized Griewank's Function &  0 \\
    $\begin{aligned}F_{1 2}\left(x\right)&=\frac{\pi}{n}\left\{10 \sin^{2}(\pi y_{i})+\sum_{i=1}^{n-1}(y_{i}-1)^{2}[1+10 \sin^{2}(\pi y_{i+1})]\right.\\&+(y_{n}-1)^{2}\Big\}+\sum_{i=1}^{n}u(x_{i},10,100,4),\\&y_i=1+\frac{1}{4}(x_i+1)\\&u\left(x_{i},a,k,m\right)=\left\{\begin{array}{ll}k(x_{i}-a)^{m},&x_{i}>a,\\0,&-a\leq x_{i}\leq a,\\k(-x_{i}-a)^{m},&x_{i}<-a.\end{array}\right.\end{aligned}$ & Generalized Penalized Function 1 &  0 \\
    $\begin{gathered}F_{13}\left(x\right)=0.1\left\{\sin^{2}(3\pi x_{1})+\sum_{i=1}^{n-1}(x_{i}-1)^{2}[1+sin^{2}(3\pi x_{i+1})]\right.\\+(x_{n}-1)^{2}[1+\sin^{2}(2\pi x_{n})]\big\}+\sum_{i=1}^{n}u(x_{i},5,100,4)\end{gathered}$ & Generalized Penalized Function 2 &  0 \\
    $F_{14}\left(x\right)=\left[\frac{1}{500}+\sum_{j=1}^{25}\frac{1}{j+\sum_{i=1}^{2}(x_{i}-a_{ij})^{6}}\right]^{-1}$ & Shekel's Foxholes Function &  0.998 \\
    $F_{15}\left(x\right)=\sum_{i=1}^{11}\left[a_{i}-\frac{x_{1}\left(b_{i}^{2}+b_{i}x_{2}\right)}{b_{i}^{2}+b_{i}x_{3}+x_{4}}\right]^{2}$ & Kowalik's Function &  0.0003075 \\
    $F_{16}\left(x\right)=4x_{1}^{2}-2.1x_{1}^{4}+\frac{1}{3}x_{1}^{6}+x_{1}x_{2}-4x_{2}^{2}+4x_{2}^{4}$ & Six-Hump Camel-Back Function &  -1.0316 \\
    $F_{17}\left(x\right)=\left(x_{2}-\frac{5.1}{4\pi^{2}}x_{1}^{2}+\frac{5}{\pi}x_{1}-6\right)^{2}+10\left(1-\frac{1}{8\pi}\right)\cos x_{1}+10$ & Branin Function &  0.398 \\
    $\begin{aligned}F_{18}\left(x\right)&=[1+(x_{1}+x_{2}+1)^{2}(19-14x_{1}+3x_{1}^{2}-14x_{2}\\&+6x_1x_2+3x_2^2)]\times[30+(2x_1-3x_2)^2(18-32x\\&+12x_1^2+48x_2-36x_1x_2+27x_2^2)]\end{aligned}$ & Goldstein-Price Function &  3 \\
    $F_{19}\left(x\right)=-\sum_{i=1}^{4}c_{i}\exp\left[-\sum_{j=1}^{4}a_{ij}\left(x_{j}-p_{ij}\right)^{2}\right]$ & Hartman's Function 1 &  -3.8628 \\
    $F_{20}\left(x\right)=-\sum_{i=1}^{4}c_{i}\exp\left[-\sum_{j=1}^{6}a_{ij}(x_{j}-p_{ij})^{2}\right]$ & Hartman's Function 2 &  -3.32 \\
    $F_{21}\left(x\right)=-\sum_{i=1}^{5}\left[(x-a_{i})(x-a_{i})^{T}+c_{i}\right]^{-1}$ & Shekel's Function 1 &  -10.1532 \\
    $F_{22}\left(x\right)=-\sum_{i=1}^{7}\left[(x-a_{i})(x-a_{i})^{T}+c_{i}\right]^{-1}$ & Shekel's Function 2 &  -10.4029 \\
    $F_{23}\left(x\right)=-\sum_{i=1}^{11}\left[(x-a_{i})(x-a_{i})^{T}+c_{i}\right]^{-1}$ & Shekel's Function 3 &  -10.5364 \\
    \hline
    \label{funcdetails}
    \end{tabular}}
\end{table}

\clearpage

\bibliographystyle{unsrt}  
\bibliography{references}

\end{document}